\documentclass[nofootinbib,preprint,tightenlines,superscriptaddress]{revtex4}

\usepackage[utf8]{inputenc}
\usepackage{amssymb,amsthm,amsmath,amstext,amsbsy}
\usepackage{url}
\usepackage{nicefrac}
\usepackage{graphicx}
\usepackage{hyperref}
\usepackage{xspace}
\usepackage{slashed}

\newcommand{\ie}{\textit{i.e.}}
\newcommand{\eg}{\textit{e.g.}}
\newcommand{\cf}{\textit{cf.}}
\newcommand{\apriori}{\textit{a priori}}
\newcommand{\aposteriori}{\textit{a posteriori}}

\newcommand{\etal}{\textit{et al.}}
\newcommand{\perse}{\textit{per se}}

\newcommand{\mathspace}{\ \ }
\newcommand{\mathtext}[1]{\mathspace\text{#1}\mathspace}

\newcommand{\MeV}{\ensuremath{\mathrm{MeV}}}

\newcommand{\fm}{\ensuremath{\mathrm{fm}}}

\newcommand{\vk}{\mathbf{k}}
\newcommand{\vp}{\mathbf{p}}
\newcommand{\vq}{\mathbf{q}}
\newcommand{\vP}{\mathbf{P}}

\newcommand{\vZero}{\mathbf{0}}

\newcommand{\vecr}{\mathbf{r}}
\newcommand{\vecp}{\mathbf{p}}
\newcommand{\vecq}{\mathbf{q}}
\newcommand{\veck}{\mathbf{k}}

\newcommand{\dd}{\mathrm{d}}

\newcommand{\dq}[1]{\!\!\frac{\mathrm{d}^3#1}{(2\pi)^3}}

\newcommand{\dn}[1]{\!\frac{\mathrm{d}#1_0}{2\pi}}

\newcommand{\dfq}[1]{\!\!\frac{\mathrm{d}^4#1}{(2\pi)^4}}

\newcommand{\gsim}{\gtrsim}

\newcommand{\ii}{\mathrm{i}}
\newcommand{\ee}{\mathrm{e}}
\newcommand{\vD}{\boldsymbol{D}}
\newcommand{\hc}{\mathrm{h.c.}}
\newcommand{\OO}{\mathcal{O}}

\newcommand{\eps}{\varepsilon}
\newcommand{\one}{\mathbf{1}}

\newcommand{\diag}{\mathrm{diag}}

\newcommand{\vdelta}{\delta^{(3)}}
\newcommand{\fvdelta}{\delta^{(4)}}

\newcommand{\bra}[1]{\langle #1|}

\newcommand{\ket}[1]{|#1\rangle}

\newcommand{\mbraket}[3]{\langle #1|#2|#3\rangle}

\newcommand{\EB}{E_\mathrm{B}}

\newcommand{\MN}{M_N}

\newcommand{\Mpi}{M_\pi}

\newcommand{\yd}{y_d}
\newcommand{\yt}{y_t}
\newcommand{\ydt}{y_{d,t}}
\newcommand{\sigmad}{\sigma_d}
\newcommand{\sigmat}{\sigma_t}
\newcommand{\sigmadt}{\sigma_{d,t}}
\newcommand{\gamd}{\gamma_d}
\newcommand{\gamt}{\gamma_t}
\newcommand{\rd}{\rho_d}
\newcommand{\rnt}{\rho_t}
\newcommand{\rnC}{r_C}

\newcommand{\LO}{\text{LO}\xspace}
\newcommand{\NLO}{\text{NLO}\xspace}
\newcommand{\NNLO}{\text{N$^2$LO}\xspace}
\newcommand{\NNNLO}{\text{N$^3$LO}\xspace}

\newcommand{\sss}{\mathrm{s}}
\newcommand{\ccc}{\mathrm{c}}
\newcommand{\fff}{\mathrm{full}}
\newcommand{\ddd}{\mathrm{diff}}

\newcommand{\Tgen}{\mathcal{T}}
\newcommand{\TS}{\Tgen_\sss}
\newcommand{\TC}{\Tgen_\ccc}
\newcommand{\TF}{\Tgen_\fff}
\newcommand{\TFf}{\Tgen_{\fff'}}

\newcommand{\TSq}{\TS^\mathrm{q}}

\newcommand{\TFq}{\TF^\mathrm{q}}
\newcommand{\TSd}{\TS^\mathrm{d}}

\newcommand{\TSda}{\TS^\mathrm{d,a}}
\newcommand{\TSdb}{\TS^\mathrm{d,b}}
\newcommand{\TSdsa}{\TS^\mathrm{d,c}}
\newcommand{\TSdsb}{\TS^\mathrm{d,d}}
\newcommand{\TFda}{\TF^\mathrm{d,a}}
\newcommand{\TFdb}{\TF^\mathrm{d,b1}}
\newcommand{\TFdc}{\TF^\mathrm{d,b2}}
\newcommand{\TFfda}{\TFf^\mathrm{d,a}}
\newcommand{\TFfdb}{\TFf^\mathrm{d,b1}}
\newcommand{\TFfdc}{\TFf^\mathrm{d,b2}}

\newcommand{\Bgen}{\mathcal{B}}
\newcommand{\BS}{\Bgen_\sss}
\newcommand{\BSda}{\BS^\mathrm{d,a}}
\newcommand{\BSdb}{\BS^\mathrm{d,b1}}
\newcommand{\BSdc}{\BS^\mathrm{d,b2}}

\newcommand{\KS}{K_\sss}
\newcommand{\KC}{K_\ccc}

\newcommand{\KCd}{\KC^{(d)}}
\newcommand{\KCt}{\KC^{(t)}}
\newcommand{\KCdt}{\KC^{(d,t)}}

\newcommand{\deltaC}{\delta_\ccc}
\newcommand{\deltaF}{\delta_\fff}

\newcommand{\ZT}{Z_{\mathcal{T}}}
\newcommand{\ZWF}{Z_{\mathcal{B}}}

\newcommand{\sktilde}[1]{\widetilde{#1}}

\newcommand{\skrowspace}{0.5em}

\newcommand{\LamNoPi}{\Lambda_{\slashed\pi}}

\begin{document}

\title{The proton--deuteron system in pionless EFT revisited}

\author{Sebastian König}
\email{koenig.389@physics.osu.edu}
\affiliation{Department of Physics, The Ohio State University, Columbus, Ohio 
43210, USA}
\affiliation{Helmholtz-Institut für Strahlen- und Kernphysik (Theorie)\\
and Bethe Center for Theoretical Physics, Universität Bonn, 53115 Bonn,
Germany}

\author{Harald W. Grießhammer}
\email{hgrie@gwu.edu; permanent address: $3$}
\affiliation{Institute for Nuclear Studies, Department of 
Physics, George Washington University, Washington DC 20052, USA}
\affiliation{Forschungszentrum Jülich, D-52425 Jülich, Germany}

\author{H.-W. Hammer}
\affiliation{Institut für Kernphysik, Technische Universität Darmstadt, 
64289 Darmstadt, Germany}
\affiliation{ExtreMe Matter Institute EMMI, GSI Helmholtzzentrum 
für Schwerionenforschung GmbH, 64291 Darmstadt, Germany}
\affiliation{Helmholtz-Institut für Strahlen- und Kernphysik (Theorie)\\
and Bethe Center for Theoretical Physics, Universität Bonn, 53115 Bonn,
Germany}

\date{\today}

\begin{abstract}
We provide a detailed discussion of the low-energy proton--deuteron system in 
pionless effective field theory, considering both the spin-quartet and doublet 
S-wave channels.  Extending and amending our previous work on the subject, we 
calculate the $^3$He--$^3$H binding energy difference both perturbatively 
(using properly normalized trinucleon wave functions) and non-perturbatively by 
resumming all $\OO(\alpha)$ Coulomb diagrams in the doublet channel.  Our 
nonperturbative result agrees well with a calculation that involves the full 
off-shell Coulomb T-matrix.  Carefully examining the cutoff-dependence in the 
doublet channel, we present numerical evidence for a new three-nucleon 
counterterm being necessary at next-to-leading order if Coulomb effects are 
included.  Indeed, such a term has recently been identified analytically.  We 
furthermore make a case for a simplified Coulomb power counting that is 
consistent throughout the bound-state and scattering regimes.  Finally, using a 
``partially screened'' full off-shell Coulomb T-matrix, we investigate the 
importance of higher-order Coulomb corrections in low-energy quartet-channel 
scattering.
\end{abstract}

\maketitle

\section{Introduction}

Effective field theories are powerful tools that can be used to carry out 
calculations in a formalism involving directly the relevant degrees of 
freedom for the physical system under consideration.  In particular, in nuclear 
systems at very low energies and momenta, pion-exchange effects cannot be 
resolved and one can hence use the so-called \emph{pionless effective field 
theory}.  This approach only includes short-range contact interactions between 
nucleons~\cite{Kaplan:1998we,vanKolck:1998bw,Chen:1999tn} and is constructed to 
reproduce the effective range expansion~\cite{Bethe:1949yr} in the two-body 
system.

The applicability of this approach, which recovers Efimov's universal approach 
to the three-nucleon problem~\cite{Efimov:1981aa,Hammer:2010kp}, stems from the 
experimental fact that the S-wave nucleon--nucleon scattering lengths in both 
the $^3S_1$ (isospin 0) channel, $a_d\approx5.42~\fm$, and in the $^1S_0$ 
(isospin 1) channel, $a_t\approx-23.71~\fm$, are significantly larger than the 
range of interaction of about $1.4~\fm$ set by the inverse pion mass.  The 
corresponding effective ranges, on the other hand, have the values $1.75$ and 
$2.73~\fm$, respectively, and are thus of the expected natural order of 
magnitude~\cite{deSwart:1995ui,Beane:2000fx}.  In 
Refs.~\cite{Bedaque:1997qi,Bedaque:1998mb,Bedaque:1999ve,Gabbiani:1999yv,
Hammer:2001gh,Bedaque:2002yg}, the formalism has been extended to the 
three-nucleon sector.  The situation there is particularly interesting because 
the triton can be interpreted as an approximate Efimov state.  Recently, a fully 
perturbative calculation of neutron--deuteron scattering up to 
next-to-next-to-leading order has been carried out by 
J.~Vanasse~\cite{Vanasse:2013sda}, using a novel technique that circumvents the 
(numerically expensive) calculation of full off-shell quantities.

Since most low-energy nuclear experiments involve more than one charged 
particle (proton), it is very important to discuss the treatment of Coulomb 
effects.  Although this interaction can be treated as a perturbative correction 
for intermediate and higher energies, it becomes strong close to threshold and 
has to be treated nonperturbatively there.  In the two-nucleon sector this was 
first discussed by Kong and Ravndal for the proton--proton 
channel~\cite{Kong:1998sx,Kong:1999sf}.  In Ref.~\cite{Ando:2007fh}, this 
analysis was extended to next-to-next-to-leading order.  A renormalization-group 
analysis of proton--proton scattering in a distorted wave basis was carried out 
in Refs.~\cite{Barford:2002je,Ando:2008jb}. Moreover, the theory was applied to 
proton--proton fusion in Refs.~\cite{Kong:2000px,Ando:2008va}.

\begingroup
\renewcommand{\thefootnote}{\it\alph{footnote}}
The three-nucleon sector with two charged particles was first discussed by 
Rupak and Kong in Ref.~\cite{Rupak:2001ci}, who calculated Coulomb-modified 
scattering phase shifts for proton--deuteron scattering in the spin-quartet 
channel.  A leading order calculation of the $^3$He nucleus including 
nonperturbative Coulomb interactions has been carried out by Ando and 
Birse~\cite{Ando:2010wq}.  Including isospin breaking effects in the 
nucleon--nucleon scattering lengths, they obtained a good description of the 
$^3$He--$^3$H binding energy difference, but they did not consider scattering 
observables.  A similar study at next-to-leading order in the pionless EFT was 
carried out using the resonating group method~\cite{Kirscher:2009aj}.  Those 
results do not include isospin breaking beyond\footnote{This crucial word is 
unfortunately missing in the journal version.} Coulomb exchange and its 
associated contact interaction and are consistent with other determinations of 
the $^3$He--$^3$H binding energy difference.
\setcounter{footnote}{0}
\endgroup

The first calculation---within the framework of pionless EFT---of 
proton--deuteron scattering in the spin-doublet channel was presented by two 
of the present authors in Ref.~\cite{Konig:2011yq}.  The current paper, which 
is largely based on results from the first author's doctoral 
thesis~\cite{Koenig:2013}, extends---and in some aspects corrects---the earlier 
results (see in particular Sec.~\ref{sec:He3-pert}).  After 
reviewing the formalism of pionless EFT in Secs~\ref{sec:Formalism} 
and~\ref{sec:Nd-IntEq}, we first focus on the bound-state sector.  Using 
properly normalized trinucleon wave functions that are discussed in 
Sec.~\ref{sec:WF}, we calculate the $^3$He--$^3$H binding-energy difference in 
first-order perturbation theory, thereby correcting our earlier calculation 
presented in Ref.~\cite{Konig:2011yq}.  In the same section we then proceed and 
perform a nonperturbative determination of the $^3$He binding energy.  We find 
that the inclusion of the full off-shell Coulomb T-matrix as done by Ando and 
Birse~\cite{Ando:2010wq} is clearly not necessary in the bound-state regime.

From the nonperturbative bound-state calculation we also find that the
doublet-channel system beyond leading order does not seem to be renormalized
correctly when Coulomb contributions are taken into account, a result that has 
recently been confirmed by analytical 
arguments~\cite{Vanasse:2014kxa}.\footnote{We note here that the possibilty of 
an $\alpha$-dependent counterterm had already been discussed between HWH and 
U.~van~Kolck shortly after the publication of Ref.~\cite{Konig:2011yq}; 
further discussions then took place between D.~Phillips, SK, HWH, and HWG.}  We 
discuss our numerical findings in some detail in Section~\ref{sec:NLO} and 
also critically review the situation in the scattering regime, arguing to 
replace the Coulomb power counting introduced by Rupak and 
Kong~\cite{Rupak:2001ci} with a simpler scheme that simply includes all 
$\OO(\alpha)$ Coulomb diagrams.  This approach has the advantage of being 
consistent throughout the bound-state and scattering regimes.  Finally, we 
investigate the importance of higher-order (in $\alpha$) Coulomb effects in 
low-energy quartet-channel scattering in Sec.~\ref{sec:HigherCoulomb}, before 
we close with a summary and an outlook.  Some details about bound states in 
pionless EFT and the ``partially-screened'' Coulomb T-matrix used in 
Sec.~\ref{sec:HigherCoulomb} are deferred to appendices.

\section{Formalism and building blocks}
\label{sec:Formalism}

We use the same formalism and notation as in Ref.~\cite{Konig:2011yq}, which we 
review here to make the present paper self-contained.  The effective 
Lagrangian for the proton--deuteron system in pionless EFT can be written as
\begin{multline}
 \mathcal{L} = N^\dagger\left(\ii D_0+\frac{\vD^2}{2\MN}\right)N
 -d^{i\dagger}\left[\sigmad+\left(\ii D_0+\frac{\vD^2}{4\MN}\right)\right]d^i
 -t^{A\dagger}\left[\sigmat+\left(\ii D_0+\frac{\vD^2}{4\MN}\right)\right]t^A
 \\[0.2cm]
 +\yd\left[d^{i\dagger}\left(N^T P^i_d N\right)+\hc\right]
 +\yt\left[t^{A\dagger}\left(N^T P^A_t N\right)+\hc\right]
 +\mathcal{L}_\mathrm{photon}+\mathcal{L}_3 \,,
\label{eq:L-Nd}
\end{multline}
with the nucleon field $N$ and two dibaryon fields $d^i$ (with spin 1 and
isospin 0) and $t^A$ (with spin 0 and isospin 1), corresponding to the deuteron
and the spin-singlet isospin-triplet virtual bound state in S-wave
nucleon--nucleon scattering.  The $\ydt$ and $\sigmadt$ are two-body coupling 
constants that have to be fixed to experimental input data (see below).  Both 
dibaryon fields are formally ghosts since their kinetic terms have a negative 
sign.  This is required to reproduce the positive values of the effective ranges 
with short-range interactions~\cite{Kaplan:1996nv}.  Despite these ``wrong'' 
signs, the Lagrangian~\eqref{eq:L-Nd} can be shown to be equivalent to the most 
general version including only nucleon fields (see, for example, 
Ref.~\cite{Bedaque:2000ab}).  One can interpret the choice of signs in
Eq.~\eqref{eq:L-Nd} as ``avoiding'' the Wigner 
bound~\cite{Phillips:1996ae,Hammer:2009zh,Hammer:2010fw}, but since the 
effective $N$--$N$ interactions (obtained by eliminating the dibaryon fields 
with the kinetic energy terms included) become energy-dependent, this is not a 
rigorous statement.

$\MN \approx 938.918~\MeV$ is the average nucleon mass.  Although the
proton--neutron mass difference is in part of electromagnetic origin, it
is safe for us to neglect this kinematic effect since $(M_p-M_n)/M_N \sim 
0.005$ enters only at high orders in the EFT power counting.\footnote{For 
example, in a neutron propagator with momentum $k$ one would have $k^2/(2M_n) = 
k^2/(2\MN + \Delta\MN) = k^2/(2\MN)\times\left[1-\Delta\MN/(2\MN)+\cdots\right]$ 
with $\Delta\MN = M_n - M_p$.}  For a more quantitative analysis, arguing that 
such corrections should be included at N$^3$LO, see Ref.~\cite{Kirscher:2011zn}.

Spin and isospin degrees of freedom are included by treating the field $N$ as a
doublet in both spaces, but for notational convenience we usually suppress the
spin and isospin indices of $N$.  The operators
\begin{equation}
 P^i_d = \frac{1}{\sqrt8}\,\sigma^2\sigma^i\tau^2 \mathtext{,}
 P^A_t = \frac{1}{\sqrt8}\,\sigma^2\tau^2\tau^A \,,
\label{eq:L-Nd-Projectors}
\end{equation}
with the Pauli matrices $\vec{\sigma}$ and $\vec{\tau}$ operating in spin and
isospin space, respectively, project out the $^3S_1$ and $^1S_0$
nucleon--nucleon partial waves.  Note that for convenience we have written the 
effective Lagrangian~\eqref{eq:L-Nd} in isospin-symmetric form by considering 
only two distinct dibaryon fields.  Alternatively, one could use states in the 
physical particle basis and consider different fields for the individual 
$n$--$p$, $n$--$n$, and $p$--$p$ ${}^1S_0$ configurations.  However, since the 
(inverse) scattering lengths and effective ranges in the $n$--$n$ and $n$--$p$ 
singlet channels are very similar, we do not need to distinguish these cases to 
the order we are working at in this paper.  In the proton--proton sector, there 
is a significant breaking of isospin symmetry due to electromagnetic effects, 
so we will treat this subsystem separately (see Sec.~\ref{sec:Coulomb-pp} 
below).

The covariant derivative
\begin{equation}
 D_\mu = \partial_\mu + \ii eA_\mu \hat{Q} \,,
\label{eq:D-mu}
\end{equation}
where $\hat{Q}$ is the charge operator ($\hat{Q}_N=(\one+\tau_3)/2$, 
$\hat{Q}_{d}=\one$, \ldots), includes the coupling to the electromagnetic field. 
 Furthermore, we have the kinetic and gauge fixing terms for the photons,
\begin{equation}
 \mathcal{L}_\mathrm{photon} = -\frac14 F_{\mu\nu}F^{\mu\nu} -\frac{1}{2\xi}
 \left(\partial_\mu A^\mu-\eta_\mu\eta_\nu\partial^\nu A^\mu\right)^2
 \mathtext{,} \eta_\mu=\text{timelike unit vector} \,,
\label{eq:L-photon}
\end{equation}
of which we only keep contributions from Coulomb photons.  These correspond to a
static Coulomb potential between charged particles, but for convenience we
introduce Feynman rules for a Coulomb-photon propagator,
\begin{equation}
 \Delta_{\mathrm{Coulomb}}(k) = \frac{\ii}{\vk^2+\lambda^2} \,,
\label{eq:FR-Coulomb-Propagator}
\end{equation}
which we draw as a wavy line, and factors $(\pm\ii e\,\hat{Q})$ for the
vertices.\footnote{Due to the sign convention chosen in the
Lagrangian~\eqref{eq:L-Nd}, dibaryon--photon vertices get an additional minus
sign.}  Following Ref.~\cite{Rupak:2001ci}, we have regulated the singularity of
the Coulomb-photon propagator at zero momentum transfer by introducing a photon
mass $\lambda$ in Eq.~\eqref{eq:FR-Coulomb-Propagator}.  This corresponds to a
screening of the Coulomb interaction in configuration space by writing it as a
Yukawa potential $\sim\ee^{-\lambda r}/r$.  In the numerical calculations that
will be discussed later on,  $\lambda$ is always taken to be small (typically
well below $1~\MeV$).  In fact, by choosing a mesh-point distribution dense
around the Coulomb peak it is possible to numerically take the zero-screening
limit $\lambda\to0$~\cite{Konig:2011yq}.

S-wave nucleon--deuteron scattering can take place in either a
spin-$\nicefrac32$ (quartet channel) or spin-$\nicefrac12$ (doublet channel)
configuration.  In the doublet channel, a three-body contact interaction is
required for renormalization already at leading order in the
EFT~\cite{Bedaque:1999ve}.  We write it here as in 
Refs.~\cite{Ando:2010wq,Griesshammer:2011md} as
\begin{equation}
 \mathcal{L}_3=\frac{\MN H(\Lambda)}{3\Lambda^2}N^\dagger\bigg(\yd^2\,
 (d^i)^\dagger d^j \sigma^i \sigma^j+\yt^2\,(t^A)^\dagger t^B \tau^A\tau^B
 - \yd\yt[(d^i)^\dagger t^A \sigma^i \tau^A + \hc] \bigg)N \,,
\label{eq:L-3}
\end{equation}
where $\Lambda$ is a momentum cutoff applied in the three-body equations
discussed below and $H(\Lambda)$ a known log-periodic function of the cutoff
that depends on a three-body parameter $\Lambda_*$.

\subsection{Full dibaryon propagators}
\label{sec:Propagators}

\begin{figure}[htbp]
\centering
\includegraphics[clip]{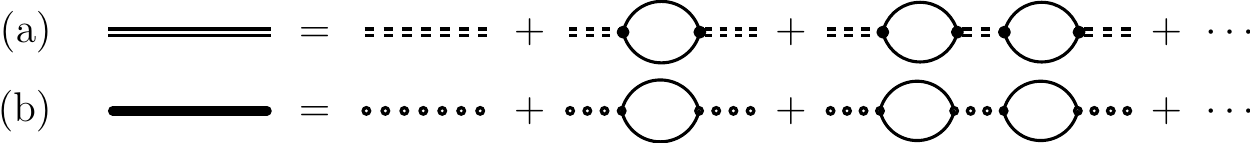}
\caption{Full dibaryon propagators in (a) the $^3S_1$ state (\ie, the deuteron)
and (b) the $^1S_0$ state.}
\label{fig:DibaryonProp}
\end{figure}

It is a well-known feature of pionless EFT that according to the standard power 
counting for systems with large S-wave scattering 
lengths~\cite{Kaplan:1998we,vanKolck:1998bw} the bare dibaryon
propagators have to be dressed by nucleon bubbles to all orders in order to get 
the full leading-order expressions.  The resulting geometric series are shown 
in Fig.~\ref{fig:DibaryonProp}.

For convenience, we also resum the effective range corrections that arise when
the dibaryons in the theory are promoted to dynamical fields by including their 
kinetic terms.  We do not go into the details of the calculations here and
simply quote the results for the renormalized propagators.  They are obtained
by demanding that the (S-wave) effective range expansions
\begin{equation}
 k\cot\delta_d = -\gamd + \frac{\rd}{2}(k^2+\gamd^2)+\,\cdots
\label{eq:ER-d}
\end{equation}
around the deuteron pole\footnote{The notation $\rd$ is used in the
quadratic term of Eq.~\eqref{eq:ER-d} because the expansion is not around
zero momentum.  The difference of $\rd$ to the effective range in a standard
expansion around zero momentum is smaller than 1\%~\cite{deSwart:1995ui}.} at
$k=\ii\gamd=\ii\sqrt{\MN E_d}$, and
\begin{equation}
 k\cot\delta_t = -\frac1{a_t} + \frac{\rnt}{2}k^2+\,\cdots
\label{eq:ER-t}
\end{equation}
for the singlet channel are reproduced.  The expansion for the singlet channel
is around zero momentum; alternatively one could also expand here around the
position of the virtual bound state.  In writing Eq.~(\ref{eq:ER-t}), however,
we have used that $\rnt=r_{0t}$ to the order we are working at and will in the
following also identify $\gamt\equiv1/{a_t}$ to make the notation more
symmetric.  After renormalization in the PDS scheme~\cite{Kaplan:1998tg},
the fully resummed propagators are
\begin{equation}
 \Delta^{ij}_d(p) \equiv \delta^{ij}\Delta_d(p) = -\frac{4\pi\ii}{\MN\yd^2}
 \cdot\frac{\delta^{ij}}{-\gamd+\sqrt{\frac{\vp^2}{4}-\MN p_0-\ii\eps}
 -\frac{\rd}{2}\left(\frac{\vp^2}{4 }-\MN p_0-\gamd^2\right)}
\label{eq:Prop-d-High}
\end{equation}
and
\begin{equation}
 \Delta^{AB}_t(p) \equiv \delta^{AB}\Delta_t(p) = -\frac{4\pi\ii}{\MN\yt^2}
 \cdot\frac{\delta^{AB}}{-\gamt+\sqrt{\frac{\vp^2}{4}-\MN p_0-\ii\eps}
 -\frac{\rnt}{2}\left(\frac{\vp^2}{ 4}-\MN p_0\right)} \,,
\label{eq:Prop-t-High}
\end{equation}
which means that we have fixed the parameters appearing in the effective
Lagrangian according to
\begin{equation}
 \sigma_{d,t} = \frac2\MN \frac{\mu-\gamma_{d,t}}{\rho_{d,t}} \mathtext{,}
 y_{d,t}^2 = \frac{8\pi}{\MN^2}\frac{1}{\rho_{d,t}} \,,
\end{equation}
where $\mu$ is the PDS renormalization scale.  These expressions are valid up to 
\NNLO since the resummation of the effective-range contributions only includes 
a subset of higher-order (\NNNLO etc.) terms.  At leading order, range 
corrections are not included and the dibaryon kinetic terms do not contribute.  
The corresponding propagators are obtained by setting $\rnt=0$ and $\rd=0$ in
Eqs.~\eqref{eq:Prop-d-High} and~\eqref{eq:Prop-t-High}, while perturbative
expressions for the \NLO and \NNLO propagators can be obtained by expanding
the equations up to linear and quadratic order in $\rd$ and $\rnt$,
respectively.  For example, one finds
\begin{multline}
 \Delta_d(p) = -\frac{4\pi}{\MN\yd^2}
 \cdot\frac{1}{-\gamd+\sqrt{\frac{\vp^2}{4}-\MN p_0-\ii\eps}}\\
 \times\left[1+\frac{\rd}{2}\frac{\vp^2/4-\MN p_0-\gamd^2}
 {-\gamd+\sqrt{\frac{\vp^2}{4}-\MN p_0-\ii\eps}}
 + \left(\frac{\rd}{2}\frac{\vp^2/4-\MN p_0-\gamd^2}
 {-\gamd+\sqrt{\frac{\vp^2}{4}-\MN p_0-\ii\eps}}\right)^{\!2}
 + \cdots \right] \,.
\label{eq:Prop-d-exp}
\end{multline}
Using these propagators then in the three-body equations (see 
below) amounts to a ``partial resummation'' of effective-range 
corrections~\cite{Bedaque:2002yg}.  For each expression, the corresponding 
deuteron wave function renormalization constant is given by the residue at the 
bound state pole:
\begin{equation}
 Z_0^{-1} = \ii\frac{\partial}{\partial p_0}
 \left.\frac{1}{\Delta_d(p)}\right|_{p_0 =-\frac{\gamd^2}{\MN},\,\vp=0} \,.
\label{eq:Z0}
\end{equation}
At leading order, one simply has
\begin{equation}
 Z_0^\LO = \gamd\rd \,,
\label{eq:Z0-LO}
\end{equation}
whereas the result from the fully resummed expression~\eqref{eq:Prop-d-High} is
\begin{equation}
 Z_0^\NNLO = \frac{\gamd\rd}{1-\gamd\rd} \,.
\label{eq:Z0-NNLO}
\end{equation}
The expressions for the perturbatively expanded propagators can then simply be
read off from the geometric series
\begin{equation}
 \frac{1}{1-\gamd\rd} = 1 + \gamd\rd + (\gamd\rd)^2 + \cdots \,.
\end{equation}

Note that since $\gamd\rd\approx0.4$, the above series converges only rather
slowly.  This fact can be taken into account by choosing an alternative
renormalization scheme, called \emph{Z-parametrization}~\cite{Phillips:1999hh},
that is constructed in such a way that it produces the fully resummed $Z_0$ as
given in Eq.~\eqref{eq:Z0-NNLO} already at \NLO.  In
Ref.~\cite{Griesshammer:2004pe} the approach has been applied to the
three-nucleon system in pionless EFT and shown to improve the overall
convergence of the theory.  In the present work, however, where as in 
Ref.~\cite{Konig:2011yq} we are primarily interested in the inclusion of 
Coulomb effects, we use the simpler scheme with the propagators as given in 
Eqs.~\eqref{eq:Prop-d-High} and~\eqref{eq:Prop-t-High}.

Furthermore we point out that, as mentioned in Ref.~\cite{Bedaque:2002yg}, the 
partial resummation of higher-order corrections in the propagators is made for 
convenience only and does certainly not improve the accuracy of the 
calculation.  We will discuss below that in fact it may introduce uncontrolled 
effects by modifying the ultraviolet behavior of the half off-shell amplitudes.
In the future it will be advisable to only perform calculations that treat 
effective-range corrections strictly perturbatively.  For the three-boson 
system this procedure has been carried out up to \NNLO by Ji and Phillips in 
Ref.~\cite{Ji:2012nj}.  Recently, an analogous calculation for the 
neutron--deuteron system has been presented in Ref.~\cite{Vanasse:2013sda}, 
using a new approach that greatly reduces the computational cost by not 
requiring to determine the full off-shell scattering amplitude.  The approach
can be adapted to also include Coulomb effects~\cite{Konig:2013cia}.
  
\subsection{Coulomb contributions in the proton--proton system}
\label{sec:Coulomb-pp}

\begin{figure}[htbp]
\centering
\includegraphics[clip]{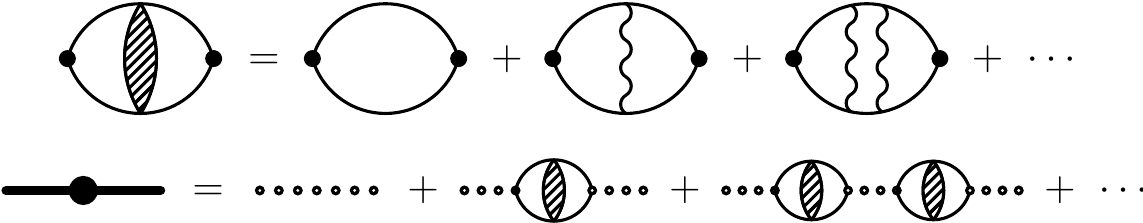}
\caption{Dressed nucleon bubble and full singlet dibaryon propagator in the 
$p$-$p$ channel.}
\label{fig:DressedBubble}
\end{figure}

The Coulomb interaction breaks the isospin symmetry that is implicit in the
dibaryon propagators considered so far.  For the $p$--$p$ part of the
singlet dibaryon we can also have Coulomb-photon exchanges inside the nucleon
bubble.  These can be resummed to all orders, yielding a dressed nucleon
bubble~\cite{Kong:1998sx,Kong:1999sf} that is subsequently used to calculate the
full singlet-dibaryon propagator in the $p$--$p$ channel.  This is shown in 
Fig.~\ref{fig:DressedBubble}.  The result for the leading order propagator
is~\cite{Ando:2010wq}
\begin{equation}
 \Delta^{AB}_{t,pp}(p) \equiv \delta^{AB}\Delta_{t,pp}(p)
 = -\frac{4\pi\ii}{\MN\yt^2} \cdot\frac{\delta^{AB}}{-1/a_C
 -\gamma\,\tilde{h}_0(p')}
\label{eq:Prop-t-pp-Low}
\end{equation}
with
\begin{equation}
 \gamma=\gamma_{\text{$p$--$p$}} = \alpha\MN
 \mathtext{,} \alpha = \frac{e^2}{4\pi} \approx \frac1{137}
 \mathtext{,} p' = \ii\sqrt{{\vp^2}/{4}-\MN p_0-\ii\eps} \,,
\end{equation}
and Euler's digamma function $\psi(x)$ in
\begin{equation}
 \tilde{h}_0(p') = \psi(\ii\eta)+\frac{1}{2\ii\eta}-\log(\ii\eta) \mathtext{,}
 \eta = \eta(p') = \frac{\gamma}{2p'} \,.
\label{eq:h-and-eta}
\end{equation}
This means that the Coulomb-modified effective range expansion\footnote{See, 
for example, Ref.~\cite{Koenig:2012bv} and references therein.} has been used 
for renormalization.  Note that since the matching is done to observables, this 
procedure is model-independent and simultaneously takes into account strong and 
electromagnetic isospin-breaking effects, which are not separated.

We denote here the $p$--$p$ S-wave scattering length simply as $a_C$.  
Corrections due to the corresponding Coulomb-modified 
effective range $r_C$ can be included in the same way as described in the 
preceding section by first resumming insertions of the kinetic-energy operators 
to all orders and then matching the result to reproduce the modified effective 
range expansion up to quadratic order.
  
To simplify the equations that we give later on, we introduce here the
propagator functions
\begin{equation}
 D_{d,t}(E;q) \equiv (-\ii)\cdot\Delta_{d,t}\left(E-\frac{q^2}{2\MN},q\right)
\label{eq:D-d-t-LO}
\end{equation}
and
\begin{equation}
 D_t^{pp}(E;q) \equiv (-\ii)\cdot\Delta_{t,pp}\left(E-\frac{q^2}{2\MN},q\right)
 \,.
\label{eq:D-t-pp}
\end{equation}

\subsection{Power counting}
\label{sec:PowerCounting}

\subsubsection{Strong sector}

Without electromagnetic effects, one can apply the standard power counting for 
pionless EFT that has been extensively discussed in the literature (see the 
reviews~\cite{Beane:2000fx,Bedaque:2002mn,Epelbaum:2008ga} and references
therein).  In this case, the typical low-energy scale $Q$ of the theory is set 
by the deuteron binding momentum $\gamd\approx45~\MeV$.  We can formally count 
the external momenta $k,p$ to be of the same order.  Since we are working in a 
setup without explicit pions, the natural breakdown scale of our theory is of 
the order of the pion mass, $\LamNoPi\sim\Mpi$.  The combination of the two 
scales yields the expansion parameter $Q/\LamNoPi \sim 1/3$ of pionless EFT.

In the three-body sector, one finds that the one-nucleon-exchange interaction 
(see \eg~Fig.~\ref{fig:nd-IntEq-Q}) has to be iterated to all 
orders in order to get the neutron--deuteron scattering amplitude (or T-matrix). 
As will be discussed in more detail in the following section, this procedure 
yields integral equations in momentum space that can be solved numerically with 
an explicit ultraviolet cutoff $\Lambda$, which has to be chosen at least as 
large as $\LamNoPi$ in order to capture all low-energy physics.  This is thus 
the scale we use below to estimate loop contributions.

We will use variations of the numerical cutoff $\Lambda$ to 
asses the numerical uncertainty and convergence of our calculation.  We stress 
here, however, that this gives typically only a lower bound on the true 
uncertainty of our results, which is determined, at any given order, by powers 
of the EFT expansion parameter $Q/\LamNoPi$.

\subsubsection{Including Coulomb effects}

The power counting has to be amended in order to incorporate Coulomb effects.  
For the $p$--$d$ system this was first done by Rupak and Kong in 
Ref.~\cite{Rupak:2001ci}.  From the form of the (Yukawa-screened) Coulomb 
potential in momentum space,
\begin{equation}
 V_{\ccc,\lambda}(q) \sim \frac{\alpha}{q^2+\lambda^2} \,,
\end{equation}
it is clear that Coulomb contributions dominate for small momentum transfers. 
As noted in Ref.~\cite{Rupak:2001ci}, they enter $\sim\alpha\MN/q$, \ie,
proportional to the Coulomb parameter $\eta$; \cf~Eq.~\eqref{eq:h-and-eta}.  
This behavior is not captured by the power counting for the strong sector, which 
consequently has to be modified in order to perform calculations including 
Coulomb effects for small external momenta.

Most importantly, one can no longer simply assume that the scale of all momenta
is set by the deuteron binding momentum, $Q\sim\gamd$.  Instead, one has to keep
track of the new scale introduced by the external momenta separately.  Rupak 
and Kong~\cite{Rupak:2001ci} generically denote this scale by $p$ and 
assume $p\ll Q$ for the power counting, meaning that one makes a simultaneous 
expansion in \emph{two} small parameters $Q/\LamNoPi$ and 
$p/(\alpha\MN)$~\cite{Rupak:2001ci}.\footnote{For a related approach in the 
pionful theory, see also Ref.~\cite{Walzl:2000cx}.}

Deducing the scaling of loops in this scheme is not straightforward anymore.
In fact, the whole discussion (see Ref.~\cite{Konig:2011yq} and the original
Ref.~\cite{Rupak:2001ci}) becomes quite cumbersome.  In the following, we
present the bottom line and will present a simpler approach in
Sec.~\ref{sec:ScatteringRevisited} that has the advantage of being consistent
throughout the scattering and bound-state regimes.

\subsubsection{Selected diagrams}

\begin{figure}[htbp]
\centering
\includegraphics[clip,width=0.85\textwidth]{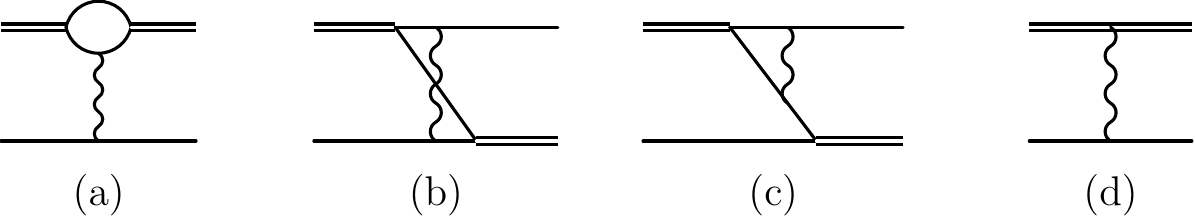}
\caption{Leading $\OO(\alpha)$ diagrams involving Coulomb photons.}
\label{fig:LeadingCoulomb}
\end{figure}

In Fig.~\ref{fig:LeadingCoulomb} we show all diagrams contributing to $p$--$d$
scattering that involve a single Coulomb-photon exchange and are thus of
order $\alpha$.  The last one, diagram~(d) has the most straightforward behavior
since it simply scales as $\alpha/p^2$.  Diagram~(a) has the same factor
because also here the Coulomb-photon propagator involves the external momentum
scale $p$, but it is further enhanced by a factor $\LamNoPi/Q$ from the nucleon
bubble, which is easy to count as there are no Coulomb-photon exchanges inside
the bubble.  This is of course in perfect agreement with the fact that the
direct coupling of the photon to the di\-bary\-on---generated by gauging the
dibaryon kinetic energy operators---only enters at \NLO in the EFT
counting.  This makes diagram~(d) both $\OO(\alpha)$ \emph{and} an
effective-range correction $\sim\rho_{d,t}$.

In both diagrams~(b) and~(c) the loops do not involve the external momentum 
$p$ but are rather dominated by the deuteron binding momentum (at least for 
diagram~(c) this is straightforward to see).  This means that compared to the 
simple one-nucleon-exchange diagram (without a photon) they are both suppressed 
by a factor $\alpha\MN/Q$.  Consistent with this one finds by direct numerical 
evaluation that they are 7\% (b) and 15\% effects (c) at the zero-momentum 
scattering threshold.

We can summarize these findings by saying that unless a diagram directly
exhibits the (regulated) Coulomb pole, it is not enhanced and thus a small
electromagnetic correction to the same diagram topology without Coulomb-photon
exchange.  The enhanced Coulomb contributions are particularly large in the
threshold region (where $p$ is very small) and should thus be iterated to all
orders.  The calculations in Refs.~\cite{Rupak:2001ci,Konig:2011yq} neglect 
both diagrams~(b) and~(c) as well as all diagrams involving more than one 
Coulomb photon, giving them an \apriori{} uncertainty of 7--15 percent.  In
Sec.~\ref{sec:ScatteringRevisited}, we will critically assess this approach.

To conclude the topic for the moment we point out again that the power counting
discussed here is specifically designed to account for Coulomb contributions
that become strong in low-momentum scattering.  At larger momenta, where
the Coulomb parameter $\alpha\MN/p$ becomes small, it would suffice to not
iterate any Coulomb diagrams but rather include them strictly perturbatively.
However, exactly because they become small we assume that it also does not spoil
the calculation to iterate them everywhere.  As we will discuss in 
Sec.~\ref{sec:He3-nonpert}, for calculations in the bound-state regime the 
counting definitely has to be modified because there all loop-momentum scales 
are set by the binding momentum of the bound state.  Based on the findings 
there, we will then argue in Section~\ref{sec:ScatteringRevisited} that the same 
counting---which is actually simpler---should also be used in the scattering 
regime.  For the moment, however, we proceed as in Ref.~\cite{Konig:2011yq} and 
include only the Coulomb diagrams~(a) and~(d).

\subsection{Three-nucleon forces}

In Sec.~\ref{sec:NLO} we will present numerical evidence that without 
refitting the three-nucleon force $H(\Lambda)$ the doublet-channel 
proton--deuteron system is not properly renormalized beyond leading order.  
Indeed, it has been shown recently~\cite{Vanasse:2014kxa} based on analytical 
arguments that a new three-nucleon counterterm is needed to renormalize the 
charged sector at next-to-leading order in a fully-perturbative framework (\ie, 
without partial resummation of effective-range corrections).

The expression in Eq.~\eqref{eq:L-3} is $SU(4)$ symmetric and can be rewritten 
as a structure of the form $(N^\dagger N)^3$.  From this, it is simple to 
construct three-nucleon interactions which can be added in charged systems by 
using the charge operator $\hat{Q}_N = (\one+\tau_3)/2$.  The case of two 
charged particles is relevant for the $p$--$d$ system: $(N^\dagger \hat{Q}_N 
N)^2 (N^\dagger N)$.  Using Fierz-transformations, one can show that its 
spin-isospin structure is different from $(N^\dagger N)^3$, and that it is up to 
a numerical constant identical to the form proposed in 
Ref.~\cite{Vanasse:2014kxa}.  This interaction acts by construction only in the 
$p$--$d$ system, and not in the $n$--$d$ one, while $(N^\dagger N)^3$ acts in 
both systems, with the same strength.  One can also write down interaction 
terms of the form 
\begin{equation}
 \mathcal{L}_3' = \frac{\MN H_{\text{$n$--$d$}}(\Lambda)}{3\Lambda^2}
 \hat{O}_{Nd,Nt}^{\text{$n$--$d$},\nicefrac12}
 + \frac{\MN H_{\text{$p$--$d$}}(\Lambda)}{3\Lambda^2}
 \hat{O}_{Nd,Nt}^{\text{$p$--$d$},\nicefrac12} \,,
\label{eq:L-3-np}
\end{equation}
with operators $\hat{O}_{Nd,Nt}^{\text{$n$--$d$},\nicefrac12}$ and
$\hat{O}_{Nd,Nt}^{\text{$p$--$d$},\nicefrac12}$ similar to the 
structure in Eq.~\eqref{eq:L-3}, but projected in such a way that they only act 
in the $n$--$d$ and $p$--$d$ doublet channels, respectively.

\section{Integral equations}
\label{sec:Nd-IntEq}

According to the power counting certain diagrams have to be resummed to all 
orders in order to calculate $N$--$d$ scattering amplitudes in pionless EFT.  
The resulting integral equations are formally Lippmann--Schwinger equations 
(with non-trivial two-body propagators).  In the following, we generically use 
the calligraphic letter $\mathcal{T}$ to denote their solutions and indicate 
with appropriate subscripts which system we are referring to.  In order to 
obtain the quantum-mechanical T-matrix one has to include an overall minus 
sign and multiply with the deuteron wave function renormalization factor, \ie,
\begin{equation}
 T(E;\vk,\vp) = -Z_0\mathcal{T}(E,\vk,\vp) \,,
\end{equation}
\cf~Appendix~\ref{sec:BoundStates}.  Throughout this paper we write 
``$\mathcal{T}$-matrix'' (or simply ``amplitude'') to indicate this 
distinction.  We denote the incoming and outgoing center-of-mass momenta by 
$\vk$ and $\vp$, respectively.

\begin{figure}[htbp]
\centering
\includegraphics[clip,width=0.5\textwidth]{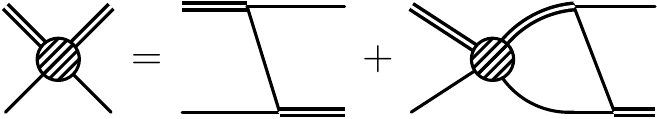}
\caption{Integral equation for the strong scattering amplitude $\TS$ in the
quartet channel.}
\label{fig:nd-IntEq-Q}
\end{figure}

\subsection{Neutron--deuteron quartet channel}

Figure~\ref{fig:nd-IntEq-Q} shows a diagrammatic representation of the
quartet-channel $n$--$d$ amplitude, which is the simplest case one can have. The
only interaction occurring here is the one-nucleon exchange diagram with 
deuteron legs on both sides.  Including all spin-, isospin- and symmetry 
factors, it is given by
\begin{equation}
 -\frac{\ii\MN\yd^2}{2}\cdot(\sigma^j\sigma^i)^\beta_\alpha\delta^b_a
 \cdot\frac{1}{\vk^2+\vk\cdot\vp+\vp^2-\MN E-\ii\eps} \,.
\label{eq:One-N-Exchange-full}
\end{equation}
Here, $i$ and $j$ are spin-1 indices and $\alpha$, $\beta$ ($a$, $b$) are 
spin-$\nicefrac12$ (isospin-$\nicefrac12$) indices.  Since the spins of all 
three nucleons taking part in the reaction have to be aligned to produce a 
total spin-$\nicefrac32$ state, the Pauli principle prohibits here a 
three-nucleon contact interaction for which the particles have to occupy the 
same point in space.  Furthermore, only the dibaryon field representing the 
deuteron can appear in the intermediate state.

Generically, we define the S-wave projected amplitude as
\begin{equation}
 \Tgen(E;k,p)
 = \frac{1}{2}\int\nolimits_{-1}^1{\dd\!\cos\theta}\,\Tgen(E;\vk,\vp)
 \mathtext{,} \theta = \theta_{\vk\cdot\vp}
 \mathtext{,} k = |\vk| \mathtext{,} p = |\vp| \,.
\end{equation}
Applying this to the one-nucleon exchange shown in
Eq.~\eqref{eq:One-N-Exchange-full} yields the projected interaction kernel
\begin{equation}
 \KS(E;k,p) \equiv \frac{1}{kp}\;
 Q_0\left(\frac{k^2+p^2-\MN E-\ii\eps}{kp}\right)
\label{eq:KS}
\end{equation}
with the Legendre function of the second kind
\begin{equation}
 Q_0(a) = \frac{1}{2}\int_{-1}^1\frac{\dd x}{x+a}
 = \frac{1}{2}\log\left(\frac{a+1}{a-1}\right) \,.
\label{eq:Q}
\end{equation}

Here and in the following, the subscript ``s'' is used to denote the strong 
part of the interaction.  The unprojected $\mathcal{T}$-matrix has the same 
spin-isospin indices as the kernel function shown in 
Eq.~\eqref{eq:One-N-Exchange-full}, $\Tgen = (\Tgen^{ij})^{\beta b}_{\alpha a}$. 
 The $n$--$d$ quartet channel is chosen by inserting $i=(1-\ii 2)/\sqrt{2}$ and 
$j=(1+\ii 2)/\sqrt{2}$ for the deuteron spin-1 indices, $\alpha=\beta=1$ for 
the nucleon spins, and $a=b=2$ to select the neutron in isospin space.  As done 
in Refs.~\cite{Gabbiani:1999yv,Bedaque:2000ab} we have used here a short-hand
notation for the spin-1 indices $i$ and $j$ that includes prefactors to be used
in a linear combination.  Written out explicitly, the strong quartet-channel
amplitude is accordingly given by
\begin{equation}
 \TSq=\frac{1}{2}\Big(\TS^{11} + \ii\left(\TS^{12} - \TS^{21}\right)
 + \TS^{22} \Big)^{12}_{12}
\end{equation}
and fulfills the integral equation 
\begin{equation}
 \TSq = -\MN\yd^2\,\KS + \TSq \otimes \left[\MN\yd^2\,D_d \KS\right] \,.
\end{equation}
As in Ref.~\cite{Konig:2011yq} we have introduced here the short-hand notation
\begin{equation}
 A \otimes B \equiv \frac1{2\pi^2}
 \int_0^\Lambda\dd q\,q^2\,A(\ldots,q)B(q,\ldots)
\label{eq:SH-int}
\end{equation}
and used the propagator function $D_d$ as defined in Eq.~\eqref{eq:D-d-t-LO}. 
As indicated in Eq.~\eqref{eq:SH-int}, we regulate all loop integrations with an
explicit momentum cutoff $\Lambda$.

\subsection{Proton--deuteron quartet channel}

\begin{figure}[htbp]
\centering
\includegraphics[clip]{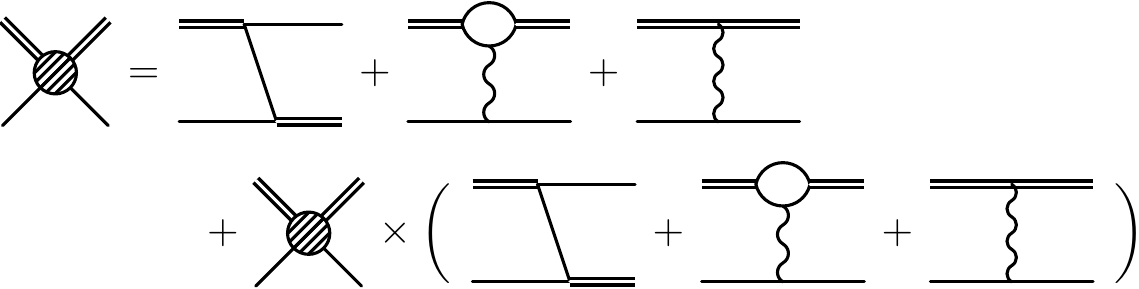}
\caption{Integral equation for the full (\ie~strong + Coulomb) scattering
amplitude $\TF$ in the quartet channel.  The diagram with the photon coupled 
directly to the deuteron only enters at \NLO and beyond.}
\label{fig:pd-IntEq-Q}
\end{figure}

Turning to the $p$--$d$ system, we repeat that according to 
Sec.~\ref{sec:PowerCounting} the dominant Coulomb contribution is the bubble 
diagram shown in Fig.~\ref{fig:LeadingCoulomb}a.  Its energy and momentum 
dependence is given by
\begin{equation}
 K_\mathrm{bubble}(E;\vk,\vp)
 = \frac{\mathcal{I}_\mathrm{bubble}(E;\vk,\vp)}{(\vk-\vp)^{2}+\lambda^2} \,,
\label{eq:K-bubble}
\end{equation}
where
\begin{equation}
 \mathcal{I}_\mathrm{bubble}(E;\vk,\vp)
 = \frac{\arctan\left(\frac{2\vp^2-\vk^2-\vk\cdot\vp}
 {\sqrt{3\vk^2-4\MN E-\ii\eps}\sqrt{(\vk-\vp)^2}}\right)
 +\arctan\left(\frac{2\vk^2-\vp^2-\vk\cdot\vp }
 {\sqrt{3\vp^2-4\MN E-\ii\eps}\sqrt{(\vk-\vp)^2}}\right)}
 {\sqrt{(\vk-\vp)^2}}
\label{eq:I-bubble-pd}
\end{equation}
is the expression for the bubble loop integral.  It can be simplified by noting
that due to the denominator in Eq.~\eqref{eq:K-bubble} the whole expression is
dominated by terms with $\vp^2\approx\vk^2$.  When the expression appears
under the $\dd q$ integral, we analogously get $\vp^2\approx\vk^2$ and can
furthermore assume that $\vq^2\approx\vk^2$ because of the pole at this position
in the deuteron propagator.  Inserting then the total center-of-mass energy,
\begin{equation}
 E = E(k) = \frac{3k^2}{4\MN}-\frac{\gamd^2}{\MN}
 \mathtext{,} k = |\vk| \,,
\label{eq:E-on-shell}
\end{equation}
we get
\begin{equation}
 \frac{\mathcal{I}_\mathrm{bubble}(E;\vk,\vp)}{(\vk-\vp)^{2}+\lambda^2}
 \approx \frac{1}{2\left|\gamd\right|}\frac{1}{(\vk-\vp)^2+\lambda^2}
\label{eq:Bubble-approx}
\end{equation}
by using the expansion $\arctan(x) = x+\OO(x^3)$---with 
$x=\sqrt{(\vk-\vp)^2}/(2|\gamd|)$ up to corrections that are further suppressed 
by powers of $\sqrt{k^2-p^2}/|\gamd|$---in Eq.~\eqref{eq:I-bubble-pd}. 
The same simplification, which effectively amounts to keeping only loop 
contributions with $q\sim p$, has also been used in Ref.~\cite{Rupak:2001ci}.  
We point out, however, that Eq.~\eqref{eq:Bubble-approx} explicitly relies on 
the energy being given by Eq.~\eqref{eq:E-on-shell} and is thus not valid in 
the bound-state regime, where $E < {-E_d} = {-\gamd^2/\MN}$.  The integral 
equation, shown diagrammatically in Fig.~\ref{fig:pd-IntEq-Q}, can then be 
written as
\begin{equation}
 \TFq = -\MN\yd^2\left(K_s -\frac12\KCd\right)
  + \TFq \otimes \left[\MN\yd^2\,D_d\left(K_s -\frac12\KCd\right)\right]
\label{eq:pd-IntEq-Q-full}
\end{equation}
with the kernel function $\KCd$ that includes the bubble diagram and, at 
next-to-leading order and beyond, also diagram~\ref{fig:LeadingCoulomb}(d),
\begin{multline}
 \KCdt(E;k,p) = -\alpha\MN\times\Bigg[\underbrace{\int_{-1}^1\dd\cos\theta\,
 \frac{\mathcal{I}_\mathrm{bubble}(E;\vk,\vp)}{(\vk-\vp)^{2}+\lambda^2}}_{\LO}
 + \underbrace{\frac{\rho_{d,t}}{2kp}\;Q\left(-\frac{k^2+p^2+\lambda^2}{2kp}
 \right)}_{\NLO}\Bigg] \\
 \approx \frac{\alpha\MN}{2kp}\;
 Q\left(-\frac{k^2+p^2+\lambda^2}{2kp}\right)
 \left(\frac{1}{|\gamd|}-\rho_{d,t}\right) \,.
\label{eq:KCdt}
\end{multline}
The variant $\KCt$ only enters in the doublet channel.

\subsection{Neutron--deuteron doublet channel}

\begin{figure}[htbp]
\centering
\includegraphics[clip]{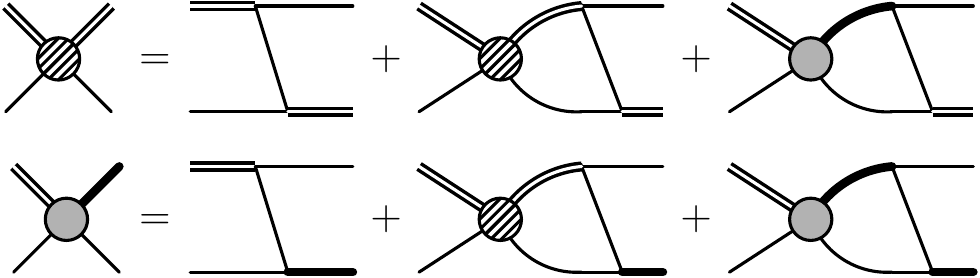}
\caption{Coupled-channel integral equation for the strong scattering amplitude
$\TS$ in the doublet channel.  The diagrams involving the three-body force have
been omitted.}
\label{fig:nd-IntEq-D}
\end{figure}

The equation for the doublet-channel amplitude is shown in
Fig.~\ref{fig:nd-IntEq-D}.  Since the spin-singlet dibaryon is now allowed to
appear in the intermediate state, we have a coupled-channel system of two
amplitudes that we call $\TSda$ and $\TSdb$.  Of these, the ``upper part''
$\TSda$ corresponds directly to the $n$--$d$ scattering process we are
interested in, whereas $\TSdb$ only enters as an off-shell quantity because
the spin-singlet dibaryon, being only a virtual bound state, cannot appear as a
true asymptotic state.

Furthermore, with the spins coupled to a total spin $\nicefrac12$, the Pauli
principle no longer prohibits a three-nucleon interaction, and indeed such a
term is needed already at leading order to renormalize the
system, as extensively discussed \eg~in the
reviews~\cite{Hammer:2010kp,Beane:2000fx}.  The corresponding diagrams have, 
however,  for simplicity been omitted in Fig.~\ref{fig:nd-IntEq-D}.  They can 
be reinstated by supplementing every one-nucleon exchange with a matching 
$N$--$d$ contact interaction from Eq.~\eqref{eq:L-3}.  With the
doublet-channel projection,\footnote{Setting $a=b=2$ formally selects a 
neutron state.  At this stage, however, in the absence of Coulomb effects and 
other isospin-breaking terms, one would find the same equations for $a=b=1$.}
\begin{subequations}
\begin{align}
\label{eq:P-D-a}
 \TSda &= \frac{1}{3}(\sigma^i)^{\alpha'}_\alpha
 (\TS^{\mathrm{a},ij})^{\beta'b}_{\alpha'a}(\sigma^j)^\beta_{\beta'}
 \Big|_{\begin {subarray}{l}a=b=2\\\alpha=\beta=1\end{subarray}} \,, \\
\label{eq:P-D-b}
 \TSdb &= \frac{1}{3}(\sigma^i)^{\alpha'}_\alpha
 (\TS^{\mathrm{b},iB})^{\beta b'}_{\alpha'a}(\tau^B)^b_{b'}
 \Big|_{\begin{subarray}{l}a=b=2\\\alpha=\beta=1\end {subarray}} \,,
\end{align}
\end{subequations}
and introducing furthermore the abbreviations
\begin{equation}
 g_{dd} = \frac{\MN\yd^2}2 \mathtext{,}
 g_{dt} = \frac{\MN\yd\yt}2 \mathtext{,}
 g_{tt} = \frac{\MN\yt^2}2 \,,
\end{equation}
the result can be written as
\begin{multline}
 \begin{pmatrix}\TSda \\[\skrowspace] \TSdb\end{pmatrix}
  = \begin{pmatrix}
  g_{dd}\left(\KS+\frac{2H(\Lambda)}{\Lambda^2}\right)\\[\skrowspace]
  -g_{dt}\left(3\KS+\frac{2H(\Lambda)}{\Lambda^2}\right)
 \end{pmatrix} \\
 +\begin{pmatrix}
 -g_{dd}D_d\left(\KS+\frac{2H(\Lambda)}{\Lambda^2}\right) &
 g_{dt}D_t\left(3\KS+\frac{2H(\Lambda)}{\Lambda^2}\right) \\[\skrowspace]
 g_{dt}D_d\left(3\KS+\frac{2H(\Lambda)}{\Lambda^2}\right) &
 -g_{tt}D_t\left(\KS+\frac{2H(\Lambda)}{\Lambda^2}\right)
 \end{pmatrix}
 \otimes\begin{pmatrix}\TSda\\[\skrowspace]\TSdb\end{pmatrix} \,.
\label{eq:IntEq-TSdb}
\end{multline}

\subsection{Proton--deuteron doublet channel: three-channel formalism}

The $p$--$d$ doublet-channel equation has a yet more complicated structure.
Due to the fact that the electromagnetic interaction does not couple to isospin
eigenstates we now need two different projections for the amplitude
$\mathcal{T}^{\mathrm{b}}$ with the outgoing spin-singlet dibaryon:
\begin{subequations}
\begin{align}
\label{eq:Proj-D-b1}
 \TFdb &= \frac{1}{3}(\sigma^i)^{\alpha'}_\alpha
 (\TF^{\mathrm{b},iB})^{\beta b'}_{\alpha'a}(\one\cdot\delta^{B3})^b_{b'}
 \Big|_{\begin{subarray}{l} a=b=1\\\alpha=\beta=1\end{subarray}} \,, \\
\label{eq:Proj-D-b2}
 \TFdc &= \frac{1}{3}(\sigma^i)^{\alpha'}_\alpha
 (\TF^{\mathrm{b},iB})^{\beta b'}_{\alpha'a}
 (\one\cdot\delta^{B1}+\ii\one\cdot\delta^{B2})^b_{b'}
 \Big|_{\begin{subarray}{l}a=1,\,b=2\\\alpha=\beta=1\end{subarray}} \,.
\end{align}
\label{eq:Proj-D-b1b2}%
\end{subequations}
The latter corresponds to the amplitude with the outgoing spin-singlet dibaryon
in a pure $p$--$p$ state.  For the diagrams where this component appears in
the intermediate state we have to insert the $p$--$p$
propagator~\eqref{eq:Prop-t-pp-Low}.

\begin{figure}[tbhp]
\centering
\includegraphics[clip]{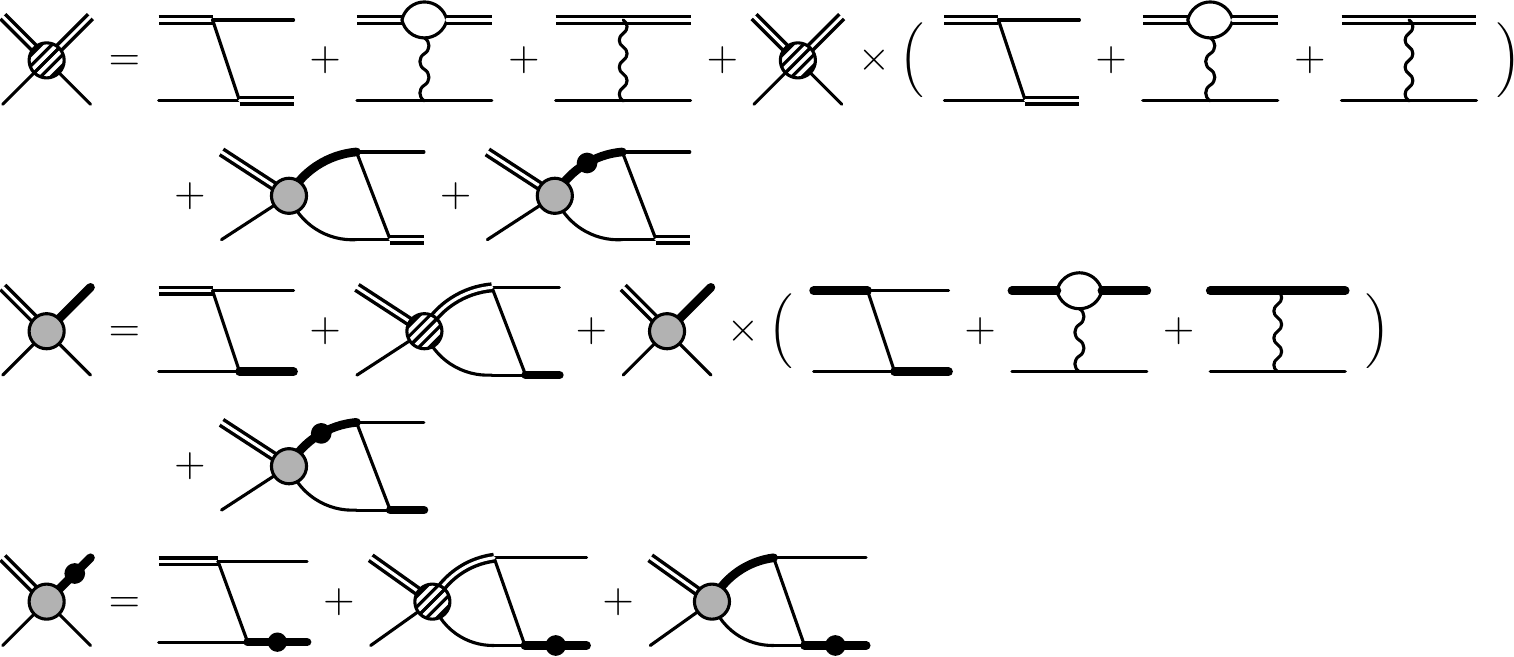}
\caption{Coupled-channel integral equation for the full (\ie, strong + Coulomb)
scattering amplitude $\TF$ in the doublet channel. The diagrams representing the
three-nucleon force have been omitted.}
\label{fig:pd-IntEq}
\end{figure}

Diagrammatically, the resulting three-channel integral equation is shown in
Fig.~\ref{fig:pd-IntEq}.  It is given by
\begin{multline}
 \begin{pmatrix}\TFda\\[\skrowspace]\TFdb\\[\skrowspace]\TFdc\end{pmatrix}
 = \begin{pmatrix}
 g_{dd}\left(\KS+\frac{2H(\Lambda)}{\Lambda^2}\right)\\[\skrowspace]
 -g_{dt}\left(\KS+\frac{2H(\Lambda)}{3\Lambda^2}\right)\\[\skrowspace]
 -g_{dt}\left(2\KS+\frac{4H(\Lambda)}{3\Lambda^2}\right)
 \end{pmatrix}
 + \begin{pmatrix}g_{dd}\KCd\\[\skrowspace]0\\[\skrowspace]0\end{pmatrix} \\
 +\begin{pmatrix}
 -g_{dd}D_d\left(\KS+\frac{2H(\Lambda)}{\Lambda^2}\right) &
 g_{dt}D_t\left(3\KS+\frac{2H(\Lambda)}{\Lambda^2}\right) & 0\\[\skrowspace]
 g_{dt}D_d\left(\KS+\frac{2H(\Lambda)}{3\Lambda^2}\right) & 
 g_{tt}D_t\left(\KS-\frac{2H(\Lambda)}{3\Lambda^2}\right) & 0\\[\skrowspace]
 g_{dt}D_d\left(2\KS+\frac{4H(\Lambda)}{3\Lambda^2}\right) &
 -g_{tt}D_t\left(2\KS+\frac{4H(\Lambda)}{3\Lambda^2}\right) & 0
 \end{pmatrix}
 \otimes \begin{pmatrix}
 \TFda\\[\skrowspace]\TFdb\\[\skrowspace]\TFdc\end{pmatrix} \\
 + \begin{pmatrix}
 -g_{dd}D_d\KCd & 0 & g_{dt}D_t^{pp}
 \left(3\KS+\frac{2H(\Lambda)}{\Lambda^2}\right)\\[\skrowspace]
 0 & -g_{tt}D_t\KCt & -g_{tt}D_t^{pp}
 \left(\KS+\frac{2H(\Lambda)}{3\Lambda^2}\right)\\[\skrowspace]
 0 & 0 & -g_{tt}D_t^{pp}\times\frac{4H(\Lambda)}{3\Lambda^2}
 \end{pmatrix}
 \otimes\begin{pmatrix}
 \TFda\\[\skrowspace]\TFdb\\[\skrowspace]\TFdc
 \end{pmatrix}
\label{eq:pd-IntEq-D-full}
\end{multline}
with the Coulomb kernel functions $\KCdt$ as defined in Eq.~\eqref{eq:KCdt}.
In writing Eq.~\eqref{eq:pd-IntEq-D-full} we have separated the terms in such a
way that the Coulomb contributions can be easily identified.

Note that the terms involving $H(\Lambda)$ in Eq.~\eqref{eq:pd-IntEq-D-full}
differ slightly from the version given in Ref.~\cite{Konig:2011yq}.
In particular, the contribution of $H(\Lambda)$ in the $p$--$p$ channel
was missing.  However, this did not affect the final results since $H(\Lambda)$
was determined numerically from the triton binding energy.

\medskip
The validity of the bubble-diagram approximation given in 
Eq.~\eqref{eq:Bubble-approx} is less clear in the doublet channel.  For 
contributions with the spin-singlet dibaryon in the intermediate state, as they
appear in Eq.~\eqref{eq:pd-IntEq-D-full}, the argument based on the deuteron
pole is actually not true.  However, for energies in the scattering regime,
Eq.~\eqref{eq:Bubble-approx} is still a good approximation (numerically, a 15\%
effect at threshold and thus compatible with neglecting the diagram shown in
Fig.~\ref{fig:LeadingCoulomb}c).

A more subtle point is that the above approximation also changes the ultraviolet
scaling of the diagram, an effect for which it is difficult to judge \apriori{} 
how important it is.  Since the only true advantage of the approximation is 
that it---quite significantly---simplifies the calculation, but certainly does 
not improve it in any physical sense, it is probably best to not use it if 
possible.  We will come back to this point in 
Sec.\ref{sec:ScatteringRevisited}.

\subsection{Pure Coulomb scattering}

In order to calculate Coulomb-subtracted phase shifts, we also need the 
amplitude for pure Coulomb scattering.  Since the electromagnetic interaction
in our approximation does not couple different channels, for both
quartet-channel and doublet-channel $p$--$d$ scattering it is given by the
simple equation
\begin{equation}
 \TC = g_{dd}\,\KCd
 - \TC \otimes \left[g_{dd}\,D_d \KCd\right] \,.
\label{eq:pd-IntEq-Q-c}
\end{equation}

\subsection{Higher-order corrections}

From Eq.~\eqref{eq:KCdt} one directly sees that the diagram with the photon
coupled directly to a dibaryon (Fig.~\ref{fig:LeadingCoulomb}d and its analog
with a spin-singlet dibaryon) is proportional to the effective range ($\rd$
or $\rnt$) and thus a correction that enters at next-to-leading order in the EFT
power counting.  This has already been mentioned in 
Sec.~\ref{sec:PowerCounting}.  Apart from that, the order of our calculation is 
determined by the expressions used for the dibaryon propagators.

The fully resummed propagators given in Eqs.~\eqref{eq:Prop-d-High}
and~\eqref{eq:Prop-t-High} have spurious deep poles that do not correspond to
actual physical bound states.  In the quartet channel, the cutoff can be chosen
low enough to avoid that pole.  Due to the larger cutoff needed in the doublet
channel, however, we cannot use the resummed propagators here.  Instead, we
follow the approach of Ref.~\cite{Bedaque:2002yg} and use the perturbative
expansions (more appropriately called ``partially resummed propagators'')
mentioned in Sec.~\ref{sec:Propagators}.  This still resums some
higher-order effective-range contributions, but removes the unphysical
pole.\footnote{This resummation procedure however affects the ultraviolet 
behavior of the propagators.  The fully resummed expressions fall off faster 
than the leading-order propagators, whereas the perturbative expansions do not 
go to zero anymore for large $p$.  This point will become important in 
Sec.~\ref{sec:NLO}.}

More precisely, at next-to-leading order we always use propagators
$D_{d,t}^\NLO$ that include a single insertion of the dibaryon kinetic energy
operator and are thus linear in the effective range.  At next-to-next-to-leading 
order (\NNLO) the propagators $D_{d,t}^\NNLO$ for the doublet-channel
calculation include corrections quadratic in the effective ranges, whereas in
the quartet-channel we use the fully-resummed expression given by
Eq.~\eqref{eq:Prop-d-High} together with a cutoff low enough to avoid the
unphysical pole.  This approach is chosen such that our quartet-channel results 
can be compared directly to those in Ref.~\cite{Rupak:2001ci}.

Since the publication of Ref.~\cite{Konig:2011yq} an agreement has been reached
in the literature~\cite{Ji:2012nj,Griesshammer:2005ga,Bedaque:2002yg} that in
the doublet channel, a second (energy-dependent) three-nucleon interaction is
needed at \NNLO for consistent renormalization.  In Ref.~\cite{Konig:2011yq} 
such a term has not been included, meaning that the doublet-channel calculation 
presented there at \NNLO is only a partial result.  In fact, as we will 
discuss in Section~\ref{sec:ScatteringRevisited}, with Coulomb effects 
included the question of correct renormalization in the double channel might 
have to be reconsidered already at next-to-leading order.

\subsection{Numerical implementation}
\label{sec:Numerics}

The integral equations presented in the previous sections have to be solved
numerically.  We do so by discretizing the integrals, using Gaussian quadrature,
principal value integration to deal with the singularity of the deuteron
propagator, and appropriate transformations of the integration domain.  We use 
here the same method as described in Ref.~\cite{Konig:2011yq}.  The regulating 
photon mass $\lambda$ is kept small in order to not modify the theory too much.  
By choosing an integration mesh-point distribution that puts emphasis on the 
Coulomb peak in the inhomogeneous parts of the integral equations and the 
low-momentum region, one obtains well-converged results and can even 
extrapolate (linearly) to the physical case $\lambda=0$.

The experimental input parameters used in the calculation are the same as in
Ref.~\cite{Konig:2011yq} and summarized in Table~\ref{tab:Params}.

\begin{table}[htbp]
\centering
 \begin{tabular}{ccc||ccc}
  Parameter & Value & Ref. & Parameter & Value & Ref. \\
  \hline\hline
  \rule{0pt}{1.1em}$\gamd$ & $45.701~\MeV$ & \cite{vanderLeun:1982aa} & $\rd$
  & $1.765~\fm$ & \cite{deSwart:1995ui} \\
  $a_t$ & $-23.714~\fm$ & \cite{Preston:1975} & $\rnt$
  & $2.73~\fm$ & \cite{Preston:1975} \\
  $a_C$ & $-7.8063~\fm$ & \cite{Bergervoet:1988zz}& $\rnC$
  & $2.794~\fm$ & \cite{Bergervoet:1988zz}\\
 \end{tabular}
\caption{Parameters used for the numerical calculation.}
\label{tab:Params}
\end{table}

\subsection{Scattering phase shifts}
\label{sec:Scatt}

From the solutions of the integral equations we can obtain the S-wave
scattering phase shifts as
\begin{equation}
 \delta(k) = \frac{1}{2\ii}
 \log\left(1+\frac{2\ii k\MN}{3\pi} Z_0\Tgen(E_k;k,k)\right) \,.
\label{eq:delta}
\end{equation}
As already mentioned, the $\mathcal{T}$-matrix is multiplied by the wave 
function renormalization constant $Z_0$ defined in Eq.~\eqref{eq:Z0}.  This 
procedure also removes the dependence of the amplitude on the coupling constant 
$\yd$, which so far we have have kept in all equations.

For the $p$--$d$ system, what has to be compared to experimental data are the
Coulomb-subtracted phase shifts
\begin{equation}
 \tilde{\delta}(k) \approx \delta_\ddd(k) \equiv \deltaF(k) - \deltaC(k) \,,
\label{eq:delta-diff}
\end{equation}
where $\deltaF(k)$ is obtained from the full integral equation including both
Coulomb and strong interactions and $\deltaC(k)$ is obtained from
Eq.~\eqref{eq:pd-IntEq-Q-c} which only includes the Coulomb interaction.
This procedure (as opposed to calculating the pure Coulomb phase shift 
analytically in a purely two-body approximation) has the advantage of properly 
taking into account the finite cutoff, the EFT expansion, and the regulating 
photon mass $\lambda$.

Results for both quartet-channel and doublet-channel $p$--$d$ phase shifts have 
been reported in Ref.~\cite{Konig:2011yq}.  Upon closer examination it turns 
out that these should actually be taken with a grain of salt.  This has already 
been hinted at when we discussed the power counting in 
Sec.~\ref{sec:PowerCounting} and the bubble-diagram 
approximation~\eqref{eq:Bubble-approx}.  Indeed, a more careful analysis of the 
bound-state regime that will be discussed in the following sections also sheds 
some new light on the scattering calculation, a point that we will finally come 
back to in Sec.~\ref{sec:ScatteringRevisited}.

\section{Trinucleon wave functions}
\label{sec:WF}

\begin{figure}[htbp]
\centering
\includegraphics[width=0.6\textwidth,clip]{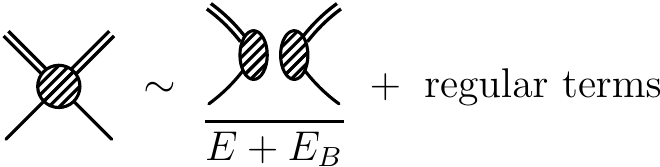}
\caption{Diagrammatic representation of the factorization of
$\mathcal{T}$-matrix at the bound-state pole.}
\label{fig:T-Factorization}
\end{figure}

The most straightforward observable in the three-nucleon bound-state regime
is the $^3\mathrm{H}$--$^3\mathrm{He}$ binding-energy shift.  It can be 
calculated in first-order perturbation theory by treating the Coulomb 
interaction as a small correction.  In order to do this we need the wave 
functions that describe the trinucleon (triton) bound state.  As illustrated 
diagrammatically in Fig.~\ref{fig:T-Factorization}, at the bound-state pole the
$\mathcal{T}$-matrix factorizes as
\begin{equation}
 \Tgen(E;k,p) = -\frac{\Bgen^\dagger(k)\Bgen(p)}{E+E_B}
 + \text{terms regular at $E=-E_B$} \,,
\label{eq:T-factorization}
\end{equation}
where the $\mathcal{B}(p)$ are what we call \emph{amputated} wave functions or
vertex factors.  For a derivation of this relation, including the sign, see
Appendix~\ref{sec:BoundStates}.

\subsection{Homogeneous equation}
\label{sec:HomEq}

For our coupled-channel problem, $\Tgen(E;k,p)$ is a matrix in channel-space,
and the wave functions will thus be vectors.  We use here the three-channel 
equation structure even in the absence of Coulomb kernel functions in order to 
separate the component that in the $p$--$d$ system (discussed below in 
Sec.~\ref{sec:PD-He3}) corresponds to the spin-singlet dibaryon being in a 
pure $p$-$p$ state.  Our interaction kernel is then given by
\begin{equation}
 \hat{K}\equiv\begin{pmatrix}
 -g_{dd}\left(\KS+\frac{2H(\Lambda)}{\Lambda^2}\right) &
 g_{dt}\left(3\KS+\frac{2H(\Lambda)}{\Lambda^2}\right) &
 g_{dt}\left(3\KS+\frac{2H(\Lambda)}{\Lambda^2}\right)\\[\skrowspace]
 g_{dt}\left(\KS+\frac{2H(\Lambda)}{3\Lambda^2}\right) &
 g_{tt}\left(\KS-\frac{2H(\Lambda)}{3\Lambda^2}\right) &
 -g_{tt}\left(\KS+\frac{2H(\Lambda)}{3\Lambda^2}\right)\\[\skrowspace]
 g_{dt}\left(2\KS+\frac{4H(\Lambda)}{3\Lambda^2}\right) &
 -g_{tt}\left(2\KS+\frac{4H(\Lambda)}{3\Lambda^2}\right) &
 -g_{tt}\times\frac{4H(\Lambda)}{3\Lambda^2}
 \end{pmatrix} \,,
\label{eq:K-B}
\end{equation}
which is found from Eq.~\eqref{eq:pd-IntEq-D-full} by factoring out all
propagators and omitting the $K_c^{(d,t)}$ entries.  Inserting the factorization
at the pole ~\eqref{eq:T-factorization} into the Lippmann-Schwinger 
equation\footnote{Note that the kernel matrix in Eq.~\eqref{eq:K-B} is not 
symmetric.  To rigorously apply the bound-state factorization, one should 
symmetrize it first by rescaling the second row with a factor $3$ and the third 
row with a factor $3/2$.  In practice, however, one can simply solve the 
homogeneous equation with the kernel as in Eq.~\eqref{eq:K-B} and insert 
the appropriate symmetrization factors whenever one calculates matrix elements 
from the solutions.} and taking the limit $E\to{-E_B}$ gives a homogeneous 
equation of the form
\begin{equation}
 \vec{\BS} = (\hat{K}\hat{D})\otimes\vec{\BS}
\label{eq:BS-IntEq}
\end{equation}
with
\begin{equation}
 \vec{\BS}\equiv\left(\BSda,\BSdb,\BSdc\right)^T
 \mathtext{,} \hat D = \diag(D_d,D_t,D_t) \,,
\end{equation}
The inhomogeneous interaction terms have dropped out here since they are
regular as $E={-E_B}$, and we have canceled the overall $\Bgen^\dagger$ terms
from the factorization.

\subsection{Normalization condition}
\label{sec:PD-NC}

In order to calculate quantities based on the wave functions $\BS$ it is
important to normalize these correctly.  To this end one has to take into 
account that the ``potential'' derived from the EFT we are using here, \ie, the
one-nucleon exchange kernel as given in Eqs.~\eqref{eq:One-N-Exchange-full}
and~\eqref{eq:KS}, is effectively energy-dependent.  The proper normalization 
condition is
\begin{equation}
 \left(\hat D\vec{\BS}\right)^T
 \otimes\frac\dd{\dd E}\left(\hat I-\hat K\right)
 \Big|_{E=-\EB^\mathrm{\,^3\mathrm{H}}}
 \otimes\left(\hat D\vec{\BS}\right) = 1 \,,
\label{eq:Triton-WF-norm}
\end{equation}
where we have defined the matrix of inverse propagators $\hat I =
\diag(I_d,I_t,I_t)$ with
\begin{equation}
 I_{d,t}(E,q,q') = \frac{2\pi^2}{q^2}{\delta(q-q')}D_{d,t}(E;q)^{-1} \,.
\end{equation}
A detailed derivation of this can again be found in 
Appendix~\ref{sec:BoundStates}.  Note, however, that the insight that
energy-dependent interactions imply a nontrivial normalization condition for
bound-state wave functions has been known since quite some time (see, for
example, the overview by Agrawala~\etal~\cite{Agrawala:1966ab} and references 
therein).

\medskip
The normalization condition derived above can be verified numerically by 
considering the residue of the $\mathcal{T}$-matrix at the bound-state pole.  
Following Hagen~\etal~\cite{Hagen:2013xga} we define
\begin{equation}
 Z = \lim\nolimits_{E\to{-E_B}}(E+E_B)\int_0^\Lambda\frac{\dd q\,q^2}{2\pi^2}
 \int_0^\Lambda\frac{\dd q'\,q'^2}{2\pi^2}
 D(E,q)\Tgen(E;q,q')D(E,q')
\label{eq:Z-T-gen}
\end{equation}
and call this quantity the $Z$-factor of the trinucleon state, or simply
\emph{trimer $Z$-factor}.  As we will discuss shortly, for our coupled-channel
system we need to sum over all components of the $\mathcal{T}$-matrix.  In
Eq.~\eqref{eq:Z-T-gen} we have written $D(E,q)$ to denote a generic dibaryon
propagator.

Inserting the factorization of the $\mathcal{T}$-matrix at the pole as given in
Eq.~\eqref{eq:T-factorization}, we find that
\begin{multline}
 Z = \int_0^\Lambda\frac{\dd q\,q^2}{2\pi^2}
 \int_0^\Lambda\frac{\dd q'\,q'^2}{2\pi^2}
 D(-E_B,q)\Bgen^\dagger(q)\Bgen(q')D(-E_B,q') \\
 = \left|\int_0^\Lambda\frac{\dd q\,q^2}{2\pi^2}D(-E_B,q)\Bgen(q)\right|^2 \,,
\label{eq:Z-B-gen}
\end{multline}
where the last equality follows by noting that the propagators are real in the
bound-state regime.  Of course, it is crucial here that the wave functions
$\Bgen(q)$ are normalized correctly, so numerically calculating $Z$ from both
Eqs.~\eqref{eq:Z-T-gen} and~\eqref{eq:Z-B-gen} and showing that the results
agree provides the means to verify our normalization
condition~\eqref{eq:Triton-WF-norm} with an explicit calculation.

We now carry out this procedure for the triton wave functions in the two-channel
formalism.  In order to implement Eq.~\eqref{eq:Z-T-gen} we also need the part
of the amplitude that describes the (unphysical) scattering of a spin-singlet
dibaryon and a nucleon.  The complete $\mathcal{T}$-matrix for the system is
then given by
\begin{equation}
 \hat{\Tgen}^{\mathrm{d}}_\mathrm{s} = \begin{pmatrix}
  \TSda & \TSdsa \\[\skrowspace] \TSdb & \TSdsb
 \end{pmatrix} \,,
\end{equation}
where the first column is determined by Eq.~\eqref{eq:IntEq-TSdb} and the
second column is a solution of the analogous equation
\begin{multline}
 \begin{pmatrix}\TSdsa \\[\skrowspace] \TSdsb\end{pmatrix}
  = \begin{pmatrix}
  -g_{dt}\left(3\KS+\frac{2H(\Lambda)}{\Lambda^2}\right)\\[\skrowspace]
  g_{tt}\left(\KS+\frac{2H(\Lambda)}{\Lambda^2}\right)
 \end{pmatrix} \\
 +\begin{pmatrix}
 -g_{dd}D_d\left(\KS+\frac{2H(\Lambda)}{\Lambda^2}\right) &
 g_{dt}D_t\left(3\KS+\frac{2H(\Lambda)}{\Lambda^2}\right) \\[\skrowspace]
 g_{dt}D_d\left(3\KS+\frac{2H(\Lambda)}{\Lambda^2}\right) & 
 -g_{tt}D_t\left(\KS+\frac{2H(\Lambda)}{\Lambda^2}\right)
 \end{pmatrix}
 \otimes\begin{pmatrix}\TSdsa\\[\skrowspace]\TSdsb\end{pmatrix}  \,.
\label{eq:IntEq-TSdb-singlet}
\end{multline}
The expression for the trimer $Z$-factor in terms of the $\mathcal{T}$-matrix is
then
\begin{equation}
 \ZT = \lim\nolimits_{E\to{-E_B}}(E+E_B)\sum\limits_{i,j=1}^2
 \left[\hat{D}_{ii} \otimes \big(\hat{\TSd}\big)_{ij}
 \otimes \hat{D}_{jj}\right] \,,
\label{eq:Z-T}
\end{equation}
where we have switched back to the short-hand notation of
Section~\ref{sec:Nd-IntEq} (with the modification that $\hat{D}=\diag(D_d,D_t)$
is only a $2\!\times\!2$ matrix here) and inserted a subscript $\mathcal{T}$ in
order to distinguish it from the expression in terms of the wave functions,
which we write as
\begin{equation}
 \ZWF = \sum\limits_{i=1}^2
 \left|\hat{D}_{ii} \otimes \big(\vec{\BS}\big)_{i}\right|^2 \,.
\label{eq:Z-WF}
\end{equation}

Numerically calculating $\ZWF$ is straightforward since it is just a simple
integral over the normalized wave functions.  Implementing the procedure to
calculate $\ZT$, on the other hand, is more delicate since in order to
numerically obtain the residue one has to rely on cancellations between large
numbers.  If, however, we approach the bound-state pole exponentially, \ie, set
\begin{equation}
 E(x) = -E_B + (E_B - E_0)\cdot\exp(-x) \,,
\label{eq:Z-exp}
\end{equation}
where $E_0$ is some energy between the deuteron binding energy $E_d$ and
$E_B$, we find that indeed $\ZT$ converges to $\ZWF$ as a function of the
parameter $x$.\footnote{The authors would like to thank P.~Hagen for suggesting
this approach.}

\begin{figure}[htbp]
\centering
\includegraphics[width=0.75\textwidth,clip]{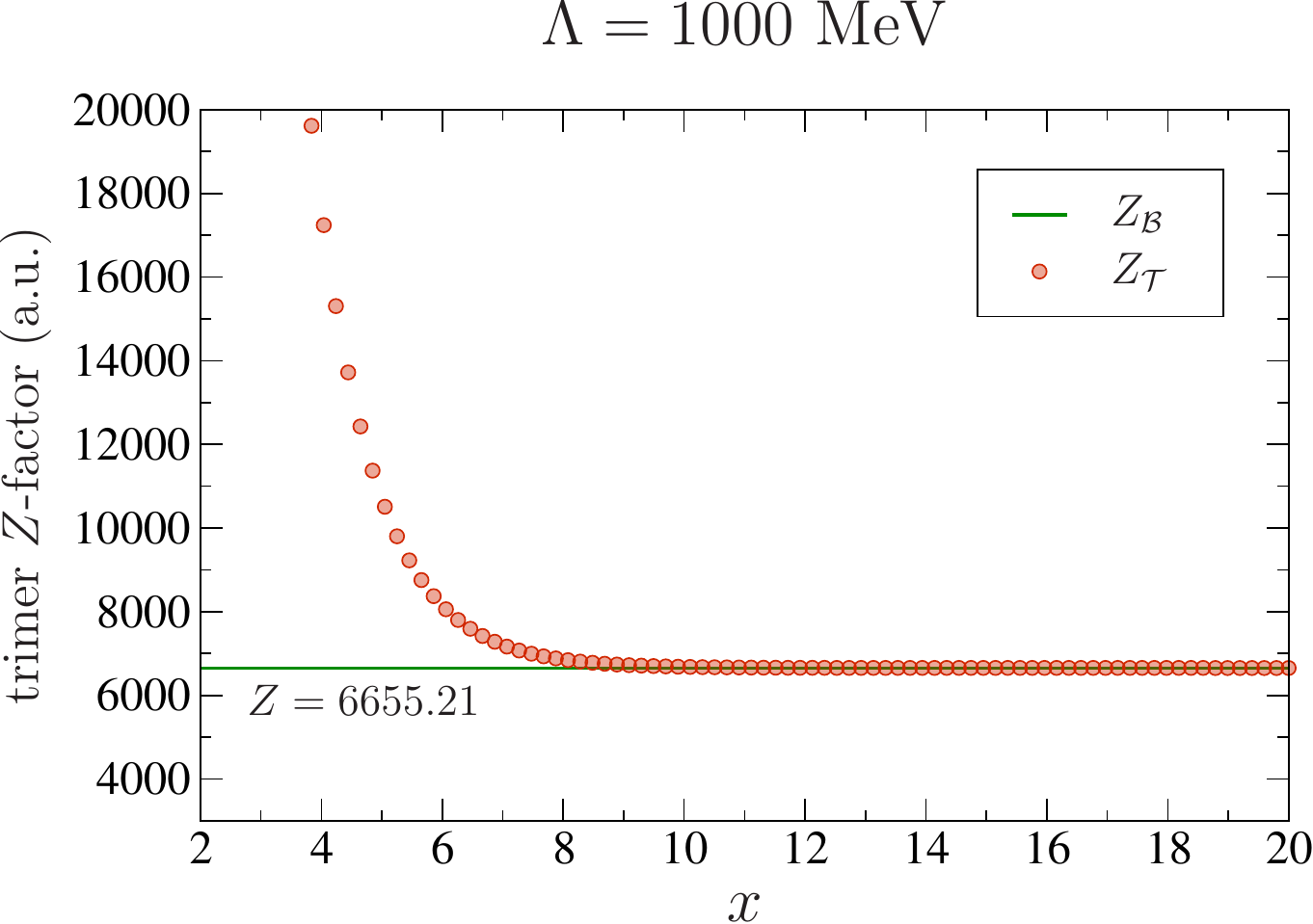}
\caption{Triton trimer $Z$ factor (in arbitrary units, a.u.) for cutoff 
$\Lambda=1000~\MeV$. $x$ is the parameter in Eq.~\eqref{eq:Z-exp}.}
\label{fig:Z-1000}
\end{figure}

We show this in Figs.~\ref{fig:Z-1000} for a cutoff $\Lambda=1000~\MeV$.  In
the calculation we have used $E_0=1.1\times E_d$, which means that at $x=20$
the distance to the pole (which is always fixed to be exactly at
$E_B=8.4818~\MeV$) is only about $10^{-8}~\MeV$.  Only if one goes even closer
to the pole ($x\gsim25$), numerical difficulties become significant and $\ZT$
starts to visibly deviate from $\ZWF$.  At $x=20$, $\ZT$ still agrees with
$\ZWF$ to within about 1.5\%.  At other cutoffs one finds the same overall 
behavior.\footnote{Note, however, that the numerical value of the $Z$-factor 
varies significantly with the cutoff.  This can be understood by noting that the 
way we have defined it in Eqs.~\eqref{eq:Z-T-gen} and~\eqref{eq:Z-B-gen}, $Z$ 
is an \emph{unrenormalized} quantity.  For the current discussion where we are 
only interested in establishing the agreement of $\ZT$ and $\ZWF$, however, 
this is of no importance.}  We take the above findings as a clear numerical 
verification underlining the correctness of our normalization
condition~\eqref{eq:Triton-WF-norm}.

\section{Helium-3 properties}
\label{sec:PD-He3}

We are now finally in a position to discuss the Helium-3 system.  In the first
part of this section we will describe how to obtain the
$^3\mathrm{H}$--$^3\mathrm{He}$ binding energy shift in a perturbative
approach based the normalized trinucleon wave functions $\BS$.  The results
presented here correct and thus supersede those given in
Ref.~\cite{Konig:2011yq}.  We will additionally describe how to obtain the
$^3\mathrm{He}$ binding energy nonperturbatively by locating the pole in the
full $\mathcal{T}$-matrix.  The results of both calculations will be shown and
discussed in Subsec.~\ref{sec:He3-LO-Results}.

\subsection{Energy shift in perturbation theory}
\label{sec:He3-pert}

In Ref.~\cite{Konig:2011yq} the $^3\mathrm{H}$--$^3\mathrm{He}$ binding-energy
difference was calculated using
\begin{equation}
 \Delta E_C^{\text{old}} = \left(\hat D\vec{\BS}\right)^T\otimes\diag(V_C,V_C,0)
 \otimes\left(\hat D\vec{\BS}\right)
\label{eq:DeltaE-VC}
\end{equation}
with the S-wave projected Coulomb potential
\begin{equation}
 V_C(E;q,q') = -\frac{4\pi\alpha}{2qq'}\;
 Q_0\!\left(-\frac{q^2+q'^2+\lambda^2}{2qq'}\right)
\label{eq:V-C}
\end{equation}
in momentum space.  The prediction for the $^3$He binding energy is then given
by
\begin{equation}
 {-\EB}^{\!\!^3\mathrm{He}} = {-\EB}^\mathrm{\!\!^3\mathrm{H}}
 + \Delta E_C^\text{old} \,.
\end{equation}

However, the naïve approach in Eq.~\eqref{eq:DeltaE-VC} is actually not
correct.  By taking the simple matrix elements of the Coulomb potential between
the wave functions $\vec{\BS}$ in the nucleon--dibaryon formalism, one
effectively couples the photon directly to the dibaryon.  However, as discussed
in Section~\ref{sec:PowerCounting}, this coupling only enters at
next-to-leading order in the EFT power counting.

More subtle, but no less important, is to realize that the expression in
Eq.~\eqref{eq:DeltaE-VC} is not renormalized correctly.  By considering the
terms that enter in the normalization condition~\eqref{eq:Triton-WF-norm} we
find that formally the wave functions $\vec{\BS}$ are proportional to the EFT
coupling constants $\yd$ and $\yt$.  In the results for physical quantities
this dependence has to drop out, just as it does in the scattering calculation
where the deuteron wave function renormalization constant cancels the
$\yd$-dependence of the $\mathcal{T}$-matrix.
                                                                        
\begin{figure}[htbp]
\centering
\begin{minipage}{0.49\textwidth}
 \centering
 \includegraphics[width=0.4275\textwidth,clip]{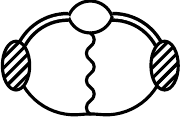}\\[0.77em]
 (a)
\end{minipage}\begin{minipage}{0.49\textwidth}
\centering
 \vspace*{0.5em}
 \includegraphics[width=0.441\textwidth,clip]{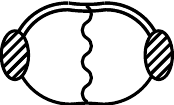}\\[0.77em]
 (b)
\end{minipage}
\caption{Diagrams contributing to the $^3\mathrm{H}$--$^3\mathrm{He}$ binding
energy difference in perturbation theory.  (a) Leading-order diagram.
(b) \NLO-correction.}
\label{fig:DeltaE}
\end{figure}

In fact, Eq.~\eqref{eq:DeltaE-VC} neglects the three-body nature of the
problem.  To correct this, what should be done is to calculate the diagram shown
in Fig.~\ref{fig:DeltaE}a at leading order and only add the contribution shown
Fig.~\ref{fig:DeltaE}b as an \NLO-correction.  Effectively, this amounts to
calculating
\begin{equation}
 \Delta E_C^{\text{new}} = \left(\hat D\vec{\BS}\right)^T\otimes\hat{K}_C
 \otimes\left(\hat D\vec{\BS}\right)
\label{eq:DeltaE-Kc}
\end{equation}
with the matrix
\begin{equation}
 \hat{K}_C = \diag(-g_{dd} K_c^{(d)}, -g_{tt} K_c^{(t)}, 0) \,.
\end{equation}
The kernel functions $K_c$ here are the same that appear in the integral
equation for the full scattering amplitude, with the leading-order contribution
given by the bubble diagram, Fig.~\ref{fig:LeadingCoulomb}a, and the direct
coupling to the dibaryon only entering at next-to-leading order.  As we will
discuss in more detail in the next section, it is important here to not use the
approximation~\eqref{eq:Bubble-approx} for the bubble diagram but rather
include the full expression given in the first line of Eq.~\eqref{eq:KCdt}.

\begin{figure}[htbp]
\centering
\begin{minipage}{0.49\textwidth}
 \centering
 \includegraphics[width=0.495\textwidth,clip]{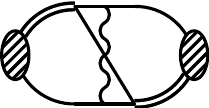}\\[0.77em]
 (a)
\end{minipage}\begin{minipage}{0.49\textwidth}
\centering
 \vspace*{0.5em}
 \includegraphics[width=0.495\textwidth,clip]{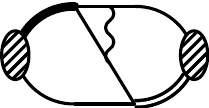}\\[0.77em]
 (b)
\end{minipage}
\caption{Additional diagrams contributing to the $^3\mathrm{H}$--$^3\mathrm{He}$
binding energy difference in perturbation theory.  (a) Box diagram.
(b) Triangle diagram.}
\label{fig:DeltaE-more}
\end{figure}

The new procedure takes into account the EFT expansion and, since $\hat{K}_C$
is proportional to $y_{d,t}^2$, $\Delta E_C^{\text{new}}$ is also renormalized
correctly.  However, the picture is still not complete.  As shown in
Fig.~\ref{fig:DeltaE-more}, there are also contributions to the energy shift
that arise from Coulomb-photon diagrams not taken into account so far.  For the
scattering calculation it was argued that they can be neglected, but we already
mentioned at the end of Section~\ref{sec:PowerCounting} that the power
counting has to be modified in the bound-state regime.  As we will discuss
shortly, these additional diagrams actually are important and should be taken
into account in the energy-shift calculation.  The complete $\hat{K}_C$ then
becomes a non-diagonal matrix.

\subsection{Nonperturbative calculation}
\label{sec:He3-nonpert}

In Ref.~\cite{Ando:2010wq}, Ando and Birse calculate the $^3$He binding energy
in pionless EFT by using a nonperturbative framework that involves the full
off-shell Coulomb T-matrix.  As already discussed by Kok~\etal{} in
Refs.~\cite{Kok:1979aa,Kok:1981aa}, the latter complication is not actually
necessary in the bound-state regime.  In this section we carry out a calculation
analogous to that of Ando and Birse, but only involving Coulomb photons.  The
resulting equation structure is still quite complex, but much simpler and in
particular easier to handle numerically than that of Ref.~\cite{Ando:2010wq}.

\subsubsection{Diagram scaling in the bound-state regime}

At this point it is important to recall from Section~\ref{sec:PowerCounting}
that the Coulomb diagrams in Figs.~\ref{fig:LeadingCoulomb}b
and~\ref{fig:LeadingCoulomb}c, which in the following we simply refer to as
the \emph{box} and \emph{triangle} diagrams, respectively, are not small
\perse, but rather that the bubble diagram, Fig.~\ref{fig:LeadingCoulomb}a, is
enhanced in the low-energy scattering regime due to the Coulomb singularity at
zero momentum transfer.  That effect is particularly prominent in the
inhomogeneous part of the integral equation, where one directly hits the
Coulomb pole in the on-shell limit, so that the expression is only rendered
finite by the regulating photon mass $\lambda$.

In the bound-state regime this is no longer the case.  There are no
inhomogeneous terms that could exhibit a Coulomb peak, and since all loops are
dominated by the binding momentum of the bound-state under consideration,
the dibaryon propagators do not further enhance integration domains of small
momentum transfer.

The consequence of all this is that if the Helium-3 binding energy is 
calculated directly from the $p$--$d$ integral equation (in an approach that 
is nonperturbative compared to the one discussed in Sec.~\ref{sec:He3-pert}), 
then all $\OO(\alpha)$ Coulomb diagrams should be treated on an equal footing 
and be included in the calculation.  Furthermore, for the integral in the 
bubble diagram we have to use the full expression as given by 
Eq.~\eqref{eq:I-bubble-pd} since the argument allowing the approximation given 
in Eq.~\eqref{eq:Bubble-approx} is not valid in the bound-state regime.

\subsubsection{Full equation structure}

\begin{figure}[htbp]
\centering
\includegraphics[clip,width=0.9\textwidth]{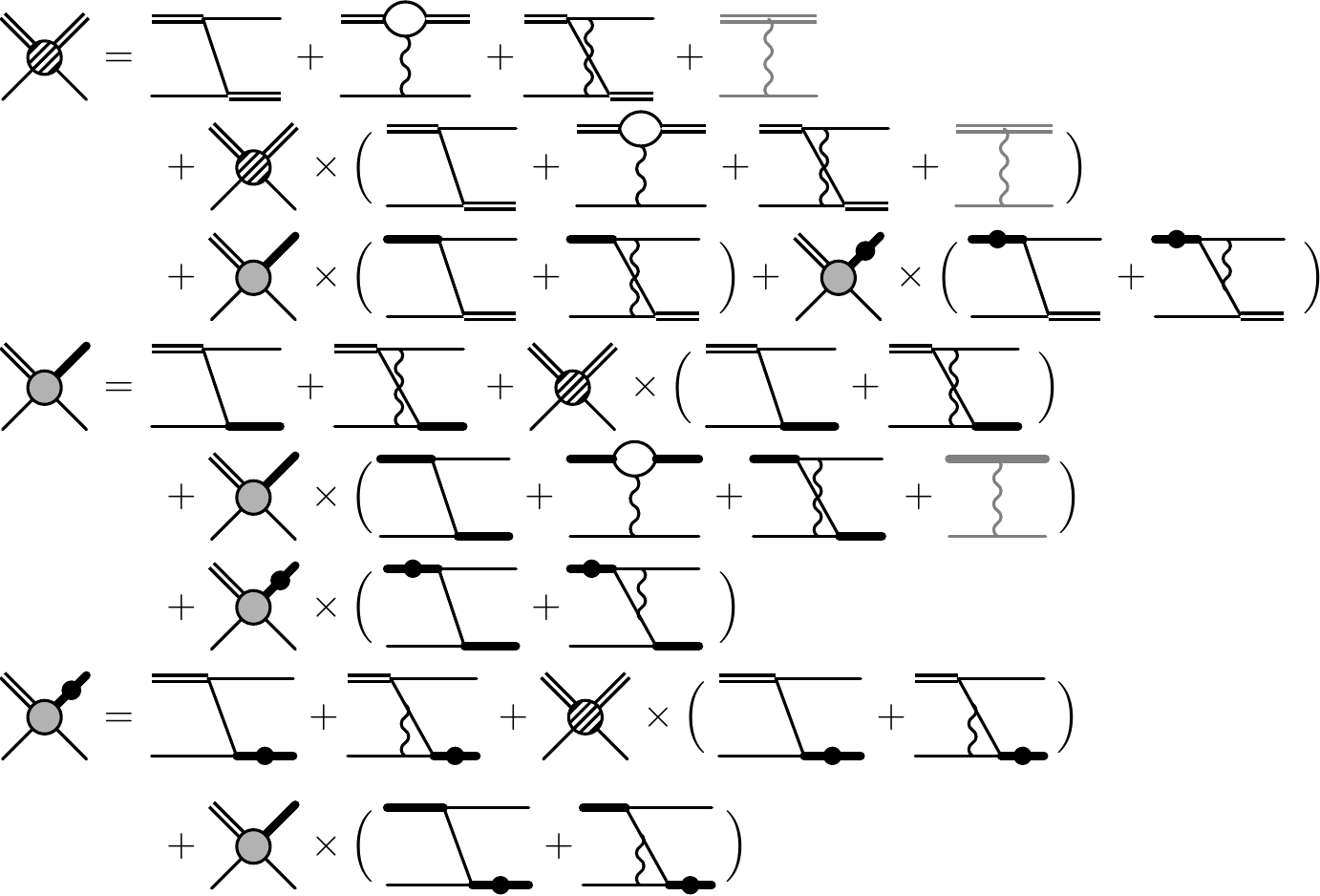}
\caption{Coupled-channel integral equation for the full scattering amplitude
$\TFf$ used for the nonperturbative $^3$He calculation.  The diagrams
representing the three-nucleon force have been omitted.  The shaded diagrams 
with the photons coupled directly to the dibaryon only enter at \NLO and 
beyond.}
\label{fig:pd-IntEq-full}
\end{figure}

The resulting full equation structure is shown diagrammatically in
Fig.~\ref{fig:pd-IntEq-full} and given by the expression
\begin{multline}
 \begin{pmatrix}\TFfda\\[\skrowspace]\TFfdb\\[\skrowspace]\TFfdc\end{pmatrix}
 = \begin{pmatrix}
 g_{dd}\left(\KS+\frac{2H(\Lambda)}{\Lambda^2}\right)\\[\skrowspace]
 -g_{dt}\left(\KS+\frac{2H(\Lambda)}{3\Lambda^2}\right)\\[\skrowspace]
 -g_{dt}\left(2\KS+\frac{4H(\Lambda)}{3\Lambda^2}\right)
 \end{pmatrix}
 + \begin{pmatrix}g_{dd}\left(\KCd+K_{\text{box}}\right)\\[\skrowspace]
 -g_{dt}K_{\text{box}}\\[\skrowspace]
 -2g_{dt}K_{\text{tri}}^{(\text{in})}\end{pmatrix} \\
 +\begin{pmatrix}
 -g_{dd}D_d\left(\KS+\frac{2H(\Lambda)}{\Lambda^2}\right) &
 g_{dt}D_t\left(3\KS+\frac{2H(\Lambda)}{\Lambda^2}\right) & 0\\[\skrowspace]
 g_{dt}D_d\left(\KS+\frac{2H(\Lambda)}{3\Lambda^2}\right) & 
 g_{tt}D_t\left(\KS-\frac{2H(\Lambda)}{3\Lambda^2}\right) & 0\\[\skrowspace]
 g_{dt}D_d\left(2\KS+\frac{4H(\Lambda)}{3\Lambda^2}\right) &
 -g_{tt}D_t\left(2\KS+\frac{4H(\Lambda)}{3\Lambda^2}\right) & 0
 \end{pmatrix}
 \otimes \begin{pmatrix}
 \TFfda\\[\skrowspace]\TFfdb\\[\skrowspace]\TFfdc\end{pmatrix} \\
 + \begin{pmatrix}
 -g_{dd}D_d\left(\KCd+K_{\text{box}}\right) &
 3g_{dt} D_t K_{\text{box}} & g_{dt}D_t^{pp}
 \left(3\KS + 3K_{\text{tri}}^{(\text{out})}
 +\frac{2H(\Lambda)}{\Lambda^2}\right)\\[\skrowspace]
 g_{dt} D_d K_{\text{box}} &
 -g_{tt}D_t\left(\KCt-K_{\text{box}}\right) & -g_{tt}D_t^{pp}
 \left(\KS + K_{\text{tri}}^{(\text{out})}
 +\frac{2H(\Lambda)}{3\Lambda^2}\right)\\[\skrowspace]
 2g_{dt} D_d K_{\text{tri}}^{(\text{in})} &
 -2g_{dt} D_t K_{\text{tri}}^{(\text{in})} &
 -g_{tt}D_t^{pp}\times\frac{4H(\Lambda)}{3\Lambda^2}
 \end{pmatrix} \\
 \otimes\begin{pmatrix}
 \TFfda\\[\skrowspace]\TFfdb\\[\skrowspace]\TFfdc
 \end{pmatrix} \,,
\label{eq:pd-IntEq-D-full-more}
\end{multline}
which, albeit quite complex, is a direct extension of
Eq.~\eqref{eq:pd-IntEq-D-full}.  We use primes in the subscript to indicate
the inclusion of the additional Coulomb contributions compared to 
Eq.~\eqref{eq:pd-IntEq-D-full}.  The first of these, $K_\text{box}$, is 
initially given by a rather complicated expression but can be simplified 
to~\cite{Hoferichter-BoxDiag:2010}
\begin{multline}
 K_{\text{box}}(E;k,p) = -\alpha\MN \\
 \times\frac12\int_{-1}^1\dd\!\cos\theta\,
 \Bigg\{\frac{\arctan\Big(\frac{2\vp^2-\vk^2-\vk\cdot\vp}
 {\sqrt{3\vk^2-4\MN E-\ii\eps}\sqrt{(\vk-\vp)^2}}\Big)
 +\arctan\Big(\frac{2\vk^2-\vp^2-\vk\cdot\vp }
 {\sqrt{3\vp^2-4\MN E-\ii\eps}\sqrt{(\vk-\vp)^2}}\Big)}
 {(\vk^2+\vp^2+\vk\cdot\vp-\MN E-\ii\eps)\sqrt{(\vk-\vp)^2}} \\
 - \frac{\lambda}{(\vk^2+\vp^2+\vk\cdot\vp-\MN E-\ii\eps)^2}
 \Bigg\} \,,
\label{eq:K-box}
\end{multline}
which is valid up to (negligible) corrections of order $\lambda^2$.  This 
expression is obtained by writing the expression obtained from the Feynman 
diagram shown in Fig.~\ref{fig:LeadingCoulomb}(b) in such a way that a Taylor 
expansion in the photon mass $\lambda$ can be carried out before performing the 
integration over one remaining Feynman parameter.  More precisely, the expansion
parameter is proportional to $\lambda/(\vk^2+\vp^2+\vk\cdot\vp-\MN 
E-\ii\eps)^{1/2}$.  Recalling that this is at least of the order of the 
deuteron binding momentum $\gamd$ (in the scattering regime) or even dominated 
by the trinucleon binding energy (in the bound-state regime), it is clear that
the approximation in Eq.~\eqref{eq:K-box} is very good.  We have checked 
numerically that this is indeed the case.\footnote{An expression for the box 
diagram without expansion in $\lambda$ can be obtained from the expression 
given in Eq.~\eqref{eq:I-box-full} in Sec.~\ref{sec:FullBox} by replacing the 
T-matrix $T_{C,\lambda}$ discussed in that section with a simple Coulomb-
photon propagator.}

\medskip
For the triangle-diagram contributions one furthermore finds
\begin{subequations}%
\begin{equation}
 K_\text{tri}^{(\text{out})}(E;k,p) = -\alpha\MN \\
 \times\frac12\int_{-1}^1\dd\!\cos\theta
 \frac{\mathcal{I}_{\text{tri}}(E;\vk,\vp)}
 {\vk^2+\vp^2+\vk\cdot\vp-\MN E-\ii\eps} \,,
\label{eq:K-tri-out}
\end{equation}
\begin{equation}
 K_\text{tri}^{(\text{in})}(E;k,p) = K_\text{tri}^{(\text{out})}(E;k,p) \,,
\label{eq:K-tri-in}
\end{equation}
\label{eq:K-tri}%
\end{subequations}%
where the superscripts indicate whether the Coulomb-photon exchange is on the
incoming (left) or outgoing (right) side of the diagram.  The loop function
appearing in Eq.~\eqref{eq:K-tri-out} is given by
\begin{multline} 
 \mathcal{I}_{\text{tri}}(E;\vk,\vp)
 = \frac{\ii}{2\sqrt{\vk^2/4+\vk\cdot\vp+\vp^2}} \\
 \times\Bigg\{
 \log\left(\frac{\ii(\vk^2/2-\vk\cdot\vp-\vp^2-\lambda^2-\MN E-\ii\eps)}
 {\sqrt{\vk^2/4+\vk\cdot\vp+\vp^2}}
 +2\sqrt{\lambda^2+3\vk^2/4-\MN E-\ii\eps}\right) \\
 -\log\left(\frac{\ii(\vk^2+\vp^2+\vk\cdot\vp-\lambda^2-\MN E-\ii\eps)}
 {\sqrt{\vk^2/4+\vk\cdot\vp+\vp^2}}+2\lambda\right)\Bigg\} \,.
\end{multline}
The spin and isospin projections for both diagrams are completely analogous to
those for the simple one-nucleon exchange diagram, with additional projection
operators from the photon vertices ensuring that contributions forbidden by
charge conservation vanish as they should.

\subsection{Leading-order results}
\label{sec:He3-LO-Results}

\subsubsection{Binding energies}

\begin{figure}[tbhp]
\centering
\includegraphics[width=0.94\textwidth,clip]{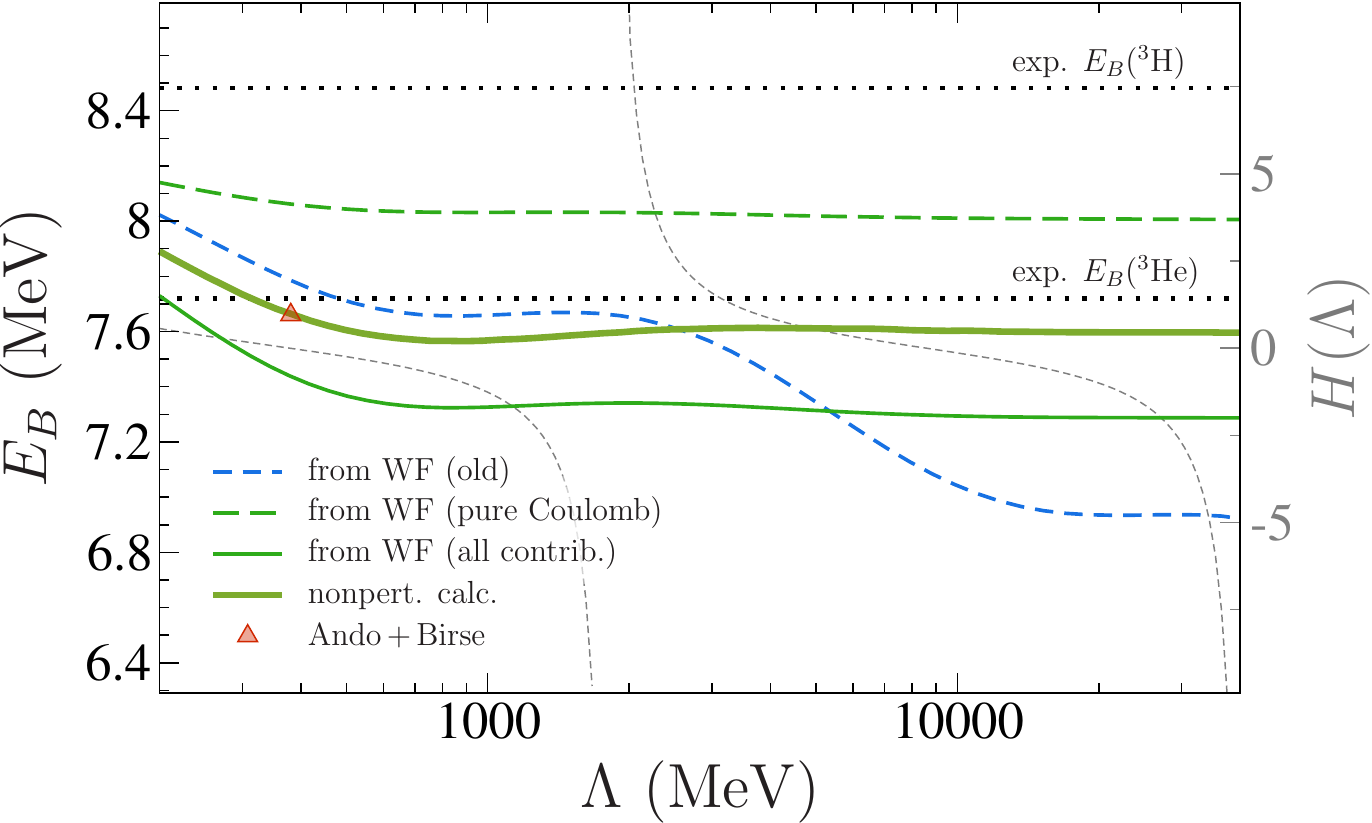}
\caption{Leading-order predictions for the $^3$He binding energy as a function
of the cutoff.  Lower dashed curve: result from the naïve calculation reported
in Ref.~\cite{Konig:2011yq}.  Upper dashed curve: corrected result according to
Eq.~\eqref{eq:DeltaE-Kc}.  Lower solid curve: corrected result including
all $\OO(\alpha)$ Coulomb diagrams.  Upper solid curve: nonperturbative result
according to Section~\ref{sec:He3-nonpert}. Red triangle: result from
Ref.~\cite{Ando:2010wq}.  Dotted lines: experimental values for the $^3$H and
$^3$He binding energies.  Thin dashed curve: cutoff-dependence of three-nucleon
force.}
\label{fig:H-En-LO-new}
\end{figure}

The leading-order results for the $^3$He binding energy (with the three-body 
force $H(\Lambda)$ fixed, for each cutoff $\Lambda$, such that the experimental 
triton binding energy is reproduced by the calculation in the $n$--$d$ channel) 
are summarized  in Fig.~\ref{fig:H-En-LO-new}.  The lower dashed curve shows 
the result of the naïve calculation reported in Ref.~\cite{Konig:2011yq}.  The 
upper dashed curve is the corrected result according to 
Eq.~\eqref{eq:DeltaE-Kc}, where only the bubble diagram is taken into account 
at leading order.  The effect of including all $\OO(\alpha)$ Coulomb diagrams 
(that are leading order in the EFT counting) is shown by the lower solid curve, 
whereas the upper solid curve shows the result obtained from the nonperturbative 
calculation described in Section~\ref{sec:He3-nonpert}.  For comparison, the 
experimental $^3$He and $^3$H binding energies are indicated as dotted lines 
and the cutoff dependence of the three-nucleon force $H(\Lambda)$ is shown as a 
thin dashed curve.

The most striking feature is that the new results, both from the perturbative
and the nonperturbative calculation, do not show the fall-off at large cutoffs
that was reported in Ref.~\cite{Konig:2011yq} and is shown here by the lower
dashed curve.  An explanation for why this effect occurs in the naïve
calculation will be given below in Section~\ref{sec:NLO}.

In fact, both new results are essentially stable beyond cutoffs of about
500~\MeV.  From Fig.~\ref{fig:H-En-LO-new} we furthermore see clearly that it
is important to take into account all $\OO(\alpha)$ Coulomb diagrams in the
perturbative calculation.  Comparing it with the result from the bubble
diagram alone shows that the box and triangle diagrams together give more than
half of the total energy shift (with the larger part coming from the triangle
contributions).  The nonperturbative result comes out closer to the experimental
value than the perturbative one, as should be expected from that more
complete calculation.  Within the typical 30\% uncertainty of a leading-order
calculation in pionless EFT, however, they agree both with one another and with
the experimental Helium-3 binding energy of about
$7.72~\MeV$.\footnote{For the perturbative calculation this statement of course
refers to the result where all $\OO(\alpha)$ diagrams are included.  The error
estimate should in this case be taken as 30\% of the \emph{energy shift} (since
that is what is calculated), which then gives a marginal agreement with the
experimental value.}

Finally, the result obtained by Ando and Birse in Ref.~\cite{Ando:2010wq} by
using the full off-shell Coulomb T-matrix lies almost exactly on
our curve.  The small energy difference of about $0.004~\MeV$ that we obtain at
their cutoff $\Lambda=380.689~\MeV$ is well beyond the accuracy of the
calculation and cannot be resolved in the plot.\footnote{Ando and Birse use the
single cutoff $\Lambda=380.689~\MeV$ for their calculation because they find
that the three-nucleon vanishes there.  We find this zero of $H(\Lambda)$ at
$\Lambda\approx377.69~\MeV$ instead.  This small discrepancy is most likely due
to differences in the numerical implementation and negligible here.}  This is in
perfect agreement with the conclusion of Kok~\etal~\cite{Kok:1979aa,Kok:1981aa}
that a Coulomb-photon approximation should be well justified in the bound-state
regime.

\subsubsection{Wave functions}

\begin{figure}[htbp]
\centering
\includegraphics[width=0.87\textwidth,clip]{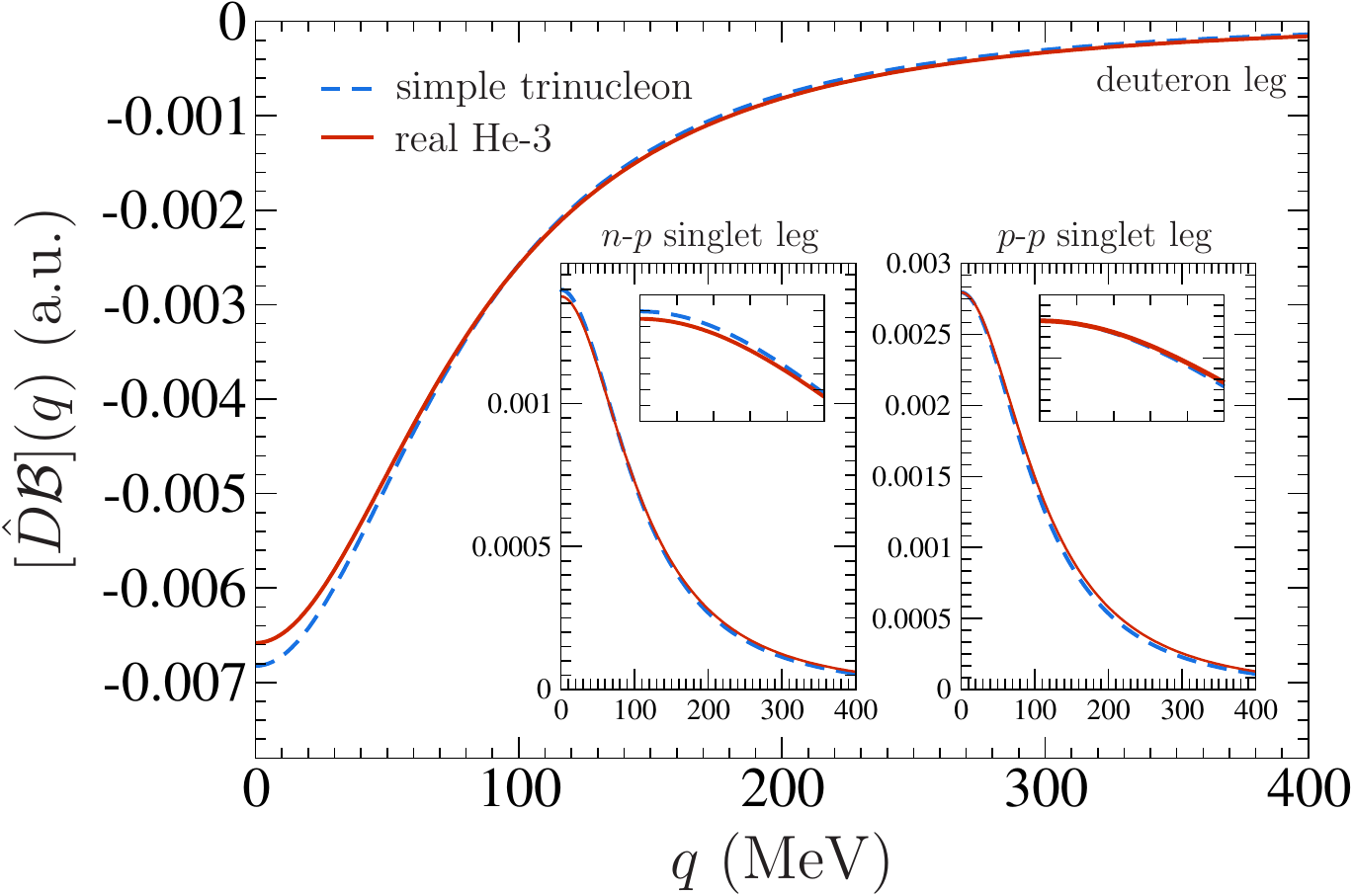}
\caption{Three-component Helium-3 wave functions (in arbitrary units, a.u.) 
calculated for a cutoff $\Lambda=400~\MeV$.  The individual components are 
labelled by the outgoing dibaryon field (``leg'') in the corresponding channel.  
Solid curves: result from full calculation involving all leading Coulomb 
diagrams.  Dashed curves: result from simple trinucleon calculation with the 
three-nucleon force fixed to give the experimental Helium-3 binding energy.}
\label{fig:He-WF}
\end{figure}

From the new equation structure~\eqref{eq:pd-IntEq-D-full-more} we can also
obtain Helium-3 wave functions.  In Fig.~\ref{fig:He-WF} we show a
representative plot, calculated at a cutoff $\Lambda=400~\MeV$.  The three 
components of the wavefunction are labelled by the outgoing dibaryon field 
(``leg'') in the corresponding channel.  The solid curves indicate the result 
obtained from the full equation including all leading Coulomb diagrams, whereas 
the dashed curves are obtained by calculating simple trinucleon wave functions 
(without Coulomb effects) for a system in which the three-nucleon force was 
fixed to reproduce the experimental $^3$He binding energy.  Both are normalized 
according to the condition~\eqref{eq:Triton-WF-norm} with the interaction 
$\hat{K}$ set to the appropriate form.

Comparing the two results, one finds that low-momentum modes are suppressed in
the wave functions from the full calculation compared to the simple trinucleon
result.  This is in good agreement with what one naïvely expects from the
repulsive Coulomb force: it is particularly strong for small relative momenta
of the charged subsystems, thus lowering the probability of the system to be in
such a state.  For the wave function component where the dibaryon is in a pure
$p$--$p$ state, this does not apply because in that case there is no Coulomb
repulsion between the dibaryon and the third nucleon.  Indeed, exactly this can
also be seen in Fig.~\ref{fig:He-WF}.

The wave functions obtained in this manner can be used as nonperturbative
input quantities for other calculations, \eg, for a consistent determination
of the Helium-3 photodisintegration in pionless EFT, which is currently work
in progress.

\section{The Coulomb problem at next-to-leading order}
\label{sec:NLO}

\begin{figure}[htbp]
\centering
\includegraphics[width=0.94\textwidth,clip]{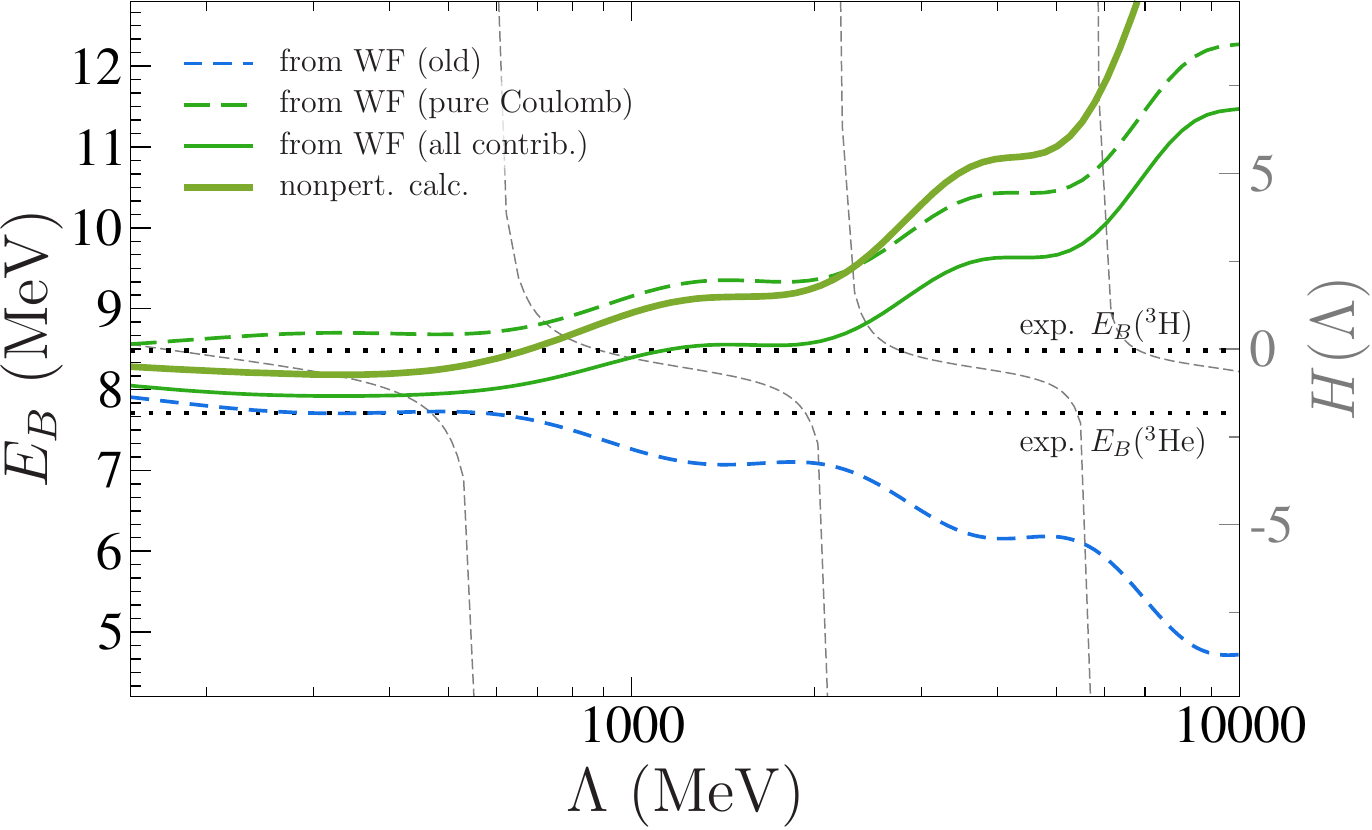}
\caption{\NLO{} results for the $^3$He binding energy as a function
of the cutoff.  The curves are as in Fig.~\ref{fig:H-En-LO-new}.}
\label{fig:H-En-NLO-new}
\end{figure}

From the promising results at leading order one might expect that going to
next-to-leading order gives both more precision and better agreement
with the experimental Helium-3 energy.  However, as shown in
Fig.~\ref{fig:H-En-NLO-new}, exactly the opposite is the case.  The results
from both the perturbative and the nonperturbative calculation are now closer
to---or even above---the triton binding energy, which certainly does not
make sense physically.  Moreover, the results are strongly cutoff-dependent
again and rise to even larger binding energies as the cutoff is increased.  From
these findings one is led to suspect that our next-to-leading order calculation 
is not renormalized properly and that a new counterterm might be needed to 
renormalize the system at this order when Coulomb effects are 
included~\cite{Koenig:2013}.  Recently, it was shown analytically that 
such a counterterm is indeed necessary~\cite{Vanasse:2014kxa}.  That 
calculation was carried out in a framework where effective-range corrections 
are included fully perturbatively and uses an analytical investigation of the 
ultraviolet behavior of the kernels that enter in the integral equations.  In 
the following we analyze the situation within the partial-resummation approach. 
As we will discuss below, the ultraviolet  behavior of the dibaryon 
propagators---and thus the behavior of the three-body amplitudes---changes from 
order to order, which complicates the situation.  Our analysis is therefore 
based largely on numerical investigations.

\subsection{Scaling of the dibaryon propagators}

The ultraviolet scaling of the dibaryon propagators can directly be inferred
from looking at their expressions.  We consider here the deuteron propagator
$D_d$ as a representative example.  At leading order, its scaling is
\begin{equation}
 D_d^\LO(E;q) \propto \frac{1}{-\gamd+\sqrt{\frac34q^2-\MN E-\ii\eps}}
 \sim \frac{1}{q} \mathtext{as} q\to\infty \,.
\end{equation}
Upon going to next-to-leading order, this is multiplied by a factor
proportional to the effective range,
\begin{equation}
 D_d^\NLO(E;q) = D_d^\LO(E;q)\left[1
 + \frac{\rho_d}{2}\cdot\frac{3q^2/4-\MN E-\gamd^2}
{-\gamd+\sqrt{\frac34q^2-\MN E-\ii\eps}} \right] \,,
\end{equation}
such that the asymptotic behavior is changed to
\begin{equation}
 D_d^\NLO(E;q) \sim \mathrm{const.} \mathtext{as} q\to\infty \,.
\end{equation}
Repeating this procedure in order to get the \NNLO and higher-order
propagators one even gets functions that are divergent ($\sim q$, $\sim q^2$,
etc.) in the ultraviolet.

\subsection{Ultraviolet behavior of the amplitude}

\begin{figure}[htbp]
\centering
\includegraphics[width=0.8\textwidth,clip]{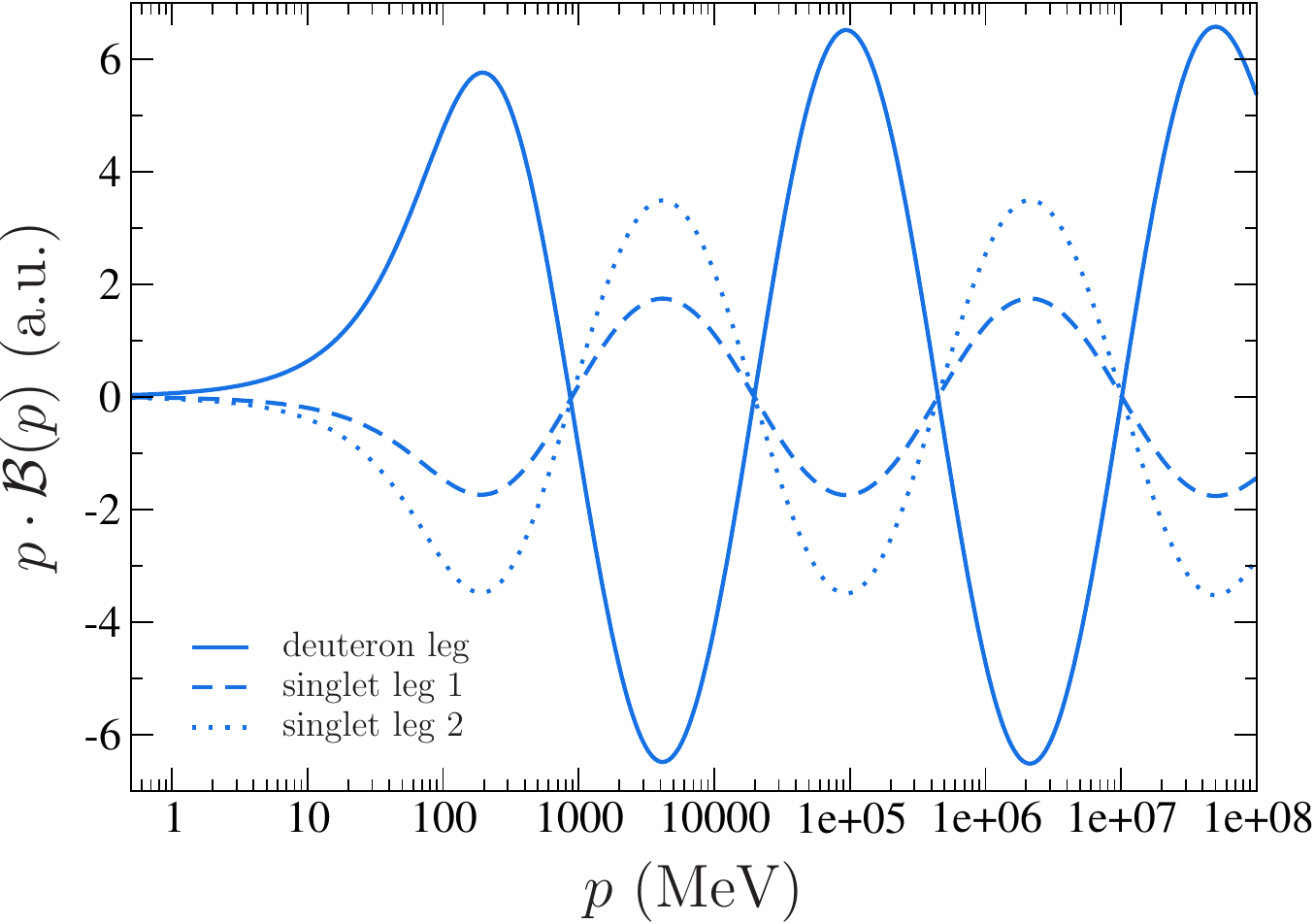}
\caption{Rescaled leading order trinucleon wave function $p\cdot\BS(p)$ (in
arbitrary units) as a function of the momentum $p$.}
\label{fig:WF-Log-LO}
\end{figure}

In order to find the behavior of the wave functions $\Bgen(p)$ introduced in
Section~\ref{sec:WF}, we first go back to the corresponding
$\mathcal{T}$-matrix.  As discussed, for example, in Ref.~\cite{Bedaque:2002yg},
for fixed $E$ and $k$ the $n$--$d$ doublet channel amplitude $\TSd(E;k,p)$ has
an asymptotic behavior determined by linear combinations of $p^{\pm\ii s_0 - 1}$
in the limit $p\to\infty$.  It is the imaginary parts with $s_0\approx1.0064$
that give rise to the log-periodic behavior of the three-nucleon force
$H(\Lambda)$ necessary to renormalize the system.  What is important for the
discussion here is the modulus of the scaling:
\begin{equation}
 |\TSd(E;k,p)| \sim \frac{1}{p} \mathtext{as} p\to\infty \,.
\end{equation}
Since the derivation of this is independent of the energy $E$, the same scaling
also applies in the bound-state regime.  In particular, it is inherited by the
wave functions, such that also
\begin{equation}
 |\BS(p)| \sim \frac{1}{p} \mathtext{as} p\to\infty \,.
\end{equation}
Indeed, as shown in Fig.~\ref{fig:WF-Log-LO}, one also finds numerically
that a plot of $p\cdot\BS(p)$ against $p$ shows a log-periodic behavior for
large $p$ with constant amplitude.

\begin{figure}[htbp]
\centering
\includegraphics[width=0.8\textwidth,clip]{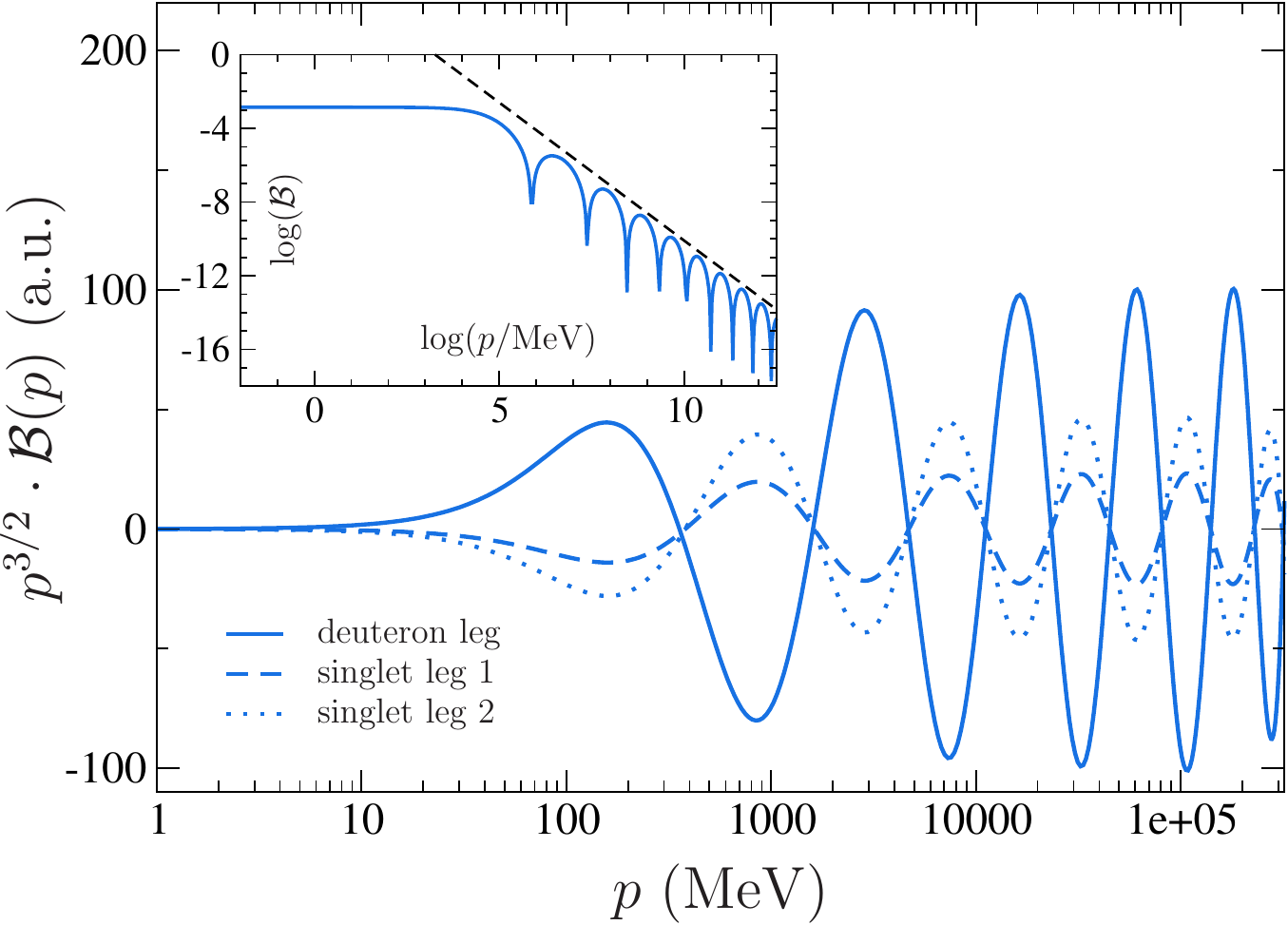}
\caption{Rescaled \NLO trinucleon wave function $p^{3/2}\cdot\BS(p)$ (in
arbitrary units) as a function of the momentum $p$.  The inlay shows the
deuteron-leg component (without rescaling) in a double-logarithmic plot; the
dashed line included there has a slope of exactly $-3/2$.  For the other two
components not included in the double-logarithmic plot, one finds exactly the
same behavior.}
\label{fig:WF-Log-NLO}
\end{figure}

The analysis above is true at leading order.  Numerically, we find at
next-to-leading order the approximate scaling
\begin{equation}
 |\BS(p)| \sim |\TSd(E;k,p)| \sim \frac{1}{p^{3/2}} \mathtext{as} p\to\infty
 \hspace{2em}\text{(\NLO)}
\end{equation}
for the wave functions (and the corresponding $\mathcal{T}$-matrix).  This
behavior is illustrated in Fig.~\ref{fig:WF-Log-NLO}, where we show the \NLO
wave functions up to large $p$, rescaled with a factor $p^{3/2}$.  As in the
leading-order case, this gives an oscillating function with approximately
constant amplitude.  Note that the oscillations are more rapid than at leading
order and that the exact log-periodicity is broken at \NLO.  This is
consistent with the behavior of the three-nucleon force $H(\Lambda)$ at
next-to-leading order (\cf~Fig.~\ref{fig:H-En-NLO-new}).  The $p^{-3/2}$
fall-off can be seen more clearly in the inlay of Fig.~\ref{fig:WF-Log-NLO},
where the deuteron-leg component of the wave function (without rescaling) is
plotted in a double-logarithmic scale.  The same scaling behavior as reported
here was found ---and explained based on analytical considerations---for the
scattering amplitude in Ref.~\cite{Platter:2006ev}.

To some extent this faster fall-off at next-to-leading order compensates the
scaling of the dibaryon propagators, but, as we shall discuss below, not by
enough to render the result convergent.

\subsection{Consequences}
\label{sec:He3-DiagramScaling}

We first consider the situation in the leading-order case.  The scaling of
the Coulomb kernel in Eq.~\eqref{eq:DeltaE-Kc} is determined there by the
large-momentum behavior of the bubble diagram without the 
approximation~\eqref{eq:Bubble-approx}.  From Eqs.~\eqref{eq:KCdt}
and~\eqref{eq:I-bubble-pd} we then have
\begin{equation}
 K_c^\LO(E;k,q) \sim \frac1{q^3} \mathtext{as} q\to\infty \,.
\label{eq:K_bub-scaling}
\end{equation}
Denoting in the following all loop momenta generically by $q$, we find
\begin{equation}
 \Big(\text{Fig.~\ref{fig:DeltaE}(a)}\Big)_{\text{at \LO}}
 \sim\; \left(\frac1q\right)^{\!2}_{\!\!\Bgen}
 \times \Big(q^3\Big)^{2}_{\!\text{loops}}
 \times \left(\frac1q\right)^{\!2}_{\!\!D}
 \times\left(\frac{1}{q^3}\right)_{\!\!\text{kernel}}
 \;\sim\; \frac1q \hspace{1.1em} \text{(\LO)}
\label{eq:DeltaE-bub-scale-LO}
\end{equation}
for the total scaling of the matrix element.  Since this is the only
contribution at leading order, the result for the energy shift converges as the
cutoff is increased.  From the same kind of analysis we can understand what led
to the cutoff dependence in the result of Ref.~\cite{Konig:2011yq}: the kernel
$V_C$ as defined in Eq.~\eqref{eq:V-C} only falls off like $1/q^2$
asymptotically such that the corresponding matrix element is logarithmically
divergent.\footnote{The same kind of divergence occurs if the bubble diagram is
approximated as in Eq.~\eqref{eq:Bubble-approx}.  However, as already mentioned
repeatedly, the approximation is not valid in the bound-state regime, and the
divergence that would appear if we were to use it there is just another
manifestation of this fact.}

Considering finally the situation at next-to-leading order, we find that we can
no longer avoid a divergent result.  For the contribution of the bubble diagram
we now obtain
\begin{equation}
 \Big(\text{Fig.~\ref{fig:DeltaE}(a)}\Big)_{\text{at \NLO}}
 \sim\; \left(\frac1{q^{3/2}}\right)^{\!2}_{\!\!\Bgen}
 \times \Big(q^3\Big)^{2}_{\!\text{loops}}
 \times \left(q^0\right)^{\!2}_{\!D}
 \times\left(\frac{1}{q^3}\right)_{\!\!\text{kernel}}
 \;\sim\; q^0 \hspace{0.9em} \text{(\NLO)} \,,
\label{eq:DeltaE-bub-scale-NLO}
\end{equation}
which gives a logarithmic divergence.  Even more problematic is the additional
contribution entering at this order.  For the kernel with the photon coupled
directly to the dibaryon we have the scaling
\begin{equation}
 \Big(\text{Fig.~\ref{fig:DeltaE}(b)}\Big)
 \sim\; \left(\frac1{q^{3/2}}\right)^{\!2}_{\!\!\Bgen}
 \times \Big(q^3\Big)^{2}_{\!\text{loops}}
 \times \left(q^0\right)^{\!2}_{\!D}
 \times\left(\frac{1}{q^2}\right)_{\!\!\text{kernel}}
 \;\sim\; q \,,
\label{eq:DeltaE-sim-scale-NLO}
\end{equation}
and thus a linear divergence.  This again also applies to the old calculation
involving $V_C$, which has the same scaling and thus diverges linearly as well.
Indeed, by comparing the dashed curves in Figs.~\ref{fig:H-En-LO-new}
and~\ref{fig:H-En-NLO-new} we see that the \NLO result has a much stronger
cutoff dependence than at leading order, and can understand this now from the
analysis carried out above.

Furthermore, we can explain as well why the new perturbative result at
\NLO looks almost like a mirror image of the old one.  If we rewrite $V_C$ in
Eq.~\eqref{eq:V-C} by factoring out a $y_{d,t}^2$, the result is just the
next-to-leading order contribution to $K_c^{(d,t)}$, but with the sign
reversed.  Due to the faster divergence of this term we barely see the effect of
the ``leading'' bubble diagram at all.

\subsection{Nonperturbative calculation}

The result found from the nonperturbative calculation also exhibits a strong
cutoff dependence at next-to-leading order, with the corresponding curve in
Fig.~\ref{fig:H-En-NLO-new} even rising somewhat faster than the perturbative
result.  From the analysis above one is led to suspect that again the scaling of
the involved diagrams is the origin of this effect.  In order to show this, we
look at what kind of diagrams are generated when the bound-state integral
equation is iterated.

\begin{figure}[htbp]
\centering
\begin{minipage}{0.49\textwidth}
 \centering
 \includegraphics[width=0.58\textwidth,clip]{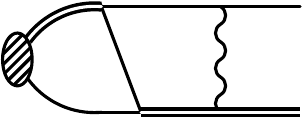}\\[0.77em]
 (a)
\end{minipage}\begin{minipage}{0.49\textwidth}
\centering
 \vspace*{0.5em}
 \includegraphics[width=0.58\textwidth,clip]{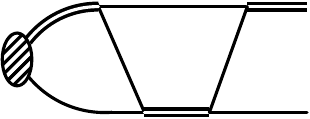}\\[0.77em]
 (b)
\end{minipage}
\caption{Two diagrams generated by iterating the bound-state integral equation.}
\label{fig:B-iteration}
\end{figure}

Figure~\ref{fig:B-iteration}(a) shows a two-loop diagram with a Coulomb-photon
exchange that is generated by iterating the equation once.  Since the
one-nucleon exchange also scales like $1/q^2$, we find a total scaling $\sim
q^{1/2}$ for the diagram by applying the same kind of analysis as in the
sections above.  This indeed indicates a divergence.

We note that a similar diagram with two subsequent nucleon exchanges, shown in 
Fig.~\ref{fig:B-iteration}(b) actually has the same behavior for large momenta. 
This means that, based on the arguments above, one has a divergence already in 
the system without Coulomb effects.  However, each nucleon exchange is always 
accompanied by a vertex from the three-nucleon force.  According to the 
renormalization prescription this, three-nucleon force is adjusted at each 
cutoff in order reproduce a three-body experimental input (the triton binding 
energy, in our case).  At \NLO, this procedure effectively also absorbs the 
divergence into $H(\Lambda)$.  The important point is now that by adding the 
Coulomb contributions into the equation after $H(\Lambda)$ is already fixed, 
the divergence problem is reintroduced.

At leading order, the situation is different because there the $1/q$-scaling of
the dibaryon propagators ensures that additional loops generated by iterating
the integral equation do not create divergences, neither due to Coulomb
photons nor due to the nucleon-exchange interaction.  The three-nucleon force
is needed in this case only to fix the oscillating behavior of the scattering
amplitude~\cite{Bedaque:1999ve}.

\subsection{Back to the scattering regime}
\label{sec:ScatteringRevisited}

\begin{figure}[htbp]
\centering
\includegraphics[width=0.85\textwidth,clip]{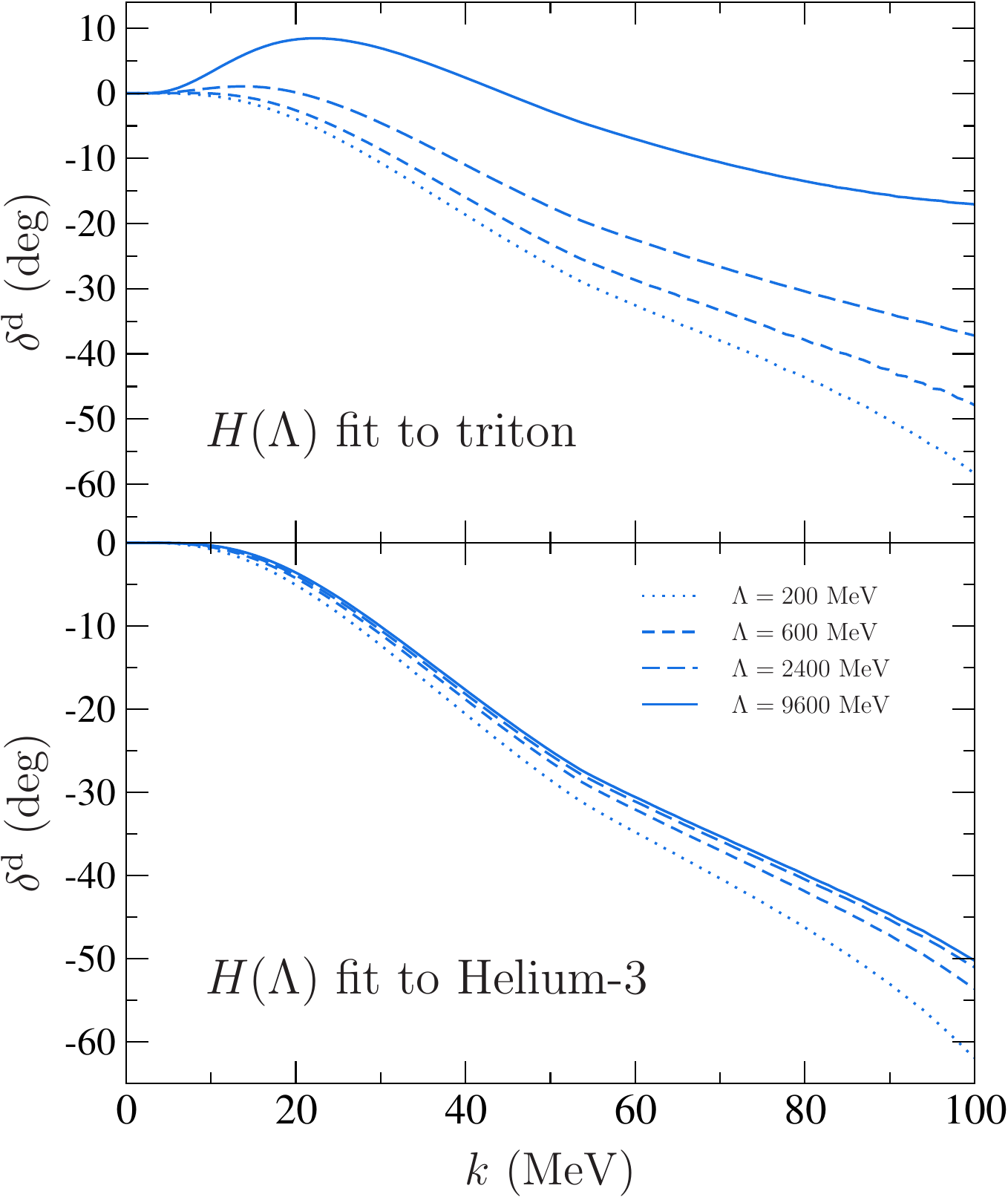}
\caption{\NLO{} $p$--$d$ scattering phase shifts for large cutoffs obtained
using Eq.~\eqref{eq:pd-IntEq-D-full-more}.  Upper panel: $H(\Lambda)$ fit to
reproduce triton binding energy in the $n$--$d$ system.  Lower panel:
$H(\Lambda)$ fit to reproduce $^3$He binding energy in the $p$--$d$ system.}
\label{fig:Phase-D-3HeFit-NLO}
\end{figure}

The above findings also raise some questions concerning the \NLO scattering
calculation in the doublet channel.  If the assertion that the system is
not renormalized correctly is true, the effect should also show up in the
cutoff-variation of the scattering phase shifts.  Indeed, this is what we find.
If we perform the phase-shift calculation with cutoffs of the same size as in
Fig.~\ref{fig:H-En-NLO-new} (up to $\Lambda \approx 10\,000~\MeV$), the result
does not seem to converge.  Instead, the curves shown for four different
cutoffs in the upper panel of Fig.~\ref{fig:Phase-D-3HeFit-NLO} move upwards
with increasing $\Lambda$.

One can of course take the stance, as we have done in Ref.~\cite{Konig:2011yq}
by varying the cutoff only between 200 and 600~\MeV, that the cutoff in an EFT
calculation should be taken of a natural order of magnitude (defined by the
scale of physics left out from the theory).  According to
Fig.~\ref{fig:H-En-LO-new}, the leading-order Helium-3 results are indeed
converged at these cutoffs.  At next-to-leading order, however, 
Fig.~\ref{fig:H-En-NLO-new} shows that the results are problematic already in 
that regime.  With this in mind, considering cutoffs far beyond the natural 
size here and in the preceding sections is primarily a tool to expose and 
analyze this behavior more clearly.

Using the full equation structure~\eqref{eq:pd-IntEq-D-full-more} to fit the
three-nucleon force such that it reproduces the experimental $^3$He energy
turns out to remove the strong cutoff dependence of the $p$--$d$ results
(without affecting the results for lower cutoffs very much), just as it does
in the $n$--$d$ system when $H(\Lambda)$ is fixed to reproduce the triton
binding energy.  This is shown in the lower panel of
Fig.~\ref{fig:Phase-D-3HeFit-NLO}, where we plot the curves analogous to those
in the upper panel obtained by re-fitting the three-nucleon force.  For
consistency we have used the same equation structure as in the bound-state
regime---including the box and triangle diagrams and without the approximation
of the bubble diagram---in the calculations of all phase-shift results shown in
Fig.~\ref{fig:Phase-D-3HeFit-NLO}.

These findings are a further indication that in the $p$--$d$ system there is a
new three-body counterterm, which at least numerically can be absorbed by
refitting $H(\Lambda)$.  Such a situation could be accounted for by adding to
the three-body Lagrangian~\eqref{eq:L-3} a piece proportional to the charge
operator (making it vanish in the triton system).  Since it is not forbidden
by symmetry, such a term should be there, and the only question is at which
order in the EFT counting it enters.  As already mentioned, it has
analytically been shown to enter at \NLO in a fully perturbative
calculation~\cite{Vanasse:2014kxa}.

\subsection{Neglected diagrams and the bubble approximation}

Although the main motivation for deriving the full equation
structure~\eqref{eq:pd-IntEq-D-full-more} with the box and triangle diagrams
included was the nonperturbative Helium-3 calculation, with the complete
expression at hand we are now also in a position to directly check the
approximations made in the scattering regime.  Based on the power counting
described in Section~\ref{sec:PowerCounting}, we neglected there the
contributions from the box and triangle diagrams, and furthermore used the
approximation given in Eq.~\eqref{eq:Bubble-approx} for the bubble diagram.

\begin{figure}[phtb]
\centering
\includegraphics[width=0.8\textwidth,clip]{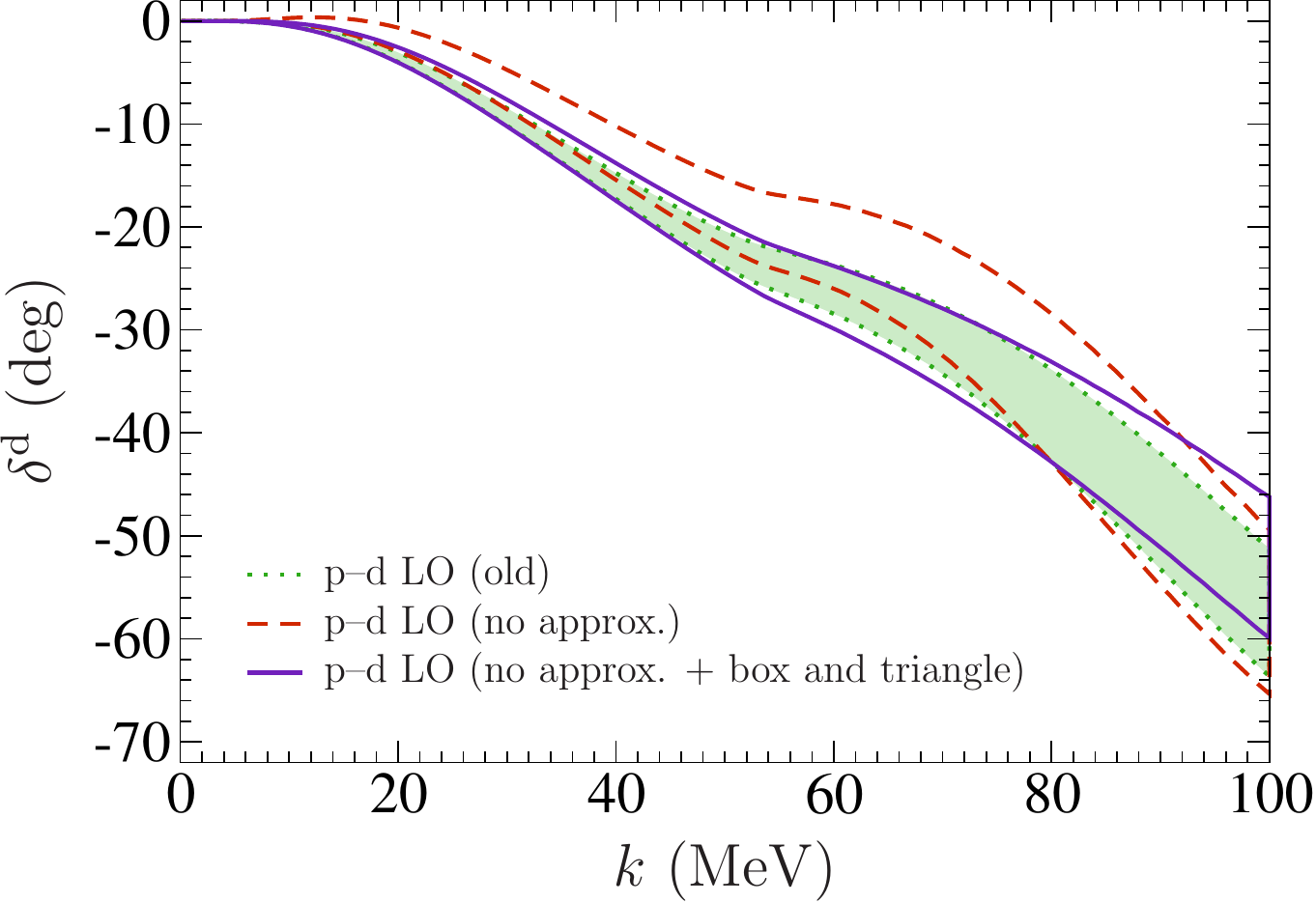}
\caption{Leading order $p$-$d$ doublet channel S-wave scattering phase shifts
as functions of the center-of-mass momentum $k$.  Dotted curve: old result 
from Ref.~\cite{Konig:2011yq}.  Dashed curve: result without approximating the 
bubble diagram.  Solid curve: same as dashed curve with additional Coulomb 
diagrams (box and triangle) included.  Error bands generated by cutoff 
variation within 200--600~\MeV.}
\label{fig:Phase-Nd-D-more-LO}
\end{figure}

To avoid interference with the problematic situation at \NLO discussed in
the previous section, we go back to the leading-order calculation and show in
Fig.~\ref{fig:Phase-Nd-D-more-LO} the old band (as given in Fig.~11 of
Ref.~\cite{Konig:2011yq}) together with the result obtained without
approximating the bubble diagram (dashed curve) and furthermore what we get
from the full equation with the box and triangle contributions included (solid
curve).  As before, all error bands were generated by varying the cutoff
between 200 and 600~\MeV, which is certainly sufficient at leading order.

The band from the full calculation overlaps well with the old result.  In
fact, if one takes the typical 30\% of a leading-order pionless EFT
calculation as a more appropriate estimate of the uncertainty, the two bands
are almost indistinguishable.  It would be tempting to interpreted this as an
\aposteriori{} confirmation of both the Coulomb power counting for the
scattering regime and the approximation used for the bubble diagram, if it
were not for the result with only the bubble approximation turned off.  Since
the corresponding band in Fig.~\ref{fig:Phase-Nd-D-more-LO} is broader and
consistently shifted upwards (for momenta $k$ below about $70~\MeV$), it turns
out that really the combination of both things---approximating the bubble
diagram and neglecting the additional contributions---is important to get the
same result as from the full calculation.  This indicates that in the latter
case there are some substantial cancellations between the effects of the
individual Coulomb diagrams.

\begin{figure}[phtb]
\centering
\includegraphics[width=0.8\textwidth,clip]{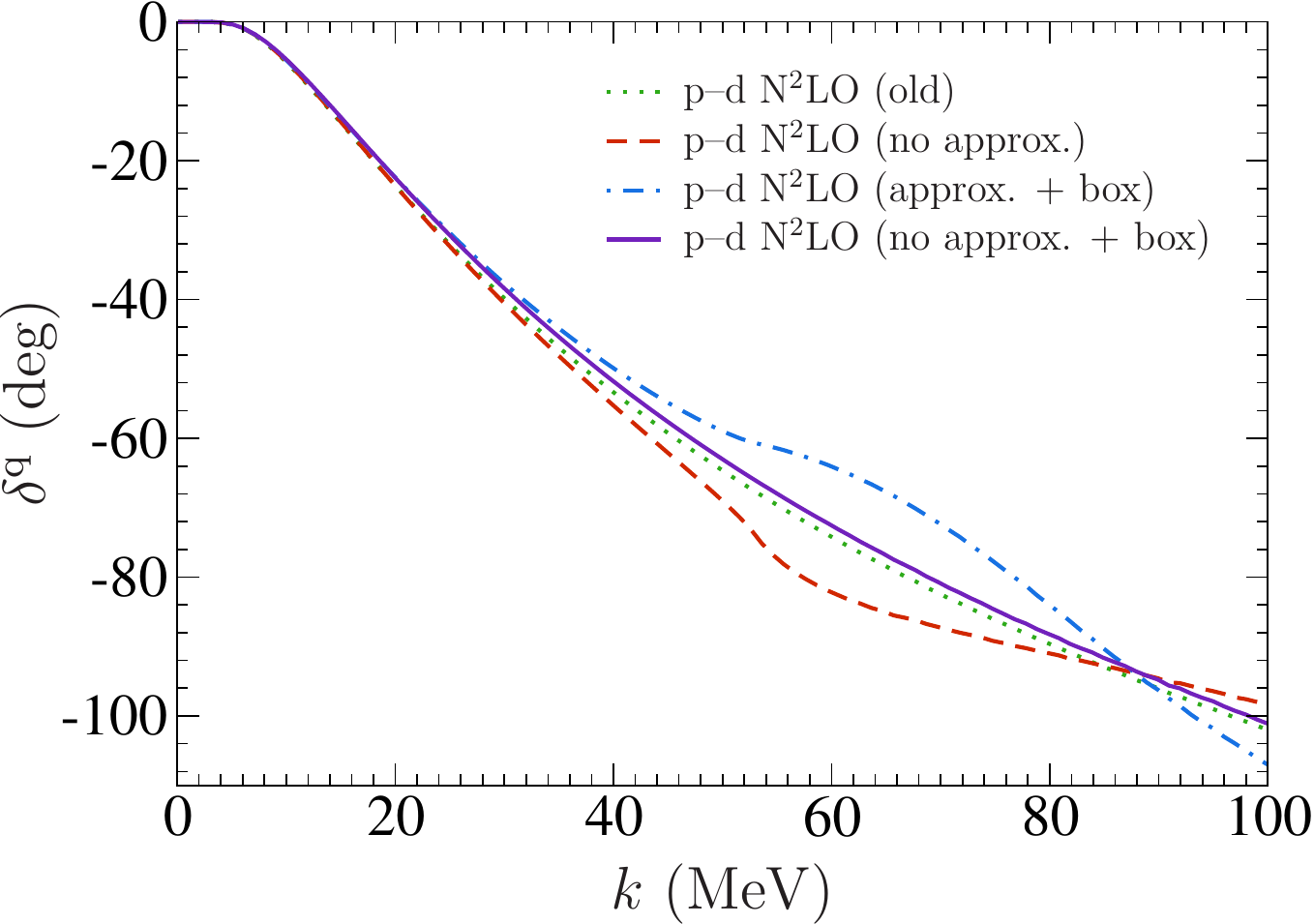}
\caption{\NNLO $p$-$d$ quartet channel S-wave scattering phase shifts
as functions of the center-of-mass momentum $k$.  Dotted curve: old result 
from Ref.~\cite{Konig:2011yq}.  Dashed curve: result without
approximating the bubble diagram.  Dash-dotted curve: result with the additional
Coulomb box diagram box included (but the bubble diagram still approximated).
Solid curve: same as dash-dotted curve, but without approximating the bubble
diagram.  All curves were calculated at a cutoff $\Lambda=140~\MeV$.}
\label{fig:Phase-Q-more-N2LO}
\end{figure}

A similar effect occurs in the quartet channel.  In this case we have
additionally calculated the result with the bubble diagram still approximated,
but the box diagram\footnote{Note that the triangle diagram does not appear in
the quartet channel since there are never two protons in the intermediate
dibaryon.} already included.  Overall, the results from the individual
calculations shown in Fig.~\ref{fig:Phase-Q-more-N2LO} differ now not so much 
at low center-of-mass momenta, but quite substantially above the deuteron 
break-up threshold ($k\gsim52~\MeV$).  Since the error bands would be as narrow 
as in Fig.~10 of Ref.~\cite{Konig:2011yq} and are not essential for what we 
want to show here, we have used a single cutoff $\Lambda=140~\MeV$ to generate 
the curves.

The cancellation between the two effects is particularly striking here: the
result obtained with both the bubble-diagram approximation turned off and the
box diagram included at the same time (solid curve) is almost identical to the
old result (dotted curve).  We have shown here the results at \NNLO where above 
the breakup threshold the effect of the cancellation is in fact somewhat larger 
than the expected uncertainty from the EFT expansion at this order.  The same 
pattern is found also at lower orders calculations.

Altogether, these findings cast some doubt on the Coulomb power counting as
discussed in Section~\ref{sec:PowerCounting}.  In particular, the good
agreement of the $\NNLO$ quartet-channel phase shifts with experimental data
that was found based on this counting scheme already in
Ref.~\cite{Rupak:2001ci} appears now to be somewhat accidental, at least at
higher energies.  Based on the findings obtained here we propose that instead of
using the old counting it might be better---also in the scattering 
regime---to simply include already at leading order all $\OO(\alpha)$ Coulomb 
diagrams if they are not, like Fig.~\ref{fig:LeadingCoulomb}(d), formally range 
corrections.  At the same time, as already alluded to below 
Eq.~\eqref{eq:Bubble-approx}, one should not use the approximation for the 
bubble diagram because it has no physical justification and turns out to have a 
more significant impact on the result than naïvely expected.

This new scheme\footnote{We would like to emphasize here that the new scheme
does not imply to do strict perturbation theory in $\alpha$.  Since we want to 
calculate scattering at low energies, the included Coulomb diagrams are still 
iterated nonperturbatively to all orders, which means that the perturbation 
expansion up to $\OO(\alpha)$ is applied to the \emph{kernels} used in the 
Lippmann--Schwinger equations.} has the advantage of being much more 
straightforward and at the same time completely consistent throughout both the 
scattering and the bound-state regime.  The price to be paid for this is that 
the calculations to be carried out are quite a bit more involved.  With 
suitable adaptive integration routines to carry out the S-wave projections of 
the Coulomb diagrams numerically, however, the integral equations can still be 
solved on ordinary desktop computers.

\section{Higher-order Coulomb contributions}
\label{sec:HigherCoulomb}

Having concluded that the original Coulomb counting scheme should be modified
raises the question how large higher-order contributions (in $\alpha$) really
are.  Very close to the zero-momentum threshold one would naïvely think them
to be quite important.

\subsection{Diagrams with full off-shell Coulomb T-matrix}

\begin{figure}[htbp]
\centering
\begin{minipage}{0.33\textwidth}
 \centering\vspace*{-0.48em}
 \includegraphics[width=0.52\textwidth,clip]{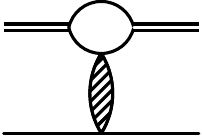}\\[0.77em]
 (a)
\end{minipage}\begin{minipage}{0.33\textwidth}
\centering
 \vspace*{0.5em}
 \includegraphics[width=0.65\textwidth,clip]{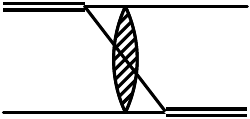}\\[0.77em]
 (b)
\end{minipage}
\begin{minipage}{0.33\textwidth}
\centering
 \vspace*{0.5em}
 \includegraphics[width=0.65\textwidth,clip]{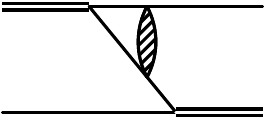}\\[0.77em]
 (c)
\end{minipage}
\caption{Diagrams involving the full off-shell Coulomb T-matrix (indicated by
the blob).}
\label{fig:LeadingCoulomb-full}
\end{figure}

Since the Coulomb interaction is strongest directly at threshold, the
scattering calculation should be a good testing ground for checking the
influence of higher-order Coulomb diagrams that we have not taken into account
so far.  Rather than simply including $\OO(\alpha^2)$ diagrams (with two
Coulomb-photon exchanges) to estimate the effect of higher-order terms, we
adopt the approach of Ando and Birse~\cite{Ando:2010wq} and directly include
the full off-shell Coulomb T-matrix in the diagrams.  Effectively, this resums
all subsequent Coulomb-photon exchanges for a given topology.  The resulting
diagrams are shown in Fig.~\ref{fig:LeadingCoulomb-full}.

In Ref.~\cite{Ando:2010wq}, Ando and Birse only consider the bound-state regime 
and argue that the finite extent of the wave functions helps to regularize the 
Coulomb singularity and thus use an unscreened interaction.  Since we cannot 
apply this approach in the scattering regime, we use here a ``partially 
screened'' Coulomb T-matrix $T_{C,\lambda}$ that is described in more detail in 
Appendix~\ref{sec:CoulombT-screened}.  Going back to an approach by 
Gorshkov~\cite{Gorshkov:1961ab,Gorshkov:1965ab}, $T_{C,\lambda}$ is defined by 
the operator equation
\begin{equation}
 \hat{T}_{C,\lambda} = \hat{V}_{C,\lambda}
 + \hat{V}_{C,\lambda}\,\hat{G}_0^{(+)}\,\hat{T}_C \,,
\end{equation}
where $\hat{T}_C$ is the exact unscreened Coulomb T-matrix and $\hat{G}_0^{(+)}$ 
is the free two-particle resolvent (Green's function).  In 
Appendix~\ref{sec:CoulombT-screened} we show that very similar to the 
unscreened Coulomb T-matrix, $T_{C,\lambda}$ can be expressed in terms of 
hypergeometric functions as
\begin{multline}
 T_{C,\lambda}(k;\vecp,\vecq) \\ = V_C(\vecp,\vecq) \Big\{1-\Delta_\lambda^{-1}
 \big[{_2F_1}\!\left(1,\ii\eta,1+\ii\eta;X_\lambda^-\right)
 -{_2F_1}\!\left(1,\ii\eta,1+\ii\eta;X_\lambda^+\right)
 \big]\Big\} \,,
\label{eq:TC-screened-hyp-2}
\end{multline}
with $\Delta_\lambda$ and $X_\lambda^\pm$ as defined in 
Eqs.~\eqref{eq:Delta-lambda} and~\eqref{eq:X-pm}, respectively.

We assume that using $T_{C,\lambda}$ in the $p$--$d$ integral equations gives an 
appropriate description for small photon masses $\lambda$ because in the limit 
$\lambda\to0$, $T_{C,\lambda}$ converges to the exact (unscreened) Coulomb 
T-matrix $T_C$.

\subsubsection{Bubble diagram}
\label{sec:FullBubble}

The most important contribution is again given by the ``full bubble diagram''
shown in Fig.~\ref{fig:LeadingCoulomb-full}a.  Its spin- and isospin structure
is exactly the same as for the leading expression with just a single
Coulomb-photon exchange.  However, the loop integral, which before could be
evaluated analytically, now has the more complicated form
\begin{multline}
 \mathcal{I}_\text{bubble}^\text{full}(E;\vk,\vp)
 = \int_0^\infty \dd q\,q^2\,
 T_{C,\lambda}\!\left(\ii\sqrt{3q^2/4-\MN E};\veck,\vecp\right) \\
 \times\int_{-1}^1\dd\!\cos\theta'\int_0^{2\pi}\frac{\dd\phi'}
 {a(b+c\cdot\cos\phi')}
\end{multline}
with
\begin{subequations}%
\begin{align}
 a &= k^2 + kq\cos\theta' + q^2 - \MN E \,, \\
 b &= p^2 + pq\cos\theta\cos\theta' + q^2 - \MN E \,, \\
 c &= pq\sin\theta\sin\theta' \,.
\end{align}
\label{eq:FullBubble-abc}%
\end{subequations}%
Here, as in Eqs.~\eqref{eq:K-box} and~\eqref{eq:K-tri-out}, $\theta$ is the
angle between the vectors $\vk$ and $\vp$, whereas $\theta'$ denotes the angle
between $\vk$ and the loop momentum $\vq$, and the azimuthal angle $\phi'$
enters through rewriting the angle between $\vp$ and $\vq$.  It can be
integrated over analytically with the result
\begin{equation}
 \int_0^{2\pi}\frac{\dd\phi'}{b+c\cdot\cos\phi'}
 = \frac{2\pi}{\sqrt{b^2-c^2}}
\end{equation}
for $b>c$, which, according to Eqs.~\eqref{eq:FullBubble-abc} is fulfilled for
the scattering-length calculation.  Only above the deuteron breakup threshold,
where the energy $E$ is positive, one would have to be more careful at this
point.

Setting
\begin{equation}
 \cos\theta' \equiv x \;\Rightarrow\; \sin\theta' = \sqrt{1-x^2} \,,
\end{equation}
the remaining angular integral has the form
\begin{equation}
 \int_{-1}^1 \frac{\dd x}{(A+B\cdot x)\sqrt{C+D\cdot x+E\cdot x^2}}
\end{equation}
with
\begin{subequations}%
\begin{align}
 A &= k^2 + q^2 - \MN E \,, \\
 B &= kq \,, \\
 C &= (p^2 + q^2 - \MN E)^2 - p^2 q^2(1-\cos^2\theta) \, \\
 D &= (p^2 + q^2 -\MN E) \times 2pq\cos\theta \, \\
 E &= p^2 q^2 \,.
\end{align}
\label{eq:FullBubble-ABC}%
\end{subequations}%
Using integration by parts, this can be done analytically.  The final result for
the full-bubble kernel function is then\footnote{Note that the expression in
Eq.~\eqref{eq:FullBubble-result} does not directly reduce to
$\mathcal{I}_\text{bubble}$ as defined in Eq.~\eqref{eq:I-bubble-pd}
when the Coulomb T-matrix is replaced with a single photon exchange.  Rather,
in that limit, the kernel function $K_\text{bub}^{\text{(full)}}(E;k,p)$ goes
over into the \LO-part of Eq.~\eqref{eq:KCdt}, including the overall prefactor
$-\alpha\MN$.}
\begin{equation}
 K_\text{bub}^{\text{(full)}}(E;k,p) = {-\frac{\MN}{2\pi^2}}
 \times\frac12\int_{-1}^1\dd\!\cos\theta
 \,\mathcal{I}_\text{bubble}^\text{full}(E;\vk,\vp)
\label{eq:K-bub-full}
\end{equation}
with
\begin{multline}
 \mathcal{I}_\text{bubble}^\text{full}(E;\vk,\vp)
 = \int_0^\infty \dd q\,q^2\,
 T_{C,\lambda}\!\left(\ii\sqrt{3q^2/4-\MN E};\veck,\vecp\right) \\
 \times\frac1{\sqrt{F}}\left[\log\!\left(\frac{A+B}{A-B}\right)
 - \log\!\left(\frac{B(2C+D)-A(D+2E)+2\sqrt{F}\sqrt{C+E+D}}
 {B(2C-D)-A(D-2E)+2\sqrt{F}\sqrt{C+E-D}}\right)\right] \,,
\label{eq:FullBubble-result}
\end{multline}
where in addition to Eqs.~\eqref{eq:FullBubble-ABC} we have defined
\begin{equation}
 F = B^2C + A^2E - ABD \,.
\end{equation}
The remaining momentum integral and the S-wave projection in
Eq.~\eqref{eq:K-bub-full} have to be carried out numerically.

Unfortunately, similar tricks cannot be used to simplify the expression for the
full box and triangle diagrams shown in Figs.~\ref{fig:LeadingCoulomb-full}b
and~c.  In those cases, the Coulomb T-matrix depends non-trivially on the loop
momentum, which means that the angular integrations can no longer be carried
out analytically.

\subsubsection{Box diagram}
\label{sec:FullBox}

The full box diagram, Fig.~\ref{fig:LeadingCoulomb-full}b, can be written as
\begin{equation}
 K_\text{box}^{\text{(full)}}(E;k,p) = {-\frac{\MN}{8\pi^3}}
 \times\frac12\int_{-1}^1\dd\!\cos\theta
 \,\mathcal{I}_\text{box}^\text{full}(E;\vk,\vp) \,.
\label{eq:K-box-full}
\end{equation}
The loop integral is
\begin{equation}
 \mathcal{I}_\text{box}^\text{full}(E;\vk,\vp) = \int\dd^3q\,
 \frac{T_{C,\lambda}\!\left(\ii\sqrt{3q^2/4-\MN E};\veck-\vecq,-\vecp\right)}
 {\left(\vecq^2-\vecq\cdot\veck+\veck^2-\MN E\right)
 \left(\vecq^2-\vecq\cdot\vecp+\vecp^2-\MN E\right)}
\label{eq:I-box-full}
\end{equation}
with
\begin{equation}
 \int\dd^3q = \int_0^\infty\dd q\,q^2\int_0^\pi\dd\theta'\int_0^{2\pi}\dd\phi'
\end{equation}
and the angles
\begin{subequations}%
\begin{align}
 \vecq\cdot\veck &= qk\cos\theta' \,, \\
 \vecq\cdot\vecp &= pq(\cos\theta\cos\theta'
 +\sin\theta\sin\theta'\cos\phi') \,.
\end{align}
\end{subequations}%
Recall that $\theta$ is the angle between $\vk$ and $\vp$.  As mentioned above, 
all four integrations in this expression (three angles plus the loop momentum) 
have do be carried out numerically.

\subsection{Comparison of quartet-channel phase shifts}

Performing the integrals for the full bubble diagram as given in 
Eq.~\eqref{eq:K-bub-full} is still feasible with standard numerical quadrature 
rules.  For the four-dimensional integration necessary to evaluate the full box 
diagram, however, we use the Vegas Monte-Carlo algorithm implemented in the 
CUBA library~\cite{Hahn:2004fe}.  To evaluate the Coulomb T-matrix we use the 
expression in terms of hypergeometric functions given in 
Eq.~\eqref{eq:TC-screened-hyp-2}, which can be implemented efficiently with the 
fast routines for ${_2F_1}$ from Ref.~\cite{Michel:2008ab}.

\begin{figure}[phtb]
\centering
\includegraphics[width=0.8\textwidth,clip]{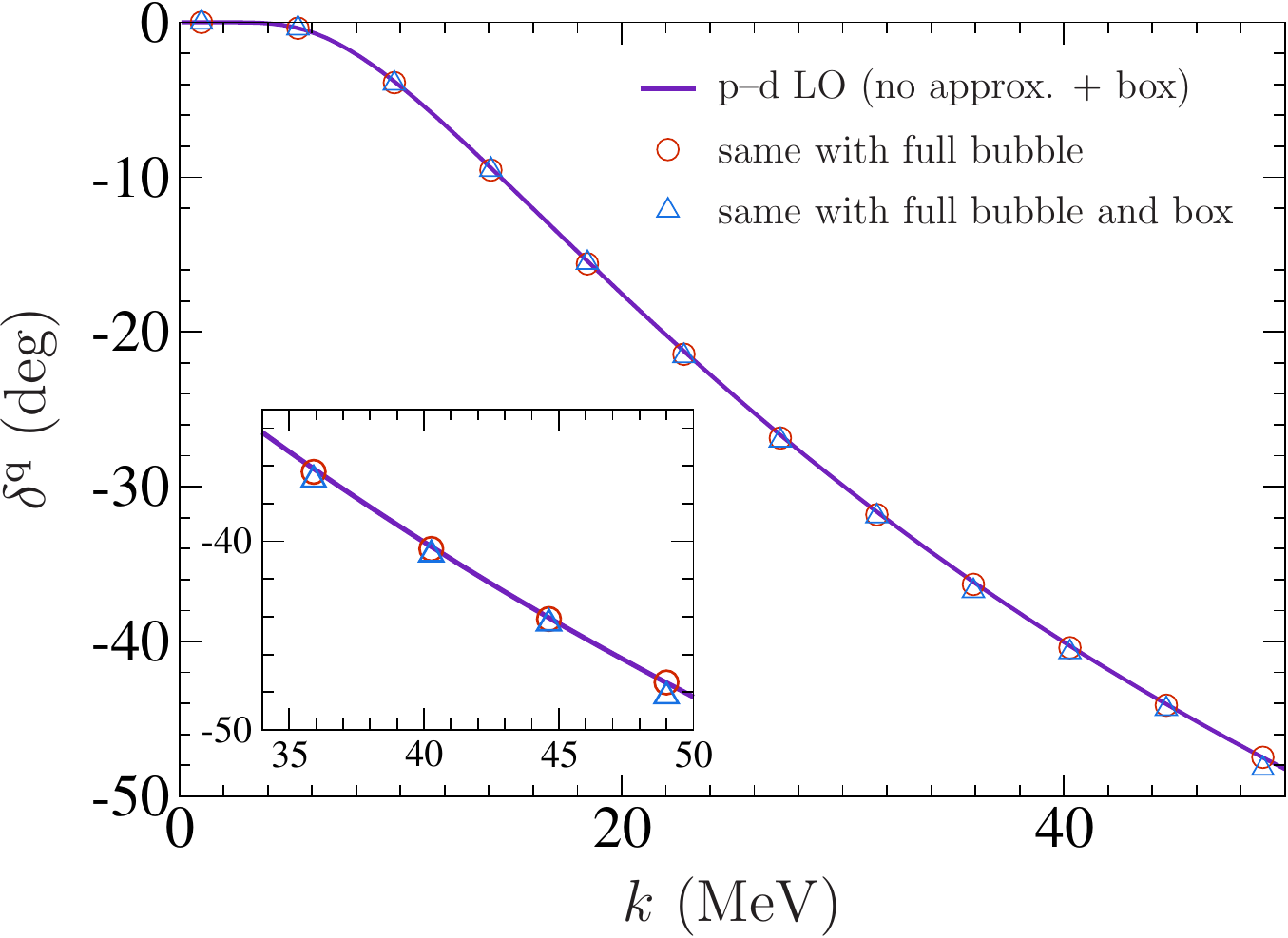}
\caption{Leading order $p$-$d$ quartet channel S-wave scattering phase shifts
as functions of the center-of-mass momentum $k$.  Solid curve: calculation with 
ordinary bubble (no approximation) and box diagrams.  Circles: calculation with 
full bubble diagram.  Triangles: calculation with full bubble and box diagrams. 
All curves were calculated at a cutoff $\Lambda=140~\MeV$.  See text for more 
details.}
\label{fig:Phase-Nd-Q-full-LO}
\end{figure}

\medskip
Results for leading-order quartet-channel phase shifts at $\Lambda=140~\MeV$ 
are shown in Fig.~\ref{fig:Phase-Nd-Q-full-LO}.  The plot demonstrates that for 
scattering momenta below the deuteron breakup threshold, higher-order Coulomb 
corrections clearly are negligible.  Only for center-of-mass momenta larger 
than about $35~\MeV$ does one see noticeable differences from the result with 
only the leading $\OO(\alpha)$ bubble and box diagrams included (solid curve in 
Fig.~\ref{fig:Phase-Nd-Q-full-LO}).  One can understand this by noting that the 
energy flowing into $T_{C,\lambda}$ is always given by $3q^2/4-\MN E$, where 
$q$ is the loop momentum.  Since in the scattering regime $E = 3k^2/(4\MN) - 
\gamd^2/\MN$, this can only be close to zero if $k\approx\sqrt{4/3}\gamd$.  For 
energies above the breakup threshold the full Coulomb T-matrix is expected to 
have a stronger influence~\cite{Kok:1979aa,Kok:1981aa}, but in that regime the 
numerical calculation is even more challenging (due to imaginary parts opening 
up) and has not been performed yet.  We can conclude, however, that for $p$--$d$
scattering at very low energies it is indeed valid to only include
$\OO(\alpha)$ Coulomb diagrams in the integral equation kernels.

\section{Summary and outlook}

In this paper we have studied several aspects of the proton--deuteron system
in pionless effective field theory.  In the bound-state regime, we have 
amended our previous~\cite{Konig:2011yq} perturbative calculation of the 
$^3$He--$^3$H binding energy difference and additionally performed a 
nonperturbative calculation of the $^3$He binding energy that agrees very well 
with the result obtained by Ando and Birse~\cite{Ando:2010wq} 
although we have only used Coulomb photons to include electromagnetic effects.  
Furthermore, at leading order both the perturbative and the nonperturbative 
calculation give results that are, within their respective uncertainties, in 
good agreement with the experimental value for the $^3$He binding energy.

At next-to-leading order, the results of both calculations exhibit a strong
cutoff dependence, indicating that the calculation is not renormalized
properly.  We have analyzed this situation by studying the ultraviolet
behavior of the diagrams entering in the calculation and argued that this
provides evidence for a new three-body counterterm being necessary to
renormalize the charged doublet-channel system at \NLO.  Indeed such a term
has been identified in Ref.~\cite{Vanasse:2014kxa}.

Re-examining the doublet-channel scattering calculation we find that the phase 
shifts presented in Ref.~\cite{Konig:2011yq} turn out not to converge when the 
cutoff in the calculation is chosen very large.  Consistent with a new 
counterterm entering at \NLO, this cutoff dependence however goes away
when one refits a single three-nucleon force to reproduce the  experimental 
$^3$He binding energy.  Since the modified fitting of the three-nucleon
force requires using the same equation structure in the bound-state and
scattering regime, we have critically reviewed the Coulomb power counting of
Rupak and Kong~\cite{Rupak:2001ci} and come to the conclusion that it might be
better to simply include all Coulomb diagrams of a given order in $\alpha$. 
This approach is simpler and more consistent because it treats the scattering
and bound-state regimes on an equal footing.

Finally, we have shown that higher-order (in $\alpha$) Coulomb contributions 
can safely be neglected in quartet-channel $p$--$d$ scattering.  With that, it 
is interesting to proceed further towards the zero-energy threshold and also 
calculate Coulomb-modified scattering lengths.  A first step in that direction 
has already been made in Ref.~\cite{Koenig:2013}.  Recently, a fully 
perturbative calculation of the quartet-channel scattering length up to \NNLO  
has been presented in Ref.~\cite{Konig:2013cia}.  In the future, we plan on 
analyzing that calculation in more detail and furthermore to extend the 
approach also to the doublet channel.

\begin{acknowledgments}
We thank Dean Lee, Daniel R. Phillips, Jared Vanasse and U. van Kolck for useful
discussions, Martin Hoferichter for help with the calculation of the box
diagram, and Philipp Hagen for suggestions concerning the trimer $Z$-factor.
HWG is particularly indebted to the Nuclear Theory groups at U.~Bonn and FZ
Jülich for their hospitality during his Sabbatical stay.  S.K. furthermore 
thanks the ``Studien\-stiftung des deutschen Volkes'' and the Bonn-Cologne 
Graduate School of Physics and Astronomy for their support.  This research 
was supported in part by the NSF under Grant No. PHY--1306250 and by the 
NUCLEI SciDAC Collaboration under DOE Grant DE-SC0008533 (SK), by the DFG 
through SFB/TR~16 ``Subnuclear structure of matter'' , by the BMBF under grant 
05P12PDFTE, by the Helmholtz Association under contract HA216/EMMI (HWH, SK), as 
well as by the US-Department of Energy under contract DE-FG02-95ER-40907, by the 
Deutsche Forschungsgemeinschaft and the National Natural Science Foundation of 
China through funds provided to the Sino-German CRC 110 ``Symmetries and the 
Emergence of Structure in QCD'', and by the EPOS network of the European 
Community Research Infrastructure Integrating Activity ``Study of Strongly 
Interacting Matter HadronPhysics3" (HWG).
\end{acknowledgments}

\appendix

\section{T-matrix factorization and wave function normalization}
\label{sec:BoundStates}

We discuss here in some detail the factorization of the $\mathcal{T}$-matrix 
given in Eq.~\eqref{eq:T-factorization} and the normalization 
condition~\eqref{eq:Triton-WF-norm} for the trinucleon wave functions.  The 
latter has already been discussed in the appendix of Ref.~\cite{Konig:2011yq}.
The following material is based on an extension of those results, originally 
presented in Ref.~\cite{Koenig:2013}.  We point out again that nontrivial 
normalization conditions for bound-state wave functions arising from 
energy-dependent interactions have been known for a long 
time~\cite{Agrawala:1966ab}.  The purpose of this appendix is to present a 
self-contained and rather pedagogical derivation, originally based on that 
given in Lurie's textbook~\cite{Lurie:1968}.

\subsection{Bethe--Salpeter equation}
\label{sec:BS-BS}

As in Ref.~\cite{Konig:2011yq} we consider a simplified nucleon--deuteron 
system where we neglect the spin-singlet dibaryon for the moment.  Our starting 
point is the Bethe--Salpeter in momentum space,
\begin{equation}
 G(k,p;P) = G_0(k,p;P) + \int\dfq{q}\int\dfq{q'}\,G(k,q;P)
 \,K(q,q';P)\,G_0(q',p;P) \,,
\label{eq:BS-ms}
\end{equation}
where $G$ denotes the full two-body nucleon--deuteron Green's function (related 
directly to the S-matrix).  $G_0$ describes the momentum-conserving free 
propagation of the individual particles,
\begin{equation}
 G_0(k,p;P) = (2\pi)^4\fvdelta(k-p)
 \cdot\Delta_d\!\left(\eta_d P+p\right) \Delta_N\!\left(\eta_N P-p\right)
\label{eq:G0-ms}
\end{equation}
with the total four-momentum $P$ and the mass ratios $\eta_{N,d}$ fulfilling 
$\eta_d+\eta_N=1$.  The interaction kernel $K$ is given by the (spin/isospin 
and S-wave projected) one-nucleon-exchange diagram,
\begin{equation}
 K(k_0,\vk,p_0,\vp;E) = \frac{-{\ii\yd^2/2}}
 {\eta_dE - \eta_NE + k_0 + p_0-\frac{(\vk+\vp)^2}{2\MN}+\ii\eps} \,,
\label{eq:1N-exchange}
\end{equation}
In order to describe the real doublet-channel system of Sec.~\ref{sec:WF}, all 
quantities defined above become matrices in channel space and one has to include 
the three-nucleon contact interaction in $K$.

Assuming the existence of a trinucleon bound state (the triton in our current
toy model) at an energy $E=-{E_B}<0$, one can show 
that~\cite{Koenig:2013,Lurie:1968}
\begin{equation}
 G(k,p;P)=\ii\frac{\psi_{B\vP}(p)\psi^\dagger_{B\vP}(k)}{E+E_B+\ii\eps}
 \ +\ \text{terms regular at $P_0 = E = -{E_B}$} \,,
\label{eq:G-factor}
\end{equation}
\ie\ $G$ factorizes at the bound state pole.  Inserting the above
factorization into Eq.~\eqref{eq:BS-ms} and multiplying by $(E+E_B)$ we find 
the homogeneous Bethe--Salpeter equation
\begin{equation}
 \int\dfq{q}\,\left[G_0^{-1}(q,p;P)-K(q,p;P)\right]\psi_{B\vP}(q)=0
\label{eq:BSE-hom-MS}
\end{equation}
after taking the limit $E\to{-E_B}$ and cancelling non-zero factors.
Equivalently, we can also obtain this in the form
\begin{equation}
 \psi_{B\vP}(p) = \Delta_d\!\left(\eta_dP+p\right)
 \Delta_N\!\left(\eta_NP-p\right)
 \cdot\int\dfq{q}\,K(q,p;P) \, \psi_{B\vP}(q)
\label{eq:BSE-hom-MS-2}
\end{equation}
from Eq.~\eqref{eq:BS-ms}.

\subsection{Three-dimensional reduction}
\label{sec:BS-3D}

We now consider a bound state at rest, $P=({-E_B},\vZero)$, and define the
\emph{amputated wave function}
\begin{equation}
 \Bgen(p_0,\vp) = \psi_{B\vZero}(p_0,\vp)
 \cdot\left[\Delta_d\!\left({-\eta_dE_B}+p_0,\vp\right)\right]^{-1}
 \cdot\left[\Delta_N\!\left({-\eta_NE_B}-p_0,\vp\right)\right]^{-1} \,,
\label{eq:BS-amputated}
\end{equation}
which fulfills the equation
\begin{equation}
 \Bgen(p_0,\vp)=\int\dfq{q}\,K(q,p;{-E_B}) \,
 \Delta_d\!\left({-\eta_dE_B}+q_0,\vq\right)
 \Delta_N\!\left({-\eta_NE_B}-q_0,\vq\right) \, \Bgen(q_0,\vq) \,.
\label{eq:BSE-hom-MS-3}
\end{equation}
Carrying out the $\dd q_0$ integration picks up the residue from the nucleon
propagator pole at $q_0={-\eta_NE_B}-\vq^2/(2\MN)+\ii\eps$.  From the
resulting right-hand side of Eq.~\eqref{eq:BSE-hom-MS-3} we then find that the
function
\begin{equation}
 \Bgen(\vp) \equiv \Bgen\!\left({-\eta_NE_B}-\frac{\vp^2}{2\MN},\vp\right)
\label{eq:BS-amputated-2}
\end{equation}
fulfills the equation
\begin{equation}
 \Bgen(\vp)=\int\dq{q}\,K\!\left(\eta_NE-\frac{\vq^2}{2\MN},\vk,
 \eta_NE-\frac{\vp^2}{2\MN},\vp;E\right)
 \Delta_d\!\left({-E_B}-\frac{\vq^2}{2\MN},\vq\right) \Bgen(\vq) \,.
\label{eq:BSE-hom-MS-amp}
\end{equation}
For future reference we also define the wave function
\begin{equation}
 \phi(\vp) = \int\dn{p}\;\psi(p_0,\vp) \,,
\label{eq:phi}
\end{equation}
for which from Eq.~\eqref{eq:BS-amputated} one immediately finds that
\begin{equation}
 \phi(\vp) = \Delta_d\!\left({-E_B}-\frac{\vp^2}{2\MN},\vp\right)
 \Bgen(\vp) \,.
\label{eq:phi-B}
\end{equation}

\subsection{Operator formalism}
\label{sec:BS-Op}

To see what exactly the factorization Eq.~\eqref{eq:G-factor} implies for the 
$\mathcal{T}$-matrix we now introduce an abstract operator notation.  The 
Bethe--Salpeter equation~\eqref{eq:BS-ms} can be written as
\begin{equation}
 G = G_0 + G K G_0 = G_0 + G_0 K G \,,
\label{eq:BS-op}
\end{equation}
where the middle and the right-hand side are equivalent.\footnote{This can be
seen, for example, by iterating both versions and noting that the results are
the same.}  Assuming again the existence of a bound state with energy 
$E={-E_B}$, we have the factorization
\begin{equation}
 G \sim \ii\frac{\ket{\psi}\bra{\psi}}{E+E_B}\,\mathtext{for}\,E\to{-E_B} \,,
\label{eq:BS-op-pole}
\end{equation}
as given explicitly in Eq.~\eqref{eq:G-factor}. Inserting this into
Eq.~\eqref{eq:BS-op}, multiplying by $(E+E_B)$, and acting on $\ket{\psi}$, we
obtain the homogeneous equation
\begin{equation}
 \ket{\psi} = G_0K\ket{\psi}
\label{eq:BS-op-hom}
\end{equation}
after taking the limit $E\to{-E_B}$ and using that $G_0$ is regular for
$E\to{-E_B}$.  This is, of course, just Eq.~\eqref{eq:BSE-hom-MS-2}.  Note that
all the operators here are in general functions of the total energy, $G=G(E)$,
$K=K(E)$, etc., but that for the sake of notational convenience we have not
written out this dependence explicitly.

\subsection{Lippmann--Schwinger equation}
\label{sec:BS-LS}

Defining the T-matrix operator $T$ via the relation
\begin{equation}
 K G = T G_0 \,,
\label{eq:T}
\end{equation}
we can rewrite Eq.~\eqref{eq:BS-op} in the form
\begin{equation}
 G = G_0 + G_0 T G_0 \,.
\label{eq:G-T}
\end{equation}
Inserting this into both sides of the original Eq.~\eqref{eq:BS-op}, we get
\begin{equation}
 G_0 + G_0 T G_0 = G_0 + G_0 K G_0 + G_0 K G_0 T G_0 \,.
\label{eq:G0-T}
\end{equation}
After cancelling the common term $G_0$ and multiplying through by $G_0^{-1}$ on
both sides, we arrive at the familiar Lippmann--Schwinger equation
\begin{equation}
 T = K + K G_0 T \,.
\label{eq:LS-op}
\end{equation}

The above relations will be used in Sec.~\ref{sec:BS-T-matrix} to find the 
proper factorization of the $\mathcal{T}$-matrix.  First, however, we consider 
again the homogeneous equation and derive the normalization condition in the 
operator formalism.

\subsection{Normalization condition}
\label{sec:BS-Norm}

Going back momentarily to the explicit momentum-space basis we again set 
$P=({-E_B},\vZero)$ and define a reduced two-body propagator $\sktilde{G}$ that 
only depends on the relative three-momenta $\vk$ and $\vp$ by
integration over the energies:
\begin{equation}
 \sktilde{G}(\vk,\vp;{-E_B}) = \int\dn{k}\int\dn{p}\;G(k,p;P) \,.
\end{equation}
By the definition~\eqref{eq:phi}, this implies
\begin{equation}
 \sktilde{G} \sim \ii\frac{\ket{\phi}\bra{\phi}}{E+E_B}
 \mathtext{for} E\to{-E_B} \,.
\label{eq:G-tilde-pole}
\end{equation}
From Eq.~\eqref{eq:BS-op} we get
\begin{equation}
 \sktilde{G} = \sktilde{G}_0+\sktilde{G_0KG} =
 \sktilde{G}_0\left(\one+{\sktilde{G}_0}^{-1}\,\sktilde{G_0KG}\right)
\end{equation}
and hence
\begin{equation}
 {\sktilde{G}}^{-1} = \left(\one+{\sktilde{G}_0}^{-1}
 \,\sktilde{G_0KG}\right)^{-1}{\sktilde{G}_0}^{-1} \,.
\label{eq:Gm1-3dBS}
\end{equation}
We now define
\begin{equation}
 {\sktilde{G}}^{-1} - {\sktilde{G}_0}^{-1} \equiv {-\sktilde{V}}
\label{eq:V-3dBS}
\end{equation}
and find
\begin{equation}
 \sktilde{G}
 = \sktilde{G}\left({\sktilde{G}_0}^{-1}-\sktilde{V}\right)\sktilde{G} \,.
\end{equation}
Inserting Eq.~\eqref{eq:G-tilde-pole} and using l'Hôpital's rule to evaluate 
the limit $E\to{-\EB}$, we readily derive the normalization condition
\begin{equation}
 \ii\bra{\phi}{\frac{\dd}{\dd E}\left({\sktilde{G}_0}^{-1}
 -\sktilde{V}\right)}\ket{\phi}\Big|_{E = -{E_B}}=1 \,,
\label{eq:Norm-3dBS}
\end{equation}
where $\sktilde{V} = {\sktilde{G}_0}^{-1} - {\sktilde{G}}^{-1}$, 
\cf.~Eq.~\eqref{eq:V-3dBS}.  A straightforward calculation shows that
\begin{equation}
 {\sktilde{G}_0}^{-1}(\vk,\vp;E) = (2\pi)^3\vdelta(\vk-\vp)
 \cdot\left[\Delta_d\left(E-\frac{\vp^2}{2\MN},\vp\right)\right]^{-1} \,,
\label{eq:G0m1-3dBS}
\end{equation}
and we also see that $\ket{\Bgen} = {\sktilde{G}_0}^{-1}\ket{\phi}$.  For the
$\sktilde{V}$ defined in Eq.~\eqref{eq:V-3dBS} we get
\begin{equation}
 \sktilde{V} = {\sktilde{G}_0}^{-1} - \left(\one+{\sktilde{G}_0}^{-1}
 \,\sktilde{G_0KG}\right)^{-1}{\sktilde{G}_0}^{-1}
 = {\sktilde{G}_0}^{-1} - \left[\sum\limits_{n=0}^\infty
 \left(-{\sktilde{G}_0}^{-1}\,\sktilde{G_0KG}\right)^n\right]
 {\sktilde{G}_0}^{-1} \,.
\label{eq:V-3dBS-exp}
\end{equation}
Furthermore, iterated application of the Bethe--Salpeter
equation~\eqref{eq:BS-op} yields
\begin{equation}
 \sktilde{G_0KG} = \sktilde{G_0KG_0} + \sktilde{G_0KG_0KG_0} + \cdots \,,
\end{equation}
such that we have a double expansion in Eq.~\eqref{eq:V-3dBS-exp}.  We write
\begin{equation}
 \sktilde{V} \equiv \sum\limits_{n=0}^\infty\sktilde{V}_n
 = \underbrace{{\sktilde{G}_0}^{-1} - {\sktilde{G}_0}^{-1}}_{\sktilde{V}_0=0}
 + \underbrace{{\sktilde{G}_0}^{-1}\sktilde{G_0KG_0}
 {\sktilde{G}_0}^{-1}}_{\sktilde{V}_1} + \cdots \,,
\end{equation}
where the index indicates the number of insertions of $K$.  This expression is 
turns out to be far simpler than it looks since all $\sktilde{V}_n$ with $n>0$ 
actually vanish.  This can be seen by induction if one uses the relation
\begin{equation}
 \sktilde{\cdots G_0}{\sktilde{G}_0}^{-1}\sktilde{G_0\cdots}
 = \sktilde{\cdots G_0\cdots} \,.
\label{eq:tilde-lemma}
\end{equation}
More details about this and a proof of Eq.~\eqref{eq:tilde-lemma} can be found 
in Refs.~\cite{Konig:2011yq,Koenig:2013}.  The final result is simply
\begin{equation}
\begin{split}
 \sktilde{V}(\vk,\vp;E) &= K\!\left(\eta_NE-\frac{\vk^2}{2\MN},\vk,
 \eta_NE-\frac{\vp^2}{2\MN},\vp;E\right) \\
 &= \frac{\ii\MN\yd^2}2\cdot\frac{1}{\vk^2+\vk\cdot\vp+\vp^2-\MN E-\ii\eps} \,.
\end{split}
\label{eq:V-3dBS-final}
\end{equation}

The only remaining step is to derive from Eq.~\eqref{eq:Norm-3dBS} the 
normalization condition for trinucleon wave functions $\Bgen(\vp)$.  This can be 
done most clearly be first establishing the connection between the quantities 
defined here and the $\mathcal{T}$-matrix elements used in the main part of this 
paper.  We are now finally in a position to do that.

\subsection{Factorization of the T-matrix}
\label{sec:BS-T-matrix}

Comparing Eqs.~\eqref{eq:BSE-hom-MS-amp} and~\eqref{eq:V-3dBS-final} with the 
interaction as given in Section~\ref{sec:Nd-IntEq} already suggests that there 
is a direct correspondence between the states $\mathcal{B}(\vp)$ introduced 
here and the trinucleon wave functions of Sec.~\ref{sec:WF}.  For the $T$ 
operator introduced in Eq.~\eqref{eq:T} we find from Eq.~\eqref{eq:G0-T} that
\begin{equation}
 \sktilde{G_0TG_0} = \sktilde{G_0KG_0} + \sktilde{G_0KG_0TG_0} \,.
\label{eq:G0-T-tilde}
\end{equation}
Applying Eq.~\eqref{eq:tilde-lemma} to the second term on the right-hand
side gives
\begin{equation}
 \sktilde{G_0KG_0TG_0}
 = \sktilde{G_0KG_0}{\sktilde{G}_0}^{-1}\sktilde{G_0TG_0} \,.
\end{equation}
Inserting into this the identity in the form
\begin{equation}
 \one = {\sktilde{G}_0}{\sktilde{G}_0}^{-1}
\end{equation}
and multiplying Eq.~\eqref{eq:G0-T-tilde} with ${\sktilde{G}_0}^{-1}$ from both
sides we find that
\begin{equation}
 {\sktilde{G}_0}^{-1}\sktilde{G_0TG_0}{\sktilde{G}_0}^{-1}
 = \left[{\sktilde{G}_0}^{-1}\sktilde{G_0KG_0}{\sktilde{G}_0}^{-1}\right]
 \sktilde{G}_0
 \left[{\sktilde{G}_0}^{-1}\sktilde{G_0TG_0}{\sktilde{G}_0}^{-1}\right] \,,
\end{equation}
where from the discussion in Section~\ref{sec:BS-Norm} we see that the 
interaction is the same as in the normalization condition:
\begin{equation}
 {\sktilde{G}_0}^{-1}\sktilde{G_0KG_0}{\sktilde{G}_0}^{-1} = \sktilde{V} \,.
\end{equation}
Comparing this now with the integral equations in Sec.~\ref{sec:Nd-IntEq}, we 
can conclude that the $\mathcal{T}$-matrix defined there is
\begin{equation}
 \ii\mathcal{T}(E;\vk,\vp)
 = \mbraket{\vk}{{\sktilde{G}_0}^{-1}\sktilde{G_0TG_0}{\sktilde{G}_0}^{-1}}{\vp}
 \,,
\label{eq:T-T}
\end{equation}
where all operators are of course functions of the energy $E$.

\medskip
As $E\to{-E_B}$ we now have, using Eq.~\eqref{eq:G-T} and noting that the
bound state cannot be in $G_0$ since it has to arise from the interaction,
\begin{equation}
 {\sktilde{G}_0}^{-1}\sktilde{G_0TG_0}{\sktilde{G}_0}^{-1} \rightarrow
 {\sktilde{G}_0}^{-1}\sktilde{G}{\sktilde{G}_0}^{-1} + \text{regular terms}
 = \ii\frac{{\sktilde{G}_0}^{-1}\ket{\phi}\bra{\phi}{\sktilde{G}_0}^{-1}}
 {E+E_B} + \text{regular terms} \,,
\end{equation}
where the second identity follows from Eq.~\eqref{eq:G-tilde-pole}.  Now,
according to Eqs.~\eqref{eq:phi-B} and~\eqref{eq:G0m1-3dBS} we have
\begin{equation}
 \ket{\phi} = \sktilde{G}_0 \ket{\mathcal{B}}
 \iff \ket{\mathcal{B}} = \sktilde{G}_0^{-1} \ket{\phi} \,,
\end{equation}
which implies
\begin{equation}
 \bra{\mathcal{B}} = \bra{\phi}\big(\sktilde{G}_0^{-1}\big)^\dagger \,.
\end{equation}
But---up to a delta function---$\sktilde{G}_0^{-1}$ is just the deuteron
propagator $\Delta_d$ which, from Eq.~\eqref{eq:Prop-d-High} we find to be a 
purely imaginary quantity.\footnote{Note the overall $\ii$ in the prefactor in
Eq.~\eqref{eq:Prop-d-High} and that remaining terms are real for $p_0<0$ and 
$\eps\to0$.}  Hence,
\begin{equation}
 \big(\sktilde{G}_0^{-1}\big)^\dagger = -\sktilde{G}_0^{-1} \,,
 \mathtext{for} E<0
\end{equation}
and
\begin{equation}
 {\sktilde{G}_0}^{-1}\sktilde{G_0TG_0}{\sktilde{G}_0}^{-1}
 = -\ii\frac{\ket{\mathcal{B}}\bra{\mathcal{B}}}{E+E_B}
 + \text{regular terms} \mathtext{as} E\to{-E_B} \,.
\end{equation}
For the $\mathcal{T}$-matrix we then find from Eq.~\eqref{eq:T-T} that
\begin{equation}
 \mathcal{T}(E;\vk,\vp) = -\frac{\mathcal{B}^\dagger(\vk)\,\mathcal{B}(\vp)}
 {E+E_B} + \text{regular terms} \mathtext{as} E\to{-E_B} \,.
\end{equation}
Finally, for the normalization condition~\eqref{eq:Norm-3dBS} written in terms
of the states $\ket{\mathcal{B}}$ we analogously find
\begin{equation}
 -\ii\bra{\mathcal{B}}\sktilde{G}_0
 \left[{\frac{\dd}{\dd E}\left({\sktilde{G}_0}^{-1}-\sktilde{V}\right)}\right]
 \!\sktilde{G}_0\ket{\mathcal{B}}\Big|_{E={-E_B}}=1 \,.
\label{eq:Norm-3dBS-B}
\end{equation}
This can be directly translated into Eq.~\eqref{eq:Triton-WF-norm} in 
Sec.~\ref{sec:WF} by considering the proper multi-channel formalism, with the 
${\sktilde{G}_0}^{-1}$ going over into the diagonal matrix of inverse 
propagators $\hat{I}$ defined below Eq.~\eqref{eq:Triton-WF-norm}.  To sort out 
the prefactors, \cf~Eq.~\eqref{eq:D-d-t-LO} and note that $\sktilde{V}$---via 
$K$ as given in Eq.~\eqref{eq:1N-exchange}---contains a factor $(-\ii)$.  
Finally, upon S-wave projection of all quantities, the operator products (in 
momentum space) become
\begin{equation}
 AB = \int\dq{q}\,A(\ldots,\vq)B(\vq,\ldots)
 \longrightarrow \frac1{2\pi^2}\int\dd q\,q^2 \,A(\ldots,q)B(q,\ldots) \,,
\end{equation}
which is exactly the operation ``$\otimes$'' that appears in 
Eq.~\eqref{eq:Triton-WF-norm}.

\section{Partially screened Coulomb T-matrix}
\label{sec:CoulombT-screened}

In this section we discuss an approximate expression for the full off-shell 
Coulomb T-matrix with Yukawa (photon-mass) screening, based on results 
originally derived by Gorshkov in the 
1960s~\cite{Gorshkov:1961ab,Gorshkov:1965ab}.
  
\subsection{Unscreened Coulomb interaction}

Closed expressions for the full off-shell Coulomb T-matrix have been known for 
a long time (see, for example, the review by Chen and Chen~\cite{Chen:1972ab}).
We consider here a quantum-mechanical two-particle system with reduced mass 
$\mu$ and charges $Z_{1,2}$.  By writing the Coulomb interaction as an operator 
$\hat{V}_C$ with
\begin{align}
 \mbraket{\vecr}{\hat{V}_C}{\vecr'}
 &= \vdelta(\vecr-\vecr') V_C(\vecr)
 \mathtext{with} V_C(\vecr)=V_C(r)\equiv\frac{\alpha Z_1Z_2}{r} \,, \\
 \mbraket{\vecp}{\hat{V}_C}{\vecq}
 &= \frac{4\pi\alpha Z_1Z_2}{(\vecp-\vecq)^2} \equiv V_C(\vecp,\vecq) \,,
\end{align}
the Lippmann-Schwinger equation for the Coulomb T-matrix $T_C$ reads
\begin{equation}
 \hat{T}_C(E) = \hat{V}_C + \hat{V}_C\,\hat{G}_0^{(+)}(E)\,\hat{T}_C(E) \,,
\label{eq:LS-C}
\end{equation}
where the energy $E$ is a free (complex) parameter and
\begin{equation}
 G_0^{(+)}(E;\vecq,\vecq') = \mbraket{\vecq}{\hat{G}_0^{(+)}(E)}{\vecq'}
 = \frac{(2\pi)^3\vdelta(\vecq-\vecq')}{E-{q^2}/{(2\mu)}+\ii\eps} \,.
\end{equation}
Introducing the (complex) momentum scale $k$ via $E={k^2}/{(2\mu)}$, the 
explicit solution of Eq.~\eqref{eq:LS-C} can be written in the Hostler
form~\cite{Chen:1972ab,Hostler:1964aa,Hostler:1964ab} as
\begin{equation}
 T_C(k;\vecp,\vecq) = V_C(\vecp,\vecq) \left\{1-2\ii\eta
 \int_1^\infty\left(\frac{s+1}{s-1}\right)^{\!\!-\ii\eta}
 \frac{\dd s}{s^2-1-\epsilon} \right\} \,,
\label{eq:TC-int}
\end{equation}
where
\begin{equation}
 \eta = \frac{\alpha\mu Z_1Z_2}{k}
\end{equation}
and
\begin{equation}
 \epsilon = \frac{(p^2-k^2)(q^2-k^2)}{k^2(\vecp-\vecq)^2} \,.
\end{equation}
This integral representation is what is used to include Coulomb effects in the 
$^3\mathrm{He}$ calculation of Ando and Birse~\cite{Ando:2010wq}. 
Alternatively, it can be recast in terms of hypergeometric functions as
\begin{multline}
 T_C(k;\vecp,\vecq) = V_C(\vecp,\vecq) \Bigg\{1-\Delta^{-1}\Big[
 {_2F_1}\!\left(1,\ii\eta,1+\ii\eta;\frac{\Delta-1}{\Delta+1}\right) \\
 -{_2F_1}\!\left(1,\ii\eta,1+\ii\eta;\frac{\Delta+1}{\Delta-1}\right)
 \Big]\Bigg\} \,,
\label{eq:TC-hyp}
\end{multline}
with the new variable $\Delta$ defined via
\begin{equation}
 \Delta^2 = 1 + \epsilon \,.
\label{eq:Delta}
\end{equation}

\subsection{Yukawa screening}

As noted in Ref.~\cite{Ando:2010wq}, the unscreened Coulomb T-matrix certainly 
cannot be used in the scattering regime.  We thus consider here an exponential 
(Yukawa) screening in configuration space,
\begin{equation}
 V_{C}(\vecr) \longrightarrow V_{C,\lambda}(\vecr)
 = \alpha Z_1Z_2\frac{\ee^{-\lambda r}}{r} \,,
\end{equation}
corresponding to the introduction of a photon mass $\lambda$ in momentum space:
\begin{equation}
 V_{C,\lambda}(\vecp,\vecq)
 = \frac{4\pi\alpha Z_1Z_2}{(\vecp-\vecq)^2+\lambda^2} \,.
\label{eq:V-Yuk}
\end{equation}

Ref.~\cite{Chen:1972ab} gives an expression for what we in the following
call the ``partially screened'' Coulomb T-matrix $\hat{T}_{C,\lambda}$,
originally derived by Gorshkov~\cite{Gorshkov:1965ab}.  It is defined by the
relation
\begin{equation}
 \hat{T}_{C,\lambda} = \hat{V}_{C,\lambda}
 + \hat{V}_{C,\lambda}\,\hat{G}_0^{(+)}\,\hat{T}_C \,,
\label{eq:LS-C-lambda}
\end{equation}
where we have not written out the energy dependence to simplify the notation.

\medskip
Note that Eq.~\eqref{eq:LS-C-lambda} is not a Lippmann--Schwinger equation because 
the operator that appears on the right-hand side is the unscreened Coulomb 
T-matrix $\hat{T}_C$.  Still, $\hat{T}_{C,\lambda}$ is an interesting object to 
study because it can be written down as a closed expression that converges to 
the unscreened Coulomb T-matrix $\hat{T}_C$ in the limit $\lambda\to0$.  Due to 
this property it should be useful in numerical calculations where one wants to 
include nonperturbative Coulomb effects in the scattering regime.  Ideally, one 
would of course like to use an expression for the exact Yukawa T-matrix in such 
an approach, but no closed solution for that quantity is known so far.  We thus 
propose to use $\hat{T}_{C,\lambda}$ as a pragmatic alternative.  Since it has 
the right behavior in the limit $\lambda\to0$, we expect it to adequately 
describe most of the nonperturbative Coulomb effects in that limit.

\medskip
Unfortunately, the expression given for $T_{C,\lambda}(k;\vecp,\vecq)$ in
Eqs.~(246) and (247) of Ref.~\cite{Chen:1972ab} is not fully
correct.\footnote{This can be seen by a straightforward dimensional analysis of
Eq.~(247) in Ref.~\cite{Chen:1972ab}.  Furthermore, the prefactor in Eq.~(246)
is written in terms of the unscreened Coulomb potential, which is clearly not
correct.}  Since in the original paper by Gorshkov~\cite{Gorshkov:1965ab} the
limit $\lambda\to0$ is taken without first giving the explicit form of the
partially screened T-matrix, we will derive it here in the following.

To this end we start from Eq.~(244) of Ref.~\cite{Chen:1972ab}, which in our
notation reads
\begin{multline}
 T_{C,\lambda}(k;\vecp,\vecq) \\ = V_{C,\lambda}(\vecp,\vecq)
 - \ii\eta\int_0^1 \frac{\dd x}{\Lambda_0(x)}
 V_{C,\lambda-\ii k \Lambda_0(x)}(x\vecp,\vecq)
 \times\exp\left\{-\ii\eta\int_x^1\frac{\dd x_1}{x_1\Lambda_0(x_1)}\right\} \,,
\label{eq:T-lambda-0-bare}
\end{multline}
with $\Lambda_0(x)$ defined as the positive root of
\begin{equation}
 \Lambda_0^2(x) = \left[1-(p/k)^2x\right](1-x) \,.
\end{equation}
$V_{C,\lambda-\ii k \Lambda_0(x)}$ is just the Yukawa 
potential~\eqref{eq:V-Yuk} with the substitution
$\lambda\longrightarrow\lambda-\ii k \Lambda_0(x)$.  To simplify the integral 
in Eq.~\eqref{eq:T-lambda-0-bare} we write~\cite{Gorshkov:1965ab}
\begin{equation}
 x = \frac{s^2-1}{s^2-(p/k)^2} \mathtext{,}
 \frac{\dd x}{\dd s} = \frac{2s\,\big(1-(p/k)^2\big)}{\big(s^2-(p/k)^2\big)^2}
\end{equation}
and find that
\begin{equation}
 \Lambda_0(x) = s(1-x) \,.
\end{equation}
Since furthermore
\begin{equation}
 1-x = \frac{1-(p/k)^2}{s^2-(p/k)^2} \,,
\end{equation}
the integral in the exponent is just
\begin{equation}
 \int_x^1\frac{\dd x_1}{x_1\Lambda_0(x_1)}
 = \int_s^\infty\frac{2\,\dd s_1}{s_1^2-1}
 = \log\!\Big(\frac{s+1}{s-1}\Big) \,.
\end{equation}
For the potential term under the integral we get
\begin{multline}
 V_{C,\lambda-\ii k \Lambda_0(x)}(x\vecp,\vecq)
 = \frac{2\pi\gamma}{\mu} \\
 \times \frac{(s^2-(p/k)^2)}{(s^2-1)(\vecq-\vecp)^2
 +\dfrac1{k^2}\Big[\lambda^2(k^2s^2-p^2)-2\ii\lambda ks(k^2-p^2)
 -(k^2-q^2)(k^2-p^2)\Big]}
\end{multline}
after a lengthy but straightforward calculation.  Adding
\begin{equation}
 0 = \lambda^2(s^2-1) - \lambda^2s^2 + \lambda^2
\end{equation}
in the denominator, we can rewrite this as
\begin{multline}
 V_{C,\lambda-\ii k \Lambda_0(x)}(x\vecp,\vecq) = \frac{2\pi\gamma}{\mu} \\
 \times \frac{(s^2-(p/k)^2)}{(s^2-1)\big[(\vecq-\vecp)^2+\lambda^2\big]
 -\dfrac1{k^2}\Big[(k^2-q^2)(k^2-p^2)\Big]
 +\dfrac1{k^2}\Big[(k^2-p^2)(\lambda^2-2\ii\lambda ks)\Big]} \,.
\end{multline}
Finally, noting that the term in the numerator cancels against the same factor
in
\begin{equation}
 \frac{\dd x}{\Lambda_0(x)} = \frac{2\,\dd s}{s^2-(p/k)^2} \,,
\end{equation}
and factoring out the Yukawa potential, we arrive at
\begin{equation}
 T_{C,\lambda}(k;\vecp,\vecq) = V_{C,\lambda}(\vecp,\vecq) \left\{1-2\ii\eta
 \int_1^\infty\left(\frac{s+1}{s-1}\right)^{\!\!-\ii\eta}
 \frac{\dd s}{s^2-1-\epsilon_\lambda+\zeta_\lambda(s)}\right\}
\label{eq:T-lambda-0-int}
\end{equation}
with
\begin{equation}
 \epsilon_\lambda = \frac{(k^2-p^2)(k^2-q^2)}
 {k^2\left[(\vecq-\vecp)^2+\lambda^2\right]}
\label{eq:epsilon-lambda}
\end{equation}
and
\begin{equation}
 \zeta_\lambda(s) = \frac{(k^2-p^2)(\lambda^2-2\ii\lambda ks)}
 {k^2\left[(\vecq-\vecp)^2+\lambda^2\right]} \,.
\label{eq:zeta-lambda}
\end{equation}
This expression is very similar to the integral form of the unscreened
Coulomb T-matrix.  The only differences are given by the new term
$\zeta_\lambda(s)$ in the denominator and the fact that all singularities, both
in the overall prefactor and under the integral, are now regulated by adding
$\lambda^2$.  In fact, one directly sees that in the limit $\lambda\to0$,
Eq.~\eqref{eq:T-lambda-0-int} converges to the unscreened expression given in
Eq.~\eqref{eq:TC-int}.

\subsection{Expression in terms of hypergeometric functions}
\label{sec:CoulombT-screened-hyp}

Something that is not noted in Refs.~\cite{Chen:1972ab}
and~\cite{Gorshkov:1965ab} is that---just like the unscreened Coulomb
T-matrix---$T_{C,\lambda}(k;\vecp,\vecq)$ can also be expressed in terms of
hypergeometric functions.  To obtain this expression we first note that the
denominator in Eq.~\eqref{eq:T-lambda-0-int} can be written as
\begin{equation}
 s^2-1-\epsilon_\lambda+\zeta_\lambda(s)
 = s^2 - (1 + d_1 \cdot D^2) - (d_2 \cdot D^2)s
\end{equation}
with
\begin{equation}
 d_1 = k^2 - q^2 - \lambda^2 \mathtext{,} d_2 = 2\ii\lambda k
\end{equation}
and
\begin{equation}
 D^2 = \frac{k^2-p^2}{k^2\left[(\vecq-\vecp)^2+\lambda^2\right]} \,.
\end{equation}
After making the transformation
\begin{equation}
 s = \frac{t+1}{t-1} \mathtext{,}
 \int_1^\infty \dd s = 2\int_1^\infty\frac{\dd t}{(t-1)^2}
\end{equation}
we get
\begin{multline}
 T_{C,\lambda}(k;\vecp,\vecq) = V_{C,\lambda}(\vecp,\vecq) \\
 \times \Big\{1 - 4\ii\eta  \int_1^\infty\frac{t^{-\ii\eta}\,\dd t}
 {-D^2(d_1+d_2)\,t^2+(4+2D^2\cdot d_1)\,t - D^2(d_1-d_2)} \Big\} \,.
\label{eq:T-lambda-0-int-t}
\end{multline}
To proceed further, we use the indefinite integral\footnote{This result has been
obtained with the help of a computer algebra software (Wolfram Mathematica).}
\begin{multline}
 \int \frac{t^\nu\,\dd t}{x_2t^2+x_1t+x_0}
 = \frac{1}{X_1}\frac{2^{-\nu}\,t^\nu}{\nu} \Bigg\{
 {_2F_1}\big(-\nu,-\nu;1-\nu;X_2^+(t)\big)\cdot X_3^+(t)^{-\nu} \\
 - {_2F_1}\big(-\nu,-\nu;1-\nu;X_2^-(t)\big)\cdot X_3^-(t)^{-\nu}\Bigg\} \,,
\label{eq:2F1-int-x-123}
\end{multline}
where
\begin{equation}
 X_1 = \sqrt{x_1^2-4x_0x_2} \mathtext{,}
 X_2^\pm(t) = \frac{x_1 \mp X_1}{x_1 + 2tx_2 \mp X_1} \mathtext{,}
 X_3^\pm(t) = \frac{tx_2}{x_1 + 2tx_2 \mp X_1} \,.
\label{eq:X-123}
\end{equation}

Evaluating this at $t=1$ is straightforward, but considering $t\to\infty$
requires a little more care.  From Eq.~\eqref{eq:X-123} one sees that
$X_2^\pm(t)$ goes to zero like $1/t$ as $t\to\infty$, such that the
hypergeometric functions simply yield one in this limit.  Since the potentially
problematic (because $\nu={-\ii\eta}$) prefactor $t^\nu$ is cancelled by the
numerator of $X_3^\pm(t)^{-\nu}$ with the remainder then going to zero as
$t\to\infty$, we can conclude that there is actually no contribution to the
integral from the upper boundary in Eq.~\eqref{eq:T-lambda-0-int-t} and that its
value is hence given by the right-hand side of Eq.~\eqref{eq:2F1-int-x-123} with
$t=1$.

Before inserting this into Eq.~\eqref{eq:T-lambda-0-int-t}, we subsequently
apply the identities~\cite{Abramowitz:1984}
\begin{equation}
 {_2F_1}(a,b;c;z) = (1-z)^{c-a-b}\,{_2F_1}(c-a,c-b;c;z)
\end{equation} 
and
\begin{equation}
 {_2F_1}(a,b;c;z) = (1-z)^{-a}\,{_2F_1}\!\left(a,c-b;c;\frac{z}{z-1}\right)
\end{equation} 
to rewrite 
\begin{multline}
 {_2F_1}(-\nu,-\nu;1-\nu;z) = (1-z)^{1+\nu}\,{_2F_1}(1,1;1-\nu;z) \\
 = (1-z)^\nu\,{_2F_1}\!\left(1,-\nu;1-\nu;\frac{z}{z-1}\right) \,.
\end{multline}
This is useful because from Eq~\eqref{eq:X-123} one finds that for
$z=X_2^\pm(t)$,
\begin{equation}
 (1-z)^\nu = 2^\nu X_3^\pm(t)^\nu \,,
\end{equation}
canceling the inverse factors of this in Eq.~\eqref{eq:2F1-int-x-123}. 
Moreover, the arguments simplify to
\begin{equation}
 \frac{z}{z-1} = -\frac12\frac{X_2^\pm(t)}{X_3^\pm(t)}
 = -\frac{x_1\mp X_1}{2t\,x_2} \,.
\end{equation}
With this, we then have
\begin{multline}
 \int_1^\infty \frac{t^\nu\,\dd t}{x_2t^2+x_1t+x_0}
 = \frac{1}{\nu X_1}\Bigg\{
 {_2F_1}\bigg(1,-\nu;1-\nu;-\frac{x_1+X_1}{2x_2}\bigg) \\
 - {_2F_1}\bigg(1,-\nu;1-\nu;-\frac{x_1-X_1}{2x_2}\bigg)\Bigg\} \,.
\label{eq:2F1-int-x-123-new}
\end{multline}

Finally, applying the above result to Eq.~\eqref{eq:T-lambda-0-int-t}, we can
write the partially-screened Coulomb T-matrix as
\begin{multline}
 T_{C,\lambda}(k;\vecp,\vecq) \\ = V_C(\vecp,\vecq) \Big\{1-\Delta_\lambda^{-1}
 \big[{_2F_1}\!\left(1,\ii\eta,1+\ii\eta;X_\lambda^-\right)
 -{_2F_1}\!\left(1,\ii\eta,1+\ii\eta;X_\lambda^+\right)
 \big]\Big\} \,,
\label{eq:TC-screened-hyp}
\end{multline}
with
\begin{equation}
 \Delta_\lambda^2 = 1 + \frac{(k^2-p^2)(k^2-q^2-\lambda^2)}
 {k^2\left[(\vecq-\vecp)^2+\lambda^2\right]} - \frac{\lambda^2(k^2-p^2)^2}
 {k^2\left[(\vecq-\vecp)^2+\lambda^2\right]^2}
\label{eq:Delta-lambda}
\end{equation}
and
\begin{equation}
 X_\lambda^\pm = \frac{2k^2\left[(\vecq-\vecp)^2+\lambda^2\right]
 (1\pm\Delta_\lambda) + (k^2-p^2)(k^2-q^2-\lambda^2)}
 {(k^2-p^2)\left[(k+\ii\lambda)^2-q^2\right]} \,.
\label{eq:X-pm}
\end{equation}
As it should, this reduces to the hypergeometric expression~\eqref{eq:TC-hyp}
for the unscreened Coulomb T-matrix in the limit $\lambda\to0$.  It is directly
clear from Eqs.~\eqref{eq:Delta-lambda} and~\eqref{eq:Delta} that
\begin{equation}
 \lim\nolimits_{\lambda\to0} \Delta_\lambda^2 = \Delta^2 \,,
\end{equation}
and a straightforward calculation then furthermore shows that
\begin{equation}
 \lim\nolimits_{\lambda\to0} X_\lambda^\pm = \frac{\Delta\pm1}{\Delta\mp1} \,.
\end{equation}

\end{document}